\documentclass[11pt]{article} 

\usepackage{amsthm,amsmath,natbib}
\RequirePackage[colorlinks,citecolor=blue,urlcolor=blue]{hyperref}
\usepackage{amsfonts}
\usepackage{amssymb}
\usepackage{color}
\usepackage[margin=1.2in]{geometry}
\usepackage{graphicx}
\usepackage{url}

\newcommand{\RR}{\mathbb{R}}
\DeclareMathOperator*{\argmin}{arg\,min}
\newtheorem{proposition}{Proposition} 

\newcommand{\tr}{\textrm{\bf tr }}
\newcommand{\Wr}{\hat{W}_r}
\newcommand{\Wtwo}{\hat{W}_2}

\begin{document}

\title{Correcting gene expression data when neither the unwanted
  variation nor the factor of interest are observed}

\author{Laurent Jacob$^1$\\
  \texttt{laurent@stat.berkeley.edu}
  \and Johann Gagnon-Bartsch$^1$\\
  \texttt{johann@berkeley.edu}
  \and Terence P. Speed$^{1,2}$\\
  \texttt{terry@wehi.edu.au}\\
  $^1$Department of Statistics, University of California, Berkeley, USA\\
  $^2$Division of Bioinformatics, WEHI, Melbourne, Australia}

\maketitle

\begin{abstract}
  When dealing with large scale gene expression studies, observations
  are commonly contaminated by unwanted variation factors such as
  platforms or batches. Not taking this unwanted variation into
  account when analyzing the data can lead to spurious associations
  and to missing important signals. When the analysis is unsupervised,
  \emph{e.g.} when the goal is to cluster the samples or to build a
  corrected version of the dataset --- as opposed to the study of an
  observed factor of interest --- taking unwanted variation into
  account can become a difficult task. The unwanted variation factors
  may be correlated with the unobserved factor of interest, so that
  correcting for the former can remove the latter if not done
  carefully. We show how negative control genes and replicate samples
  can be used to estimate unwanted variation in gene expression, and
  discuss how this information can be used to correct the expression
  data or build estimators for unsupervised problems. The proposed
  methods are then evaluated on three gene expression datasets. They
  generally manage to remove unwanted variation without losing the
  signal of interest and compare favorably to state of the art
  corrections.
\end{abstract}

\section{Introduction}
\label{sec:intro}

Over the last few years, microarray-based gene expression studies
involving a large number of samples have been carried
out~\citep{Cardoso2007MINDACT,Network2008Comprehensive}, with the hope
to help understand or predict some particular \emph{factors of
  interest} like the prognosis or the subtypes of a cancer. Such large
gene expression studies are often carried out over several years, may
involve several hospitals or research centers and typically contain
some \emph{unwanted variation} (UV) \emph{factors}. These factors can
arise from technical elements such as batches, different technical
platforms or laboratories, or from biological signals which are
unrelated to the factor of interest of the study such as heterogeneity
in ages or different ethnic groups. They can easily lead to spurious
associations. When looking for genes which are differentially
expressed between two subtypes of cancer, the observed differential
expression of some genes could actually by caused by laboratories if
laboratories are partially confounded with subtypes. When doing
clustering to identify new subgroups of the disease, one may actually
identify any of the UV factors if their effect on gene expression is
stronger than the subgroup effect. If one is interested in predicting
prognosis, one may actually end up predicting whether the sample was
collected at the beginning or at the end of the study because better
prognosis patients were accepted at the end of the study. In this
case, the classifier obtained would have little value for predicting
the prognosis of new patients outside the study. Similar problems
arise when doing a meta-analysis, \emph{i.e.}, when trying to combine
several smaller studies rather than working on a large heterogeneous
study~: in a dataset resulting from the merging of several studies the
strongest effect one can observe is generally related to the
membership of samples to different studies.

A very important objective is therefore to remove these UV factors
without losing the factor of interest. The problem can be more or less
difficult depending on what is actually considered to be observed and
what is not. If both the factor of interest and all the UV factors
(say technical batches or different countries) are known, the problem
essentially boils down to a linear regression~: the expression of each
gene is decomposed as an effect of the factor of interest plus an
effect of the unwanted variation factor. When the variance of each
gene is assumed to be different within each batch, this leads to
so-called location and scale adjustments such as used in
dChip~\citep{Li2003DNA}. \cite{Johnson2007Adjusting} shrink the
unwanted variation and variance of all genes within each batch using
an empirical Bayes method. This leads to the widely used ComBat method
which generally give good performances in this
case. \cite{Walker2008Empirical} propose a version of ComBat which
uses replicate samples to estimate the batch effect. When replicate
samples are available an alternative to centering, known as
ratio-based method~\citep{Luo2010comparison} is to remove the average
of the replicate samples within each batch rather than the average of
all samples. Assuming that the factor of interest is associated with
the batch, this should ensure that centering the batches does not
remove the signal associated with the factor of interest. Note that
this procedure is closely related to some of the methods we introduce
in this paper.

Of course there is always a risk that some unknown factors also
influence the gene expression. Furthermore it is sometimes better when
tackling the problem with linear models to consider UV factors which
are actually known as unknown. Their effect may be strongly non-linear
because they don't affect all samples the same way or because they
interact with other factors, in which cases modeling them as known may
give poor results. When the UV factors are modeled as unknown, the
problem becomes more difficult because one has to estimate UV factors
along with their effects on the genes and that many estimates may
explain the data equally well while leading to very different
conclusions. ICE~\citep{Kang2008Accurate} models unwanted variation as
the combination of an observed fixed effect and an unobserved random
effect. The covariance of the random effect is taken to be the
covariance of the gene expression matrix. The risk of this approach is
that some of the signal associated with the signal of interest may be
lost because it is included in the covariance of the gene expression
matrix. SVA~\citep{Leek2007Capturing} addresses the problem by first
estimating the effect of the factor of interest on each gene then
doing factor analysis on the residuals, which intuitively can give
good results as long as the UV factors are not too correlated with the
factor of interest. \cite{Teschendorff2011Independent} propose a
variant of SVA where the factor analysis step is done by independent
component analysis~\citep{Hyvärinen2001Independent} instead of
SVD. The same model as \cite{Leek2007Capturing} is considered in a
recent contribution of \cite{Gagnon-Bartsch2012Using} coined RUV-2,
which proposes a general framework to correct for UV in microarray
data using negative \emph{control genes}. These genes are assumed not
to be affected by the factor of interest and used to estimate the
unwanted variation component of the model. They apply the method to
several datasets in an extensive study and show its very good behavior
for differential analysis, in particular comparable performances to
state of the art methods such as ComBat~\citep{Johnson2007Adjusting}
or SVA~\citep{Leek2007Capturing}. \cite{Sun2012Multiple} recently
proposed LEAPS, which estimates the parameters of a similar model in
two steps~: first the effect of unwanted variation is estimated by SVD
on the data projected along the factor of interest, then the unwanted
variation factors responsible for this effect and the effect of the
factor of interest are estimated jointly using an iterative coordinate
descent scheme. A sparsity-inducing penalty is added to the effect of
the factor of interest in order to make the model
identifiable. \cite{Yang2012Accounting} adopt a related approach~:
they also use the sparsity-inducing penalty, do not have the
projection step and relax the rank constraint to a trace constraint
which makes the problem jointly convex in the unwanted variation and
effect of the factor of interest. \cite{Listgarten2010Correction}
model the unwanted variation as a random effect term, like ICE. The
covariance of the random effect is estimated by iterating between a
maximization of the likelihood of the factor of interest (fixed
effect) term for a given estimate of the covariance and a maximization
of the likelihood of the covariance for a given estimate of the fixed
effect term. This is also shown to yield better results than ICE and
SVA.

Finally when the factor of interest is not observed, the problem is
even more difficult. It can occur if one is interested in unsupervised
analyses such as PCA or clustering. Suppose indeed that one wants to
use a large study to identify new cancer subtypes. If the study
contains several technical batches, includes different platforms or
different labs or any unknown factor, the samples may cluster
according to one of these factors hence defeating the purpose of using
a large set of samples to identify more subtle subtypes. One may also
simply want to ``clean'' a large dataset from its UV without knowing
in advance which factors of interest will be studied. Accordingly in
the latter case, any knowledgeable person may want to start form the
raw data and use the factor of interest once it becomes known to
remove UV. \cite{Alter2000Singular,Nielsen2002Molecular} use SVD on
gene expression to identify the UV factors without requiring the
factor of interest, \cite{Price2006Principal} do so using axes of
principal variance observed on SNP data. These approaches may work
well in some cases but relies on the prior belief that all UV factors
explain more variance than any factor of interest. Furthermore it will
fail if the UV factors are too correlated with the factor of
interest. If the factor of interest is not observed but the unwanted
variation factor is assumed to be an observed batch, an alternative
approach is to project the data along the batch factors, equivalently
to center the data by batch. This is conceptually similar to using one
of the location and scale adjustment methods such as~\cite{Li2003DNA}
or \cite{Johnson2007Adjusting} without specifying the factor of
interest. \cite{Benito2004Adjustment,Marron2007Distance} propose a
distance weighted discrimination (DWD) method which uses a supervised
learning algorithm to finds a hyperplane separating two batches and
project the data on this hyperplane.  As with the dChip and ComBat
approaches, assuming that the unwanted variation is a linear function
of the observed batch may fail if other unwanted variation factors
affect gene expression or if the effect of the batch is a more
complicated --- possibly non-linear --- function, or involves
interaction with other factors. In addition, like the SVD approach,
these projections may subsequently lead to poor estimation of the
factor of interest if it is correlated with the batch effect~: if one
of the batches contains most of one subtype and the second batch
contains most of the other subtype the projection step removes a large
part of the subtype signal. Finally,
Oncomine~\citep{Rhodes2004ONCOMINE,Rhodes2004Large-scale,Rhodes2007Oncomine}
regroups a large number of gene expression studies which are processed
by median centering and normalizing the standard deviation to one for
each array. This processing does not explicitly take into account a
known unwanted variation factor or try to estimate it. It removes
scaling effects, \emph{e.g.} if one dataset or part of a dataset has
larger values than others, but it does not correct for multivariate
behaviors such as the linear combination of some genes is larger for
some batch. On the other hand, it does not run the risk to remove
biological signal of this form.

The contribution of \cite{Gagnon-Bartsch2012Using} suggests that
negative controls can be used to estimate and remove efficiently
sources of unwanted variation. Our objective here is to propose ways
to improve estimation in the unsupervised case, \emph{i.e}, when the
factor of interest is not observed. We use a random effect to model
unwanted variation. As we discuss in the paper, this choice is crucial
to obtain good estimators when the factor of interest is not
observed. In this context, two main difficulties arise~:
\begin{itemize}
\item In the regression case, and assuming the covariance of the
  random term is known, introducing such a random term simply turns
  the ordinary least square problem into a generalized least square
  one for which a closed form solution is also available. For some
  unsupervised estimation problems --- such as clustering --- dealing
  with the random effects can be more difficult~: the dedicated
  algorithm --- such as $k$-means --- may not apply or adapt easily to
  the adding of a random effect term. Our first contribution is to
  discuss general ways of adapting dedicated algorithms to the
  presence of a random effect term.
\item As always when using random effects, a major difficulty is to
  estimate the covariance of the random term. Following
  \cite{Gagnon-Bartsch2012Using}, we can use negative control genes to
  estimate this covariance, but this amounts to assuming that the
  expression of these genes is really not influenced by the factor of
  interest. A second contribution of this paper is to propose new ways
  to estimate the unwanted variation factors using \emph{replicate
    arrays}. Replicate arrays correspond to the same biological
  sample, but typically differ for some unwanted variation
  factors. For example in a dataset involving two platforms, some
  samples could be hybridized on the two platforms. More generally in
  a large study, some control samples or mixtures of RNAs may be
  re-hybridized at several times of the study, under different
  conditions. We still don't know what the factor of interest is, but
  we know it takes the same value for all replicates of the same
  sample. Changes among these replicates are therefore only caused by
  unwanted variation factors, and we intend to use this fact to help
  identify and remove this unwanted variation.
\end{itemize}
A third contribution of this paper is to assess how various correction
methods, including the ones we introduce, perform in an unsupervised
estimation task on three gene expression datasets subject to unwanted
variation. We show that the proposed methods generally allow us to
solve the unsupervised estimation problem, with no explicit knowledge
of the unwanted variation factors.

Section~\ref{sec:fixedRuv} recalls the model and estimators used in
\cite{Gagnon-Bartsch2012Using}. Section~\ref{sec:unsupRand} presents
our methods addressing the case where the factor of interest is
unobserved using a random effect term. We first make the assumption
that the unwanted variation factors, or equivalently the covariances
of the random terms are known. Section~\ref{sec:Westimation} discusses
how estimation of this unwanted variation factor can be improved. In
Section~\ref{sec:result}, the performance of the proposed methods are
illustrated on three gene expression datasets~: the gender data of
\cite{Vawter2004Gender} which were already used to evaluate the RUV-2
method of \cite{Gagnon-Bartsch2012Using}, the TCGA glioblastoma
data~\citep{Network2008Comprehensive} and MAQC-II data from rat liver
and blood~\citep{Lobenhofer2008Gene,Fan2010Consistency}. We finish
with a discussion in Section~\ref{sec:discussion}.

\paragraph{Notation}

For any matrix $A \in \RR^{m,n}$, $\|A\|_F \stackrel{\Delta}{=}
\left(\sum_{i=1}^m\sum_{j=1}^ na_{ij}^2\right)^{\frac{1}{2}}$ denotes
the Frobenius norm of $A$. In addition for a positive definite matrix
$B \in \RR^{m,m}$ we define $\|A\|_B \stackrel{\Delta}{=}
\|B^{-\frac{1}{2}}A\|_F$. We refer to the problem where the factor of
interest is observed as the \emph{supervised} problem, and to the
problem where it is unobserved as the \emph{unsupervised} problem.

\section{Existing fixed effect estimators~: RUV-2 and naive RUV-2}
\label{sec:fixedRuv}

The RUV (Removal of Unwanted Variation) model used by
\cite{Gagnon-Bartsch2012Using} was a linear model, with one term
representing the effect of the factors of interest on gene expression
and another term representing the effect of the unwanted variation
factors~:
\begin{equation}
  \label{eq:ruv}
  Y = X\beta + W\alpha + \varepsilon,
\end{equation}
with $Y \in\RR^{m\times n}$, $X\in\RR^{m\times p}$,
$\beta\in\RR^{p\times n}$,
$W\in\RR^{m\times k}$, $\alpha\in\RR^{k\times n}$ and
$\varepsilon\in\RR^{m\times n}$. $Y$
is the observed matrix of expression of $n$ genes for $m$ samples, $X$
represents the $p$ factors of interest, $W$ the $k$ unwanted variation
factors and $\varepsilon$ some noise, typically $\varepsilon_j \sim
\mathcal{N}(0,\sigma^2_\varepsilon I_m),\, j=1,\ldots,n$. Both
$\alpha$ and $\beta$ are modeled as fixed, \emph{i.e.},
non-random~\citep{Robinson1991That}.

A similar model was also used in~\cite{Leek2007Capturing} and
\cite{Listgarten2010Correction}. The latter used a mixed effect model
where $\alpha$ was random. All these methods addressed the supervised
problem, \emph{i.e.}, the case where $X$ is an observed factor of
interest and the goal is to estimate its effect $\beta$ knowing that
some other factors influence gene expression. They differ in the way
parameters $\beta$ and $W\alpha$ are estimated. For the sake of
clarity, we now recall the approach of \cite{Gagnon-Bartsch2012Using}
from which our unsupervised estimators are derived.

\subsection{Supervised RUV-2}

The method of \cite{Gagnon-Bartsch2012Using} exploits the fact that
some genes are known to be negative controls, \emph{i.e.}, not to be
affected by the factor of interest. Formally we suppose that
$\beta_c=0$ where $\beta_c$ is the restriction of $\beta$ to its
column in some index $c$ denoting the negative control genes.

The objective in \cite{Gagnon-Bartsch2012Using} is to test the
hypothesis that $\beta_j = 0$ for each gene $j$ in order to identify
differentially expressed genes. They use an intuitive two-step
algorithm coined \emph{RUV-2} to estimate $W\alpha$ and $\beta$~:

\begin{enumerate}
\item Use the columns of $Y$ corresponding to control genes
  \begin{equation}
    \label{eq:ctlCols}
    Y_c = W\alpha_c + \varepsilon_c,
  \end{equation}
  to estimate $W$. Assuming iid noise $\varepsilon_j \sim
  \mathcal{N}(0,\sigma^2_\varepsilon I_m),\, j\in c$, the
  $(W\alpha_c)$ matrix maximizing the likelihood of~\eqref{eq:ctlCols}
  is $\argmin_{W\alpha_c,\textrm{rank}W\alpha_c\geq k}\|Y_c
  -W\alpha_c\|_F^2$. By the Eckart-Young
  theorem~\citep{Eckart1936Approximation}, this argmin is reached for
  $\widehat{W\alpha_c} = U\Lambda_kV^\top$, where $Y_c = U\Lambda V^\top$
  is the singular value decomposition (SVD) of $Y_c$ and $\Lambda_k$
  is the diagonal matrix with the $k$ largest singular values as its
  $k$ first entries and $0$ on the rest of the diagonal. We can for
  example use $\Wtwo = U\Lambda_k$.
\item Plug $\Wtwo$ back in the model~\eqref{eq:ruv} and do a
  regression to get $\hat{\alpha}, \hat{\beta}$. Assuming iid noise
  $\varepsilon_j \sim \mathcal{N}(0,\sigma^2_\varepsilon I_m),\,
  j=1,\ldots,n$, this estimates $(\alpha,\beta)$ by maximizing the
  likelihood of~\eqref{eq:ruv} after doing a plug-in of $\Wtwo$.
\end{enumerate}

The hypothesis $\beta_j = 0$ can be tested using the regression model
in step 2. More generally, RUV-2 can yield a corrected expression
matrix $\tilde{Y} = Y - \Wtwo\hat{\alpha}$.

\subsection{Naive unsupervised RUV-2}
\label{sec:unsupRuv}

If we assume that in addition $X$ is not specified, a simple approach
discussed in \cite{Gagnon-Bartsch2012Using} and called naive RUV-2 can
be used to remove $W\alpha$. This is to find an estimate of $W$,
\emph{e.g.} doing factor analysis on control genes as in RUV-2, and
then simply project the data in the orthogonal space of $\Wtwo$. More
precisely,
\begin{enumerate}
\item As in RUV-2, use the columns of $Y$ corresponding to control
  genes to estimate $W$ by maximizing the likelihood
  of~\eqref{eq:ctlCols}~: $\Wtwo = U\Lambda_k$.
\item Estimate $\alpha$ by doing a full regression of $Y$ against
  $\Wtwo$~: $\hat{\alpha} = (\Wtwo^\top \Wtwo)^{-1}\Wtwo^\top
  Y$. This amounts to maximizing the likelihood of~\eqref{eq:ruv}
  after plugging in $\Wtwo$ and taking $X\beta = 0$.
\item Once $W$ and $\alpha$ are estimated, $\Wtwo\hat{\alpha}$ can
  be removed from $Y$. The relevant unsupervised procedure to estimate
  $X\beta$ (clustering algorithm, PCA...) can be applied to the
  corrected expression matrix $Y-\Wtwo\hat{\alpha}$.
\end{enumerate}

This approach is referred to as \emph{naive RUV-2} in
\cite{Gagnon-Bartsch2012Using} and is expected to work well as long as
$X$ and $W$ are not too correlated. We study two directions that
should improve estimation when $X$ and $W$ are not expected to be
orthogonal. The first one in Section~\ref{sec:unsupRand} is to use a
random $\alpha$ version of~\eqref{eq:ruv}. The second one in
Section~\ref{sec:Westimation} is to explore new estimators of the
unwanted variation factors $W$. We expect our estimators of $X\beta$
to be more sensitive to the quality of the estimated unwanted
variation than in the supervised case~: since $X$ is not observed
anymore there is a higher risk of removing the factor of interest from
$Y$ by including it in $\Wtwo$.

\section{Random $\alpha$ model for unsupervised estimation}
\label{sec:unsupRand}

The supervised RUV-2 and naive RUV-2 estimators presented in
Section~\ref{sec:fixedRuv} model both the variation caused by the factor
of interest and the unwanted variation as a fixed effect. Modelling
$\alpha$ as a random quantity leads to smoother corrections which can
give better estimates of $\beta$ in the supervised case. This is the
model chosen by~\cite{Listgarten2010Correction}, and if we use control
genes to estimate $W$ as discussed in Section~\ref{sec:fixedRuv} this
leads to a random $\alpha$ version of
RUV-2~\citep{Gagnon-Bartsch2012Removing}. We find that random $\alpha$
models can be especially helpful when $X$ is unobserved. This section
discusses random $\alpha$ based estimation of $X\beta$, how it can be
done and why it can work better than the fixed $\alpha$ model of naive
RUV-2. We start by introducing the random $\alpha$ model in the case
where $X$ is observed.

\subsection{Supervised random $\alpha$ model}

If one endows the columns of $\alpha$ with a normal distribution
$\alpha_j \stackrel{iid}{\sim}
\mathcal{N}(0,\sigma_\alpha^2I_k),\, j=1,\ldots,n$, then
\eqref{eq:ruv} can be re-written
\begin{equation}
  \label{eq:ruvRand}
  Y = X\beta + \tilde{\varepsilon},
\end{equation}
where $\tilde{\varepsilon}_j \sim \mathcal{N}(0,\Sigma),\,
j=1,\ldots,n$ for $\Sigma \stackrel{\Delta}{=} \sigma_\alpha^2
WW^\top + \sigma^2_\varepsilon I_m$. The maximum likelihood estimator
of $\beta$ when $\Sigma$ is known is called the \emph{generalized
  least square}
estimator~\citep{Robinson1991That,Freedman2005Statistical} and is
given by~:
\begin{equation}
\label{eq:gls}
\hat{\beta} = \argmin_\beta\|Y-X\beta\|^2_\Sigma =
(X^\top\Sigma^{-1}X)^{-1} X^\top \Sigma^{-1} Y,
\end{equation}

In this sense the random $\alpha$ versions of RUV discussed
in~\cite{Gagnon-Bartsch2012Removing} are generalized least square
estimators made feasible by appropriate estimators of $\Sigma$.

Estimator~\eqref{eq:gls} leads to a smoother correction of the
unwanted variations because the scaling by $\Sigma^{-1}$ only shrinks
directions of $Y$ which correspond to unwanted variation
proportionally to the amount of variance observed in this
direction. By contrast in the fixed effect version of RUV-2, the
regression step against $(X,W)$ fits as much signal as possible on
every unwanted variation direction regardless how much variance was
actually observed along this direction in the control genes.

Using the posterior likelihood of model \eqref{eq:ruv} with prior
$\alpha_j \sim \mathcal{N}(0,\sigma_\alpha^2I_k),\,
j=1,\ldots,n$, we obtain a well known~\citep{Robinson1991That}
equivalent formulation of estimator~\eqref{eq:gls}~:

\begin{equation}
\label{eq:glsPost}
\hat{\beta} = \argmin_\beta\min_\alpha \left\{
  \|Y-X\beta-W\alpha\|^2_F + \frac{\sigma_\varepsilon^2}{\sigma_\alpha^2}\|\alpha\|_F^2 \right\}.
\end{equation}

In the supervised case this equivalence only serves to highlight the
relationship between~\eqref{eq:gls} and the fixed effect estimator
since both \eqref{eq:gls} and \eqref{eq:glsPost} can be solved in
closed form. In the unsupervised case, \eqref{eq:gls} can become hard
to solve and we will use a similar identity to help maximize the
likelihood.

\subsection{Unsupervised random $\alpha$ model}

We now consider an unsupervised version of~\eqref{eq:ruvRand}. $X$ is
not observed anymore and becomes a parameter~: we want to estimate
$X\beta$. This makes the estimation problem more difficult for several
reasons. First, this obviously increases the number of parameters to
be estimated. Second, this may introduce a product $X\beta$ in the
objective, making it non-convex in $(X, \beta)$ --- unless the problem
can be formulated in terms of the $X\beta$ matrix as a
parameter. Third, this may lead to the introduction of discrete
constraints which make the resulting optimization problem even
harder. Finally as we discussed in Section~\ref{sec:intro}, existing
algorithms which were designed to overcome these three difficulties
may not adapt easily to non identity covariance matrices $\Sigma$,
even in the case where this matrix is known.

Typical unsupervised estimation problems include clustering, where $X$
is constrained to be a membership matrix and $\beta$ is a real matrix
whose rows are the cluster centers, PCA where $X\beta$ is only
constrained to have a low rank, and sparse dictionary
learning~\citep{Mairal2010Sparse} where the $\ell_2$ norm of the
columns of $X$ are constrained to be $1$ and the $\ell_1$ norm of
$\beta$ is penalized. We first consider the estimation of $X\beta$
assuming that $\Sigma$ is known. Estimation of $\Sigma$ is discussed
in Section~\ref{sec:Westimation}.

\subsubsection{Optimization}

A direct approach could be to solve the maximum likelihood equation
of~\eqref{eq:ruvRand} for $X\beta$~:
\begin{equation}
  \label{eq:uML}
  \min_{X\beta \in \mathcal{M}} \|Y - X\beta\|_\Sigma^2,
\end{equation}
where $\mathcal{M}$ is a set of matrices which depends on the
unsupervised problem. Solving~\eqref{eq:uML} is difficult in general
because of the structure of $\mathcal{M}$ which can be discrete, or at
least non-convex. Dedicated algorithms often exist for the case where
$\Sigma = I_m$, but do not always generalize to non-spherical
$\Sigma$.  The purpose of this section is to
reformulate~\eqref{eq:uML} to make it solvable using the dedicated
$\Sigma = I_m$ algorithm.

We first discuss why~\eqref{eq:uML} is difficult in the specific case
of clustering to fix the ideas. In this case the problem can be stated
as~:
\begin{equation}
  \label{eq:ckmeans}
  \min_{X\in \mathcal{C},\, \beta\in\RR^{k\times n}} \|Y - X\beta\|_\Sigma^2,
\end{equation}
where $\mathcal{C}$ denotes the set of cluster membership matrices
\[
\mathcal{C} \stackrel{\Delta}{=} \bigg\{M \in \{0,1\}^{m\times k},
\sum_{j=1}^k M_{i,j} = 1, i=1,\ldots,m \bigg\}.
\]

Problem~\eqref{eq:ckmeans} corresponds to a non-spherical version of
the $k$-means objective. It is hard because $\mathcal{C}$ is discrete,
and because the objective is not jointly convex in $(X, \beta)$. The
$k$-means algorithm~\citep{Hastie2001elements} finds a local minimum
in the case where $\Sigma=I_m$ by iterating between fixing $\beta$ to
minimize~\eqref{eq:ckmeans} with respect to $X$ and fixing $X$ to
minimize with respect to $\beta$. However this classical algorithm
cannot be simply adapted to solve~\eqref{eq:ckmeans} with a general
$\Sigma$ because when fixing $\beta$ the solution for the rows of $X$
are coupled by $\Sigma$ whereas for the diagonal $\Sigma$ of $k$-means
they are independent, and simply amount to assigning each sample to
the cluster whose current center is the closest. A workaround is to
estimate $X$ by iteratively solving~\eqref{eq:ckmeans} for each row
--- keeping $\beta$ and the other rows fixed. It is also possible to
relax $\mathcal{C}$ by constraining the elements of $X$ to be in
$[0,1]$ instead of $\{0,1\}$, and solve in $X$ using a projected
gradient descent. We observed empirically that for a fixed $\Sigma$
this approach does well at minimizing the objective, but that the
solution can be very sensitive to mediocre estimates of
$\Sigma$. Instead we use a reformulation of~\eqref{eq:uML} which
allows us to use the algorithm for $\Sigma=I_m$.

The following proposition allows us to remove the $\Sigma$ in
problem~\eqref{eq:uML} at the expense of introducing a particular
correction term~:
\begin{proposition}
\label{prop:prop1}
Let $R \in \RR^{m\times n},\, W\in\RR^{m\times k},\, \nu > 0$, then
  \begin{equation}
    \label{eq:prop1}
    \min_{\alpha\in\RR^{k\times n}} \left\{ \|R - W\alpha\|^2_F + \nu
    \|\alpha\|^2_F\right\}
    = \|R\|^2_{S(W, \nu)},
  \end{equation}
  where $S(W, \nu) \stackrel{\Delta}{=} \nu^{-1} \left(WW^\top + \nu
    I_m \right)$.
\end{proposition}
This identity can be thought of as recapitulating normal Bayesian
regression~: the left hand side is the negative posterior log
likelihood of $R_j | \alpha_j \stackrel{iid}{\sim}
\mathcal{N}(W\alpha_j, I)$ with prior $\alpha_j \stackrel{iid}{\sim}
\mathcal{N}(0,\nu^{-1}I)$, summed over all columns $j$. Using a
Bayesian argument, this can also be written as the negative posterior
log likelihood of the $\alpha_j | R_j$. The $\alpha_j | R_j$ term
disappears by minimization over $\alpha_j$ and the remaining prior
$R_j$ term is the right hand side of~\eqref{eq:prop1}, with covariance
$S(W, \nu)$ being the sum of the covariance of the conditional mean
$\textrm{Cov}[\mathbb{E}[R_j|\alpha_j]] = \textrm{Cov}[W\alpha_j] =
\nu^{-1}WW^\top$ and the mean of the conditional variance
$\mathbb{E}[\textrm{Cov}[R_j|\alpha_j]] = I$. The identity can also be
verified by simply carrying the optimization over $\alpha$ on the left
hand side, as detailed in Appendix~\ref{app:prop1}.

We now detail how this identity can be used to solve~\eqref{eq:uML}
given a positive definite covariance matrix $\Sigma$. Defining $W =
U\left(\Lambda - \nu I\right)^{-\frac{1}{2}}U^\top$, where $\Sigma =
U\Lambda U^\top$ is the spectral decomposition of $\Sigma$ and for any
$\nu$ smaller than the smallest eigenvalue of $\Sigma$, we obtain
$S(W, \nu) = \Sigma$. Setting $R=Y-X\beta$, the right hand side
of~\eqref{eq:prop1} becomes the objective of~\eqref{eq:uML}, which we
need to solve for $X\beta$ in order to solve our unsupervised
estimation problem --- but can be difficult as we illustrated with the
clustering example~\eqref{eq:ckmeans}. Proposition~\eqref{eq:prop1}
suggests that we can instead solve~:
\begin{equation}
  \label{eq:reformulate}
  \min_{X\beta \in \mathcal{M}}\,\min_{\alpha\in\RR^{k\times n}} \left\{\|Y -
  W\alpha - X\beta\|^2_F + \nu
  \|\alpha\|^2_F\right\},
\end{equation}

which generalizes~\eqref{eq:glsPost} to the case where $X$ is not
observed. In the supervised case, closed form solutions were available
for both \eqref{eq:glsPost} and \eqref{eq:gls}. Now that $X$ is
unobserved Proposition~\ref{prop:prop1} implies that the formulations
\eqref{eq:uML} and \eqref{eq:reformulate} are still minimized by the
same set of $X\beta$, but no closed form is available for either of
them in general. However the $X\beta$ part of \eqref{eq:reformulate}
involves $\min_{X\beta \in \mathcal{M}} \|\tilde{Y} - X\beta\|^2_F$ on
a corrected matrix $\tilde{Y} \stackrel{\Delta}{=} Y - W\alpha$. The
\eqref{eq:reformulate} formulation may therefore be minimized
numerically provided that a dedicated algorithm, such as $k$-means for
clustering, is available to solve $\min_{X\beta \in \mathcal{M}}
\|\tilde{Y} - X\beta\|^2_F$, and regardless how difficult it is to
minimize numerically $\|Y - X\beta\|^2_\Sigma$ for general $\Sigma$
under the constraints $\mathcal{M}$.

Note that more generally, defining $W = U\left(\Lambda - \xi
  I\right)^{-\frac{1}{2}}U^\top$ for $\xi < \nu$ , we obtain $S(W,
\nu) = \Sigma + (\nu - \xi)I$, \emph{i.e.}, a ridged version of
$\Sigma$. In practice as discussed in Section~\ref{sec:Westimation},
we use various estimators of $\Sigma$. In all cases, this estimator is
based on a limited amount of data which can give it a high variance
and make it poorly conditioned. For these two reasons, ridging is a
good thing and in practice we use $\xi=0$ and some $\nu > 0$ which we
choose using a heuristic we propose in Section~\ref{sec:result}.

A joint solution for $(X\beta,\alpha)$ is generally not available for
\eqref{eq:reformulate}. A first naive solution is to set $X\beta=0$,
maximize over $\alpha$, and then apply the relevant unsupervised
estimation procedure, \emph{e.g.}, $k$-means to $Y-W\alpha$. More
generally a possible way of maximizing the likelihood of $X\beta$ is
to alternate between a step of optimization over $\alpha$ for a given
$X\beta$, which corresponds to a ridge regression problem, and a step
of optimization over $X\beta$ for a given $\alpha$ using the relevant
unsupervised estimation procedure for $\Sigma = I_m$. Each step
decreases the objective $\|Y - X\beta - W\alpha\|_F +
\nu\|\alpha\|_F^2$, and even if this procedure does not converge in
general to the maximum likelihood of $X\beta$, it may yield better
estimates than the naive version.

Finally note that Proposition~\ref{prop:prop1} only holds for $\nu >
0$. For $\nu = 0$ the objective on the left hand side
of~\eqref{eq:prop1} is the joint negative log-likelihood of the fixed
effect model~\eqref{eq:ruv}. Fixing $X\beta=0$ and optimizing over
$\alpha$ in this case yields the naive RUV-2 algorithm
of~\cite{Gagnon-Bartsch2012Using}. However it still makes sense to
optimize the joint likelihood of $(X\beta, \alpha)$ by a similar
iterative scheme. In this case the step where we optimize over
$\alpha$ becomes an ordinary linear regression.

\subsubsection{Benefit of random $\alpha$ for unsupervised estimation}

\begin{figure}
  \centering
  \includegraphics[width=.29\linewidth]{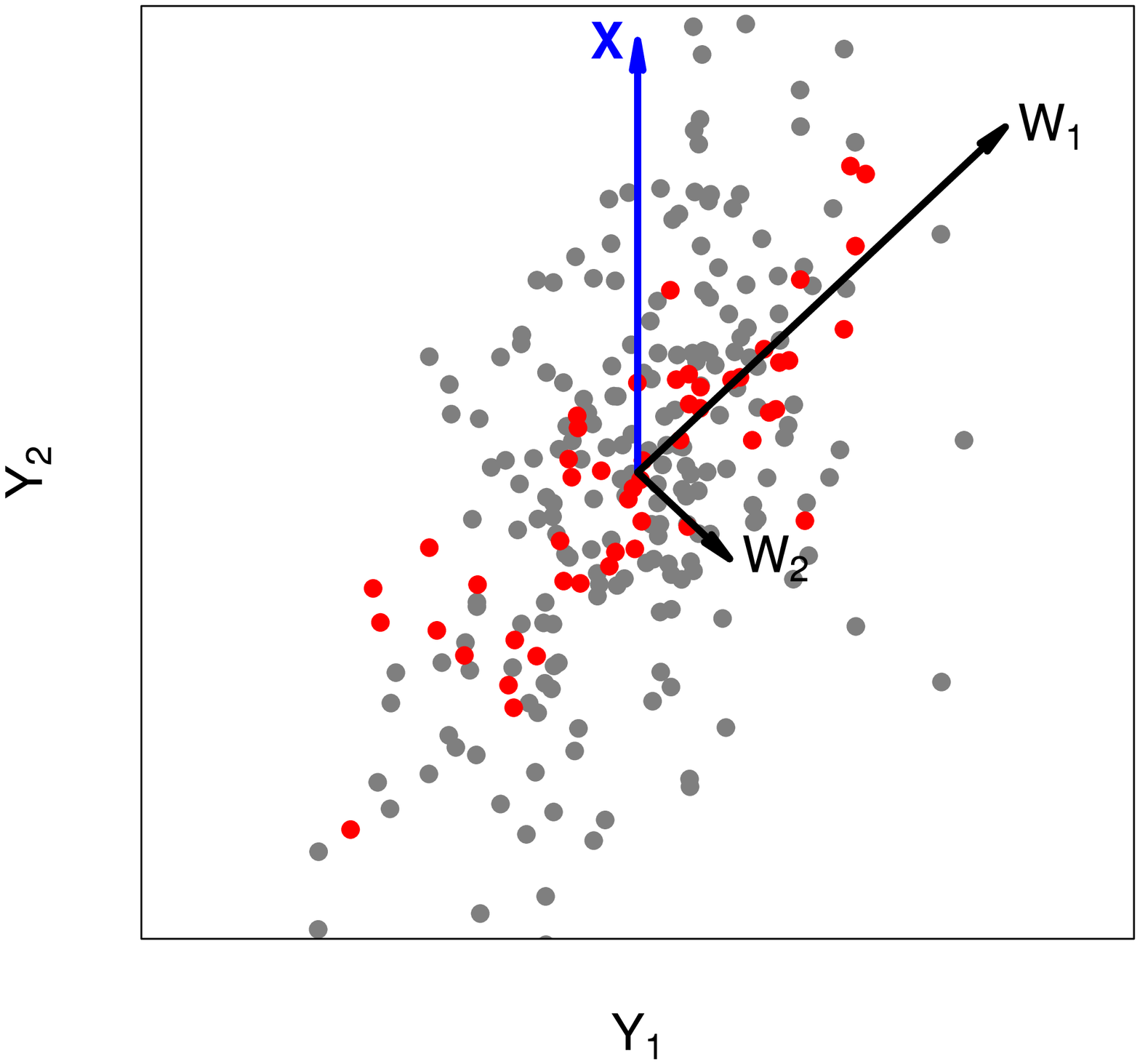}
  \includegraphics[width=.29\linewidth]{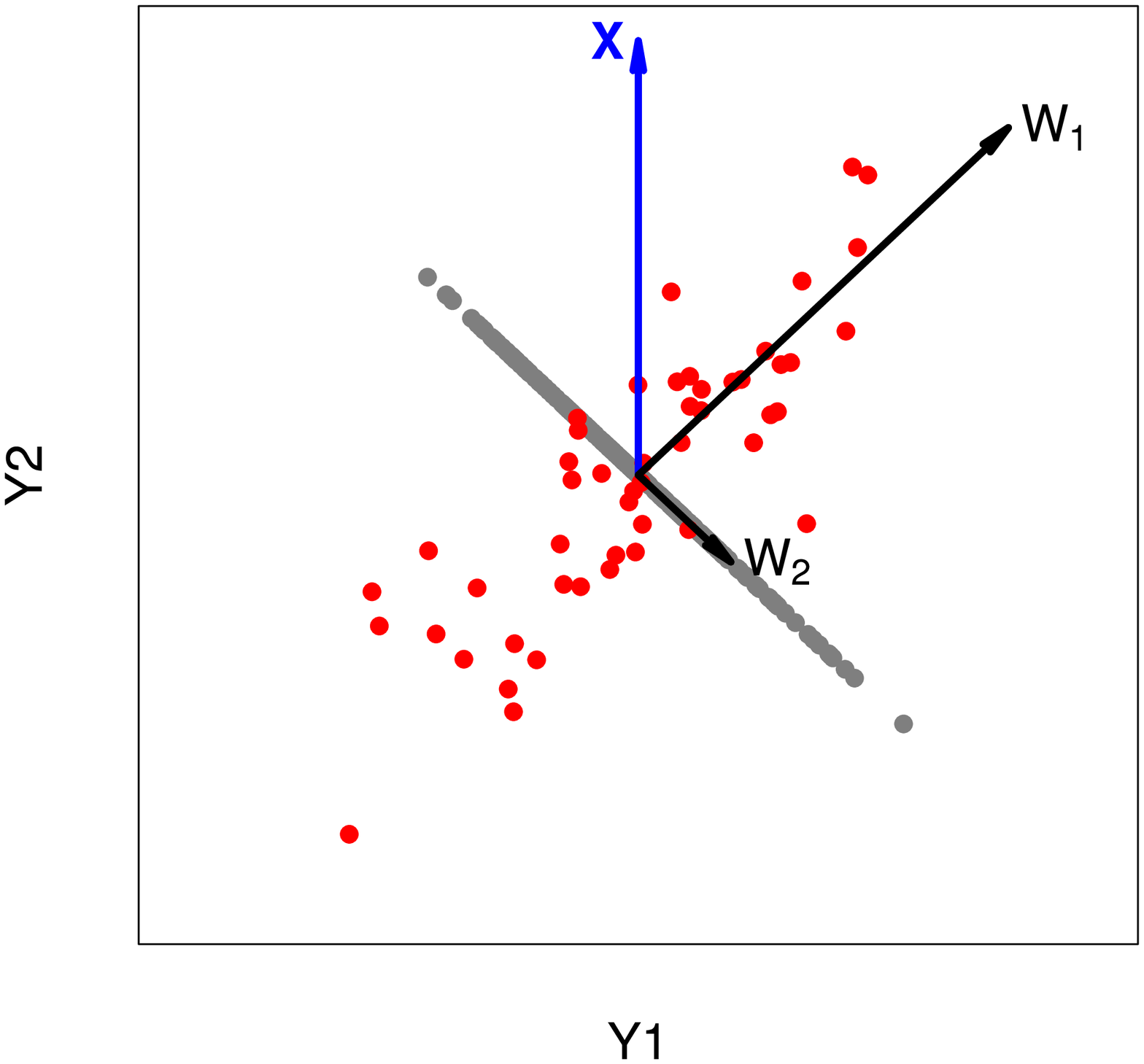}
  \includegraphics[width=.29\linewidth]{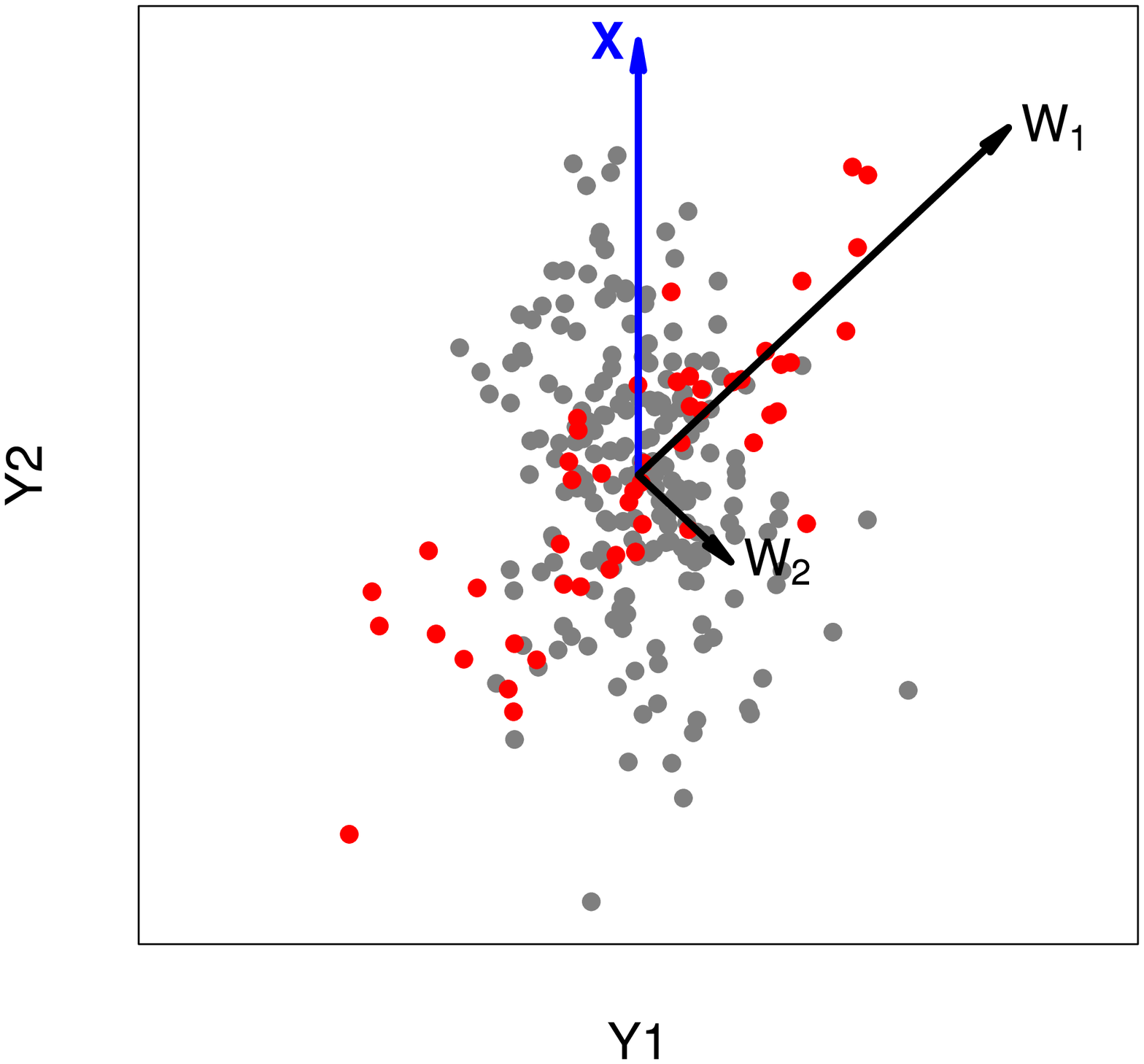}
  \caption{Naive RUV-2 (fixed $\alpha$) and random $\alpha$ based
    corrections, with $m=2$.}
  \label{fig:fixedVsRand}
\end{figure}

If $W$ is estimated as in RUV-2, the naive method of setting
$X\beta=0$, maximizing over $\alpha$ and applying the relevant
unsupervised estimation procedure yields a random-$\alpha$ version of
naive RUV-2, \emph{i.e.}, one in which the regression step is replaced
by a ridge regression. The benefit of this procedure compared to naive
RUV-2 may not be obvious, as the only technical difference is to
replace an ordinary regression by a ridge regression. But since
$\alpha$ is estimated with $X\beta=0$, the difference can be important
if $X$ and $W$ are correlated. Figure~\ref{fig:fixedVsRand} shows an
example of such a case. The left panel represents genes for $m=2$. Red
dots are control genes, gray ones are regular genes. The largest
unwanted variation $W_1$ has some correlation with the factor of
interest $X$. In naive RUV-2 with $k=1$, the correction projects the
samples in the orthogonal space of $W_1$, which can remove a lot of
the signal coming from the factor of interest. This is illustrated on
the center panel which shows the data corrected by naive RUV-2,
\emph{i.e.}, by projecting the genes on the orthogonal space of
$W_1$. The projection removes all effect coming from $W_1$ but also
reduces a lot the association of genes with $X$. Note that this is
true regardless of the amount of variance actually caused by $W_1$~:
the result would be the same with an almost spherical unwanted
variation $\|W_1\| \simeq \|W_2\|$ because once $W$ is identified, the
projection step of naive RUV-2 does not take into account any variance
information. On the other hand, the projection does not account at all
for the unwanted variation along $W_2$. By contrast, the random
$\alpha$ correction shown on the right panel of
Figure~\ref{fig:fixedVsRand} takes the variance into account. The
ridge regression only removes a limited amount of signal along each UV
direction, proportional to the amount of variance that was observed in
the control genes. In the extreme case where $W_1$ is equal to $X$,
the projection of RUV-2 removes all association of the genes with
$X$. Provided that the amount of variance along $W_1$ is correctly
estimated, the random $\alpha$ model only removes the variation caused
by $W_1$, leaving the actual association of the genes with $X$.

To summarize, given an estimate of $W$, we consider either a fixed or
a random $\alpha$ model to estimate $X\beta$. For each of these we
have either the naive option of estimating $\alpha$ for $X\beta = 0$
or to minimize the joint likelihood of $(X\beta, \alpha)$ by iterating
between the estimation of $X\beta$ and $\alpha$. The naive fixed
$\alpha$ procedure corresponds to the naive RUV-2
of~\cite{Gagnon-Bartsch2012Using}. We expect the other alternatives to
lead to better estimates of $X\beta$.

\subsection{Correlation on samples vs correlation on genes}

In model~\eqref{eq:ruvRand} we arbitrarily chose to endow $\alpha$
with a distribution, which is equivalent to introducing an $m\times m$
covariance matrix $\Sigma$ on the rows of the $Y - X\beta$
residuals. If we choose instead to model the rows of $W$ as iid normal
vectors with spherical covariance, we introduce a $n\times n$
covariance matrix $\Sigma'$ on the columns of the $Y - X\beta$
residuals. In the supervised case and if we consider a random $\beta$
as well, the maximum a posteriori estimator of $\beta$ incorporates
the prior encoded in $\Sigma'$ by shrinking the $\beta_j$ of
positively correlated genes towards the same value and enforcing a
separation between the $\beta_j$ of negatively correlated genes. This
approach was used in \cite{Desai2012Cross-}. As an example if $\alpha
\in \RR^{1\times n}$ is a constant vector, \emph{i.e.}, if $\Sigma'$ is a
constant matrix with an additional positive value on its diagonal, the
maximum a posteriori estimator of $\beta$ boils down to the
``multi-task learning'' estimator~\citep{Evgeniou2005Learning,
  Jacob2009Clustered} detailed in Appendix~\ref{app:mtl}. This model
as is does not deal with any unwanted variation factor which may
affect the samples. Using two noise terms in the regression model, one
with correlation on the samples and the other one on the genes would
lead to the same multi-task penalty shrinking some $\beta_j$ together,
but with a $\|.\|^2_\Sigma$ loss instead of the regular Frobenius loss
discussed in Appendix~\ref{app:mtl}.

Note that this discussion assumes $\Sigma'$ is known and used to
encode some prior on the residuals of the columns of $Y-X\beta$. If
however $\Sigma'$ needs to be estimated, and estimation is done using
the empirical covariance of $Y - X\beta$, the estimators of $\beta$
derived from~\eqref{eq:ruvRand} and from the model with an $n\times n$
covariance $\Sigma'$ on the $Y - X\beta$ residuals become very
similar, the only difference being that in one case the estimator of
$\alpha | W, Y-X\beta$ is shrinked and in the other case the estimator
of $W | \alpha, Y-X\beta$ is shrinked.

\section{Estimation of $W$}
\label{sec:Westimation}

When $X$ is observed it is usual to introduce the classical mixed
effect or generalized least square
regression~\citep{Robinson1991That,Freedman2005Statistical} for a fixed
covariance $\Sigma$ first and to discuss the practical estimation in a
second step. Similarly for each of the options discussed in
Section~\ref{sec:unsupRand}, we rely on a particular estimate of
$W$. A common approach coined feasible generalized least squares
(FGLS) in the regression case~\citep{Freedman2005Statistical} is to
start with an ordinary least squares step to estimate $\beta$ and then
use some estimator of the covariance on
$Y-X\beta$. \cite{Listgarten2010Correction} used a similar approach to
estimate their covariance matrix. An alternative is to use the
restricted maximum likelihood (REML) estimator which corrects for the
fact that the same data are used to estimate the mean $X\beta$ and the
covariance, whereas the empirical covariance matrix maximizes the
covariance assuming $X\beta$ is known. \cite{Listgarten2010Correction}
mention that they tried both ML and REML but that this did not make a
difference in practice.

In this section, we discuss estimation procedures for $W$ which are
relevant in the case where $X$ is not observed. As a general comment,
recall that fixed $\alpha$ models such as naive RUV-2 require
$\textrm{rank}(W) < \textrm{rank}(Y)$ otherwise the entire $Y$ is
removed by the regression step, but random $\alpha$ models allow
$\textrm{rank}(W) \geq \textrm{rank}(Y)$ because they lead to
penalized regressions.

So far, we considered the estimate $\Wtwo$ used in supervised RUV-2,
which relies on the SVD of the expression matrix restricted to its
control genes. This estimate was shown to lead to good performance in
supervised estimation~\citep{Gagnon-Bartsch2012Using}. Unsupervised
estimation of $X\beta$ may be more sensitive to the influence of the
factor of interest $X$ on the control genes~: in the case of fixed
$\alpha$ models, if the estimated $W$ is very correlated with $X$ in
the sense of the canonical correlation
analysis~\citep{Hotelling1936Relation}, \emph{i.e.}, if there exists a
linear combination of the columns of $X$ which has high correlation
with a linear combination of the columns of $W$, then most of the
association of the genes with $X$ will be lost by the
correction. Random $\alpha$ models are expected to be less sensitive
to the correlation of $W$ with $X$ but could be more sensitive to poor
estimates of the variance carried by each direction of unwanted
variation.

This suggests that unsupervised estimation methods could benefit from
better estimates of $W$. We present two directions that could lead to
such estimators.

\subsection{Using replicates}

We first consider ``control samples'' for which the factor of interest
$X$ is $0$.  In practice, one way of obtaining such control samples is
to use replicate samples, \emph{i.e.}, samples that come from the same
tissue but which were hybridized in two different settings, say across
time or platform. The profile formed by the difference of two such
replicates should therefore only be influenced by unwanted variation
-- those whose levels differ between the two replicates. In
particular, the $X$ of this difference should be $0$. More generally
when there are more than two replicates, one may take all pairwise
differences or the differences between each replicate and the average
of the other replicates. We will denote by $d$ the indices of these
artificial control samples formed by differences of replicates, and we
therefore have $X^d = 0$ where $X^d$ are the rows of $X$ indexed by
$d$.

\subsubsection{Unsupervised estimation of $W\alpha$}
\label{sec:shot}

A first intuition is that such samples could be used to identify
$\alpha$ the same way~\cite{Gagnon-Bartsch2012Using} used control
genes to estimate $W$. Therefore, we start by presenting how control
samples can be used to estimate $W\alpha$ in an unsupervised fashion,
and then discuss how the procedure yields a new estimator of $W$. We
consider the following algorithm~:

\begin{itemize}
\item Use the rows of $Y$ corresponding to control samples
  \begin{equation}
    \label{eq:ctlRows}
    Y^d = W^d\alpha + \varepsilon^d,
  \end{equation}
  to estimate $\alpha$. Assuming iid noise $\varepsilon_j \sim
  \mathcal{N}(0,\sigma^2_\varepsilon I_m),\, j=1,\ldots,n$, the
  $(W^d\alpha)$ matrix maximizing the likelihood of~\eqref{eq:ctlRows}
  is $\argmin_{W^d\alpha,\textrm{rank}W^d\alpha\geq k}\|Y^d
  -W^d\alpha\|_F^2$. By the same argument used in
  Section~\ref{sec:fixedRuv} for the first step of RUV-2, this argmin
  is reached for $\hat{W^d\alpha} = PE_kQ^\top$, where $Y^d =
  PEQ^\top$ is the singular value decomposition (SVD) of $Y^d$ and
  $E_k$ is the diagonal matrix with the $k$ largest singular values as
  its $k$ first entries and $0$ on the rest of the diagonal. We can
  use $\hat{\alpha} = E_k Q^\top$.
\item Plugging $\hat{\alpha}$ in~\eqref{eq:ctlCols}, the maximum
  likelihood of $W$ is now solved by a linear regression, $\hat{W}=
  Y_c \hat{\alpha}^\top_c (\hat{\alpha}_c\hat{\alpha}_c^\top)^{-1}$.
\item Once $W$ and $\alpha$ are estimated, $\hat{W}\hat{\alpha}$ can
  be removed from $Y$.
\end{itemize}

$X$ is not required in this procedure which in itself constitutes an
unsupervised correction for $Y$. 

\subsubsection{Comparison of the two estimators of $W$}
\label{sec:shotW}

This procedure also yields an estimator of $W$, which can be plugged
in any of the procedures we discussed in
Section~\ref{sec:unsupRand}. The estimator $\Wtwo$ we considered so
far was obtained using the first left singular vectors of the control
genes $Y_c$, which can also be thought of as a regression of the
control genes on their first right singular vectors, \emph{i.e.}, the
main variations $E_k Q^\top$ observed in the control genes. By
contrast the estimator that we discuss here is obtained by a
regression of the control genes against the main variations observed
in the control genes for the control samples formed by differences of
replicates. Assuming our control genes happen to be influenced by the
factor of interest $X$, \emph{i.e.}, $\beta_c \neq 0$, the estimator
of $W$ solely based on control genes may have more association with
$X$ than it should, whereas the one using differences of replicate
samples should not be affected. On the other hand, restricting
ourselves to the variation observed in differences of replicates may
be too restrictive because we don't capture unwanted variation when no
replicates are available.

To make things more precise, let us assume that the control genes are
actually influenced by the factor of interest $X$ and that $\beta_c
\sim \mathcal{N}(0,\sigma_{\beta_c}^2)$. In this case we have
$\mathbb{E}[Y_c Y_c^\top] = XX^\top\sigma_{\beta_c}^2 + WW^\top
\sigma_\alpha^2 + I_m \sigma_\varepsilon^2$, so if we use $Y_c$ to
estimate $W$ or $\Sigma$ as we do for $\Wtwo$ the estimate will be
biased towards $X$. Let us now consider the estimator $\Wr$ obtained
by the replicate based procedure. To simplify the analysis we assume
that $k = d$ and therefore $\hat{\alpha} = Y^d$ in the procedure
described in Section~\ref{sec:shot}. Consequently $\hat{W}\hat{\alpha}
= Y_c(Y_c^d)^\top\left(Y_c^d (Y_c^d)^\top\right)^{-1}Y^d$. Define $\Wr
\stackrel{\Delta}{=} Y_c(Y_c^d)^\top\left(Y_c^d
  (Y_c^d)^\top\right)^{-\frac{1}{2}}$. Assuming $X^d$ is indeed equal
to $0$ we can develop~:
\begin{displaymath}
  \Wr = (X\beta_c + W\alpha_c + \varepsilon_c)(W^d\alpha_c +
  \varepsilon_c^d)^\top \left(W^d\alpha_c\alpha_c^\top(W^d)^\top +
    W^d\alpha_c(\varepsilon_c^d)^\top +
    \varepsilon_c^d\alpha_c^\top(W^d)^\top + \varepsilon_c^d(\varepsilon_c^d)^\top\right)^{-\frac{1}{2}}.
\end{displaymath}
We now make some heuristic asymptotic approximations in order to get a
sense of the behavior of $\Wr$. $\alpha_c\alpha_c^\top$ and
$\varepsilon_c^d(\varepsilon_c^d)^\top$ are Wishart variables which by
the central limit theorem are close to $cI_m\sigma_\alpha^2$ and
$cI_m\sigma_\varepsilon^2$ respectively if the number of control genes
is large enough regardless how good the control genes are,
\emph{i.e.}, how small $\sigma_{\beta_c}$ is. In addition dot products
between independent multivariate normal variables are close to $0$ in
high dimension so we approximate $\beta_c\alpha_c^\top,
\beta_c\varepsilon_c^\top$ and $\alpha_c\varepsilon_c^\top$ by
$0$. The approximations involving $\beta_c$ depend in part how good
the control genes are, but can still be valid for larger
$\sigma_{\beta_c}$ if the number of control gene is large enough. We
further assume that $\sigma_\varepsilon \ll \sigma_\alpha$ and that
the control samples are independent from the samples for which we
estimate $W$ and ignore the $cI_m\sigma_\varepsilon^2$ and
$\varepsilon_c(\varepsilon_c^d)^\top$ terms. Implementing all these
approximations yields $\Wr \simeq \sigma_\alpha
c^{\frac{1}{2}}W(W^d)^\top\left(W^d(W^d)^\top\right)^{\frac{1}{2}}$. Writing
$W^d = A\Delta B^\top$ for the SVD of $W^d$, we obtain $\Wr \simeq
\sigma_\alpha c^{\frac{1}{2}}WB_rA_r^\top$, where $r$ is the rank of
$W^d$ and $A_r, B_r$ contain the first $r$ columns of $A$ and $B$
respectively. This suggests that if $W^d$ has rank $k$ $\Wr$ is a good
estimator of $W$ in the sense that it is not biased towards $X$ even
if the control genes are influenced by $X$. If $W^d$ is column rank
deficient, the $B_r$ mapping can delete or collapse unwanted variation
factors in $\Wr$. The effect is easier to observe on the estimator of
$\Sigma$~: $\Wr\Wr^\top \simeq \sigma^2_\alpha cWB_rB_r^\top W^\top$.
Consider for example the following case with $3$ unwanted variation
factors and $3$ replicate samples with unwanted variation $(1,0,3)$,
$(0,1,3)$ and $(1,1,3)$. The $W^d$ and corresponding $B_rB_r^\top$
obtained by taking differences between replicates $(1,2)$, $(1,3)$ and
$(3,2)$ are
\begin{displaymath}
  W^d = \left(
    \begin{array}{c c c}
      1 & -1 & 0\\
      0 & -1 & 0\\
      1 & 0 & 0
    \end{array}
\right)
\qquad\textrm{ and }\qquad
  B_rB_r^\top = \left(
    \begin{array}{c c c}
      1 & 0 & 0\\
      0 & 1 & 0\\
      0 & 0 & 0
    \end{array}
\right),
\end{displaymath}
so the $B_rB_r^\top$ factor removes the third factor from the estimate
of $\Sigma$. This is because the $3$ replicates have the same value
for the third factor. Similarly if two factors are perfectly
correlated on the replicate samples, \emph{e.g.}, the first two
factors for $(1,1,0)$, $(0,0,1)$ and $(1,1,1)$, the $W^d$ and
corresponding $B_rB_r^\top$ for the same differences between
replicates $(1,2)$, $(1,3)$ and $(3,2)$ are
\begin{displaymath}
  W^d = \left(
    \begin{array}{c c c}
      1 & 1 & -1\\
      0 & 0 & -1\\
      1 & 1 & 0
    \end{array}
\right)
\qquad\textrm{ and }\qquad
  B_rB_r^\top = \left(
    \begin{array}{c c c}
      1/2 & 1/2 & 0\\
      1/2 & 1/2 & 0\\
      0 & 0 & 1
    \end{array}
\right),
\end{displaymath}
which collapses the first two factors into an average factor and
leaves the third one unchanged.

Finally, another option in the context of random $\alpha$ models is to
combine the control gene based and replicate based estimators of $W$
by concatenating them. In terms of $\Sigma$, this amounts to summing
the two estimators of the covariance matrix. This may help if, as in
our first example, some factors are missing from $\Wr$ because all
pairs of replicates have the same value for these factors. In this
case, combining it with $\Wtwo$ could lead two an estimate containing
less $X$ but still containing all the unwanted variation factors.

\subsection{Using residuals}
\label{sec:Wupdate}

We already mentioned that in the case where $X$ is observed, a common
way of estimating $W$ known as FGLS is to first do an ordinary
regression of $Y$ against $X$, then compute the empirical covariance
on the residuals $Y-X\hat{\beta}$. The estimators of $W$ that we
discussed work around the estimation of $X\beta$ by using genes for
which $\beta$ is known to be $0$ or samples for which $X$ is known to
be $0$. Once we start estimating $X\beta$, \emph{e.g.}, by iterating
over $X\beta$ and $\alpha$ as described at the end of
Section~\ref{sec:unsupRand} we can use a form of FGLS and re-estimate
$W$ using $Y - \hat{X\beta}$. If the current estimator of $X\beta$ is
correct, this amounts to making all the genes control genes, and all
the samples control samples.

\subsection{Using a known $W$}
\label{sec:knownW}

Finally in some cases we may want to consider that $W$ is
observed. For example, if the dataset contains known technical
batches, involves different platforms or labs, $W$ could encode these
factors instead of being estimated from the data. In particular if the
corresponding $W$ is a partition of the samples, then naively
estimating $\alpha$ by regression using $X\beta=0$ and removing
$W\hat{\alpha}$ from $Y$ corresponds to mean-centering the groups
defined by $W$. In most cases however, this procedure or its shrunken
equivalent doesn't yield good estimates of $X\beta$. This was also
observed by~\cite{Gagnon-Bartsch2012Using} in the supervised case. One
reason is that this $W$ only accounts for known unwanted variation
when other unobserved factors can influence the gene expression. The
other one is that this approach leads to a linear correction for the
unwanted variation in the representation used in $W$. If we know that
gene expression is affected by the temperature of the scanner, setting
a column of $W$ to be this temperature leads to a linear correction
whereas the effect of the temperature may be quadratic, or involve
interactions with other factors. In this case, estimating $W$
implicitly allows us to do a non-linear correction because the
estimated $W$ could fit any non-linear representation of the observed
unwanted variation which actually affects gene expression.

\section{Result}
\label{sec:result}

We now evaluate our unsupervised unwanted variation correction methods
on several microarray gene expression datasets. We focus on
clustering, a common unsupervised analysis performed on gene
expression data~\citep{Speed2003Statistical} typically to identify
subtypes of a disease~\citep{Perou2000Molecular}. Note that the
proposed method is in no way restricted to clustering, which is simply
used here to assess how well each correction did at removing unwanted
variation without damaging the signal of interest.

Because of the unsupervised nature of the task, it is often difficult
to determine whether the partition obtained by one method is better
than the partition obtained by another method. We adopt the classical
strategy of turning supervised problems into unsupervised ones~: for
each of the three datasets we consider in this section, a particular
grouping of the samples is known and considered to be the factor of
interest which must be recovered. Of course, the correction methods
are not allowed to use this known grouping structure.

In order to quantify how close each clustering gets to the objective
partition, we adopt the following squared
distance~\citep{Bach2007DIFFRAC} between two given partitions
$\mathcal{C}=(c_1,...,c_k)$ and $\mathcal{C}'=(c'_1,...,c'_k)$ of the
samples into $k$ clusters :
\begin{equation}
  \label{eq:clScore}
  d(\mathcal{C},\mathcal{C}') = k - \sum_{i,j=1}^k \frac{|c_i \cap c'_j|^2}{|c_i||c'_j|}.
\end{equation}
This score ranges between $0$ when the two partitionings are
equivalent, and $k-1$ when the two partitions are completely
different. To give a visual impression of the effect of the
corrections on the data, we also plot the data in the space spanned by
the first two principal components, as PCA is known to be a convex
relaxation of $k$-means~\citep{Zha2001Spectral}. For each of the
correction methods that we evaluate, we apply the correction method to
the expression matrix $Y$ and then estimate the clustering using a
$k$-means algorithm.

In addition to the methods we introduced in
Sections~\ref{sec:unsupRand} and~\ref{sec:Westimation} of this paper,
we consider as baselines~:
\begin{itemize}
\item An absence of correction,
\item A centering of the data by level of the known unwanted variation factors.
\item The naive RUV-2 procedure of~\cite{Gagnon-Bartsch2012Using}.
\end{itemize}

Centering of the data by level of the known unwanted variation factors
means that if several unwanted variation factors are expected to
affect gene expression, we center each group of samples having the
same value for all of the unwanted variation factors. Of course this
implies that unwanted variation factors are known and this may not
correct the effect of unobserved factors.

On each dataset, we evaluate three basic correction methods~: the
unsupervised replicate-based procedure described in
Section~\ref{sec:shot}, the random $\alpha$ model with $\hat{W} =
\Wtwo$ and the random $\alpha$ model with $\hat{W}$ combining $\Wtwo$
and $\Wr$. For each of these three methods, we also evaluate an
iterative version which alternates between estimating $\alpha$ using a
ridge regression and estimating $X\beta$ using the sparse dictionary
learning technique of~\citep{Mairal2010Sparse}. The sparse dictionary
estimator finds $(X,\beta)$ minimizing the objective~:
\begin{displaymath}
  \frac{1}{2}\|(Y-\widehat{W\alpha}) - X\beta\|_F^2 + \lambda \|\beta\|_1,
\end{displaymath}
under the constraint that $X\beta$ has rank $p$ and the columns of $X$
have norm $1$. The problem is not jointly convex in $(X,\beta)$ so
there is no guarantee of obtaining a global minimizer. For the
iterative methods, we also re-estimate $W$ using the
$Y-\widehat{X\beta}$ residuals as discussed in
Section~\ref{sec:Wupdate} every $10$ iterations. Many other methods
are possible such as iterative fixed $\alpha$ (for different
estimators of $W$), random $\alpha$ using the $\Wr$ estimator
introduced in Section~\ref{sec:shot} with or without iterations, with
or without updating $W$, or replacing the sparse dictionary estimator
by other matrix decomposition matrix techniques such as convex
relaxations of $k$-means. We restrict ourselves to just the set of
methods mentioned for the sake of clarity and because we think they
illustrate the benefit of the different options that we discuss in
this paper.

Some of the methods require the user to choose some hyper-parameters~:
the ranks $k$ of $W\alpha$ and $p$ of $X\beta$, the ridge $\nu$ and
the strength $\lambda$ of the $\ell_1$ penalty on $\beta$. In order to
decrease the risk of overfitting and to give good rules of thumb for
application of these methods to new data, we tried to find relevant
heuristics to fix these hyperparameters and to use them consistently
on the three benchmarks. Some methods may have given better
performances on some of these datasets for different choices of the
hyperparameters. We discuss empirical robustness of the methods to
their hyperparameters for each dataset. The rank $k$ of $W\alpha$ was
chosen to be close to $m/4$, or to the number of replicate samples
when the latter was smaller than the former. For methods using $p$, we
chose $p=k$. For random $\alpha$ models, we use $k=m$~: the model is
regularized by the ridge $\nu$. We do not combine it with a
regularization of the rank. Since $\nu$ is a ridge parameter acting on
the eigenvalues of $W^\top W$, we chose it to be $\sigma_1(W^\top
W)\times 10^{-3}$, where $\sigma_1(A)$ is the largest eigenvalue of a
positive semidefinite matrix $A$. Special care must be taken for the
iterative methods~: their naive counterparts use $X\beta=0$ and
therefore minimize $\|Y-W\alpha\|_F^2$, whereas the iterative
estimators minimize $\|Y-X\beta-W\alpha\|_F^2$ which can lead to
corrections of smaller norm $\|W\alpha\|_F$. In order to make them
comparable, we choose $\lambda$ such that $\|W\alpha\|_F$ is close to
the one obtained with the non-iterative algorithm.

All experiments were done using R. For the experiments using sparse
dictionary algorithms, we used the SPAMS
software~\footnote{http://spams-devel.gforge.inria.fr/} developed by
the first author of \cite{Mairal2010Sparse} under GPL license and for
which an R interface is available.

\subsection{Gender data}

We first consider a 2004 dataset of patients with neuropsychiatric
disorders. This dataset was published by~\cite{Vawter2004Gender} in
the context of a gender study~: the objective was to study differences
in gene expression between male and female patients affected by these
neuropsychiatric disorders. It was already used
in~\cite{Gagnon-Bartsch2012Using} to study the performances of RUV-2~:
the number of genes from the X and Y chromosomes which were found
significantly expressed between male and female patients was used to
assess how much each correction method helped. This gender study is an
interesting benchmark for methods aiming at removing unwanted
variation as it expected to be affected by several technical and
biological factors~: two microarray platforms, three different labs,
three tissue localizations in the brain. Most of the $10$ patients
involved in the study had samples taken from the anterior cingulate
cortex (a), the dorsolateral prefontal cortex (d) and the cerebellar
hemisphere (c). Most of these samples were sent to three independent
labs~: UC Irvine (I), UC Davis (D) and University of Michigan, Ann
Arbor (M). Gene expression was measured using either HGU-95A or
HGU-95Av2 Affymetrix arrays with $12,600$ shared between the two
platforms. Six of the $10\times 3\times 3$ combinations were missing,
leading to $84$ samples. We use as control genes the same $799$
housekeeping probesets which were used
in~\cite{Gagnon-Bartsch2012Using}. We use all possible differences
between any two samples which are identical up to the brain region or
the lab, leading to $106$ differences. Note that using all these
replicates leads to non-iid data in the replicate based methods~: each
sample is dependent on all the differences in which it is
involved. The discussion in Section~\ref{sec:shotW} suggests that this
could bias the estimation of $W$. However we seem to obtain reasonable
results for this dataset. As a pre-processing, we also center the
samples per array type.

\begin{figure}
  \centering
  \includegraphics[width=.6\linewidth]{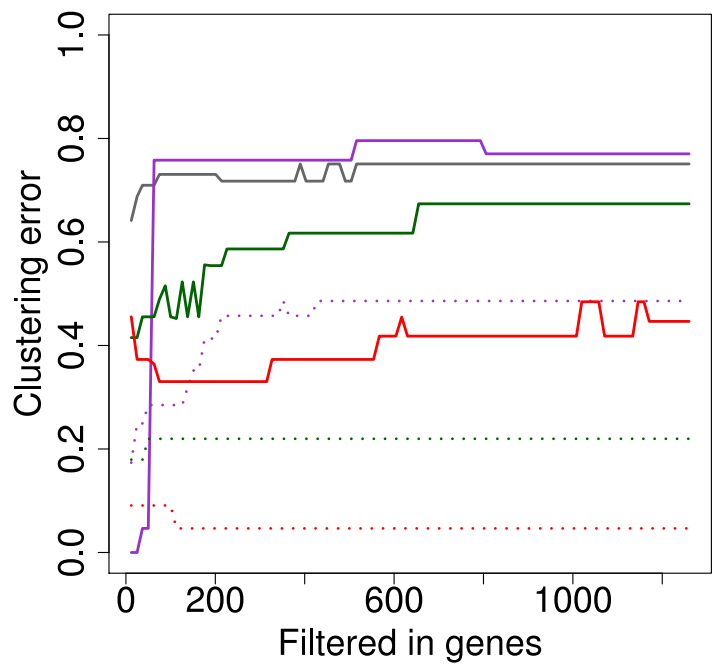}
  \caption{Clustering error against number of genes selected (based on
    variance) before clustering. Gray line~: naive RUV-2, purple
    lines~: replicate-based corrections, green lines~: random $\alpha$
    model using $\Wtwo$, red lines~: random $\alpha$ model using a
    combination of $\Wtwo$ and $\Wr$. Full lines~: no iteration,
    dotted lines~: with iterations.}
  \label{fig:gender_all}
\end{figure}

\begin{figure}
  \centering
  \includegraphics[width=.49\linewidth]{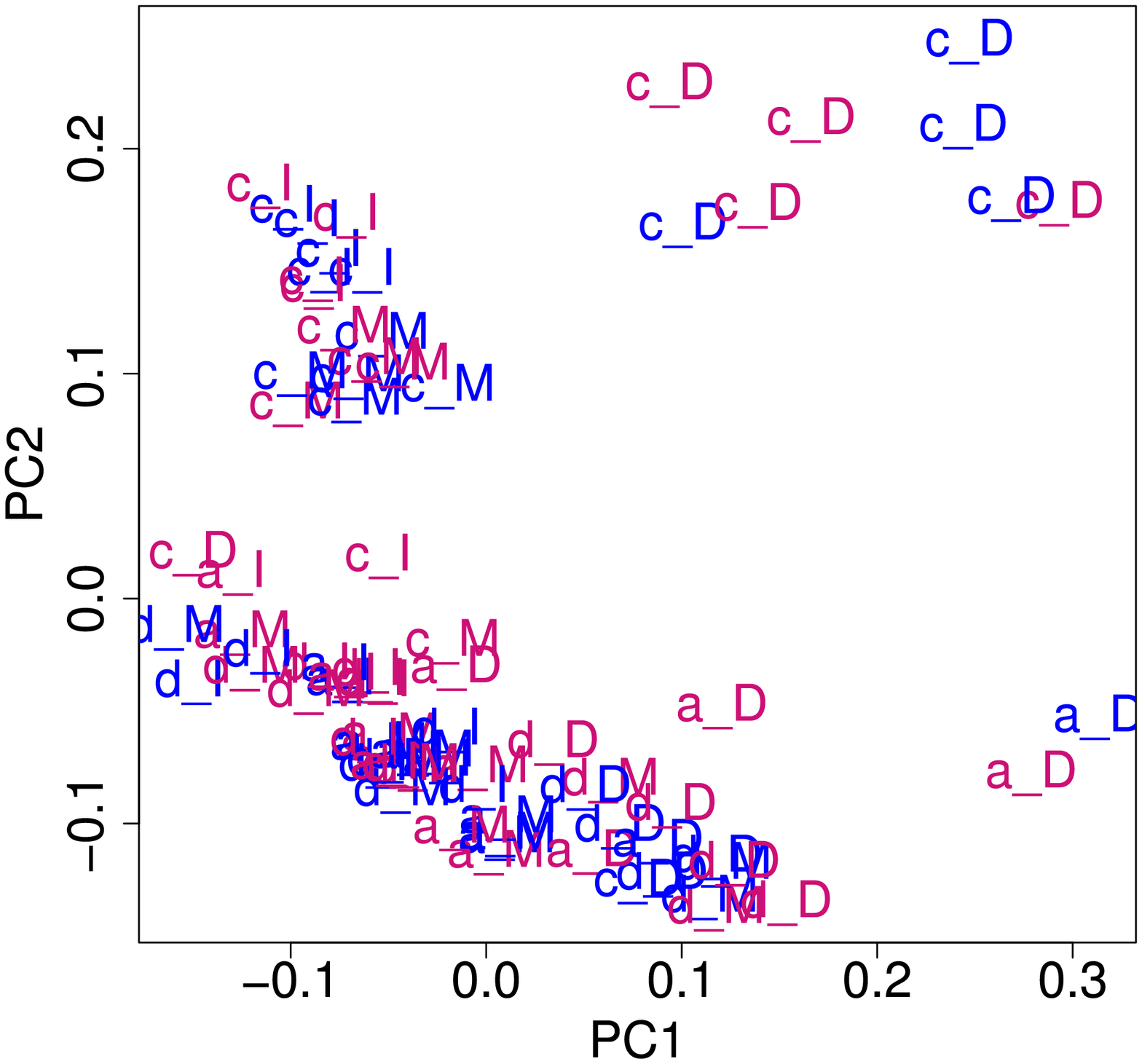}
  \includegraphics[width=.49\linewidth]{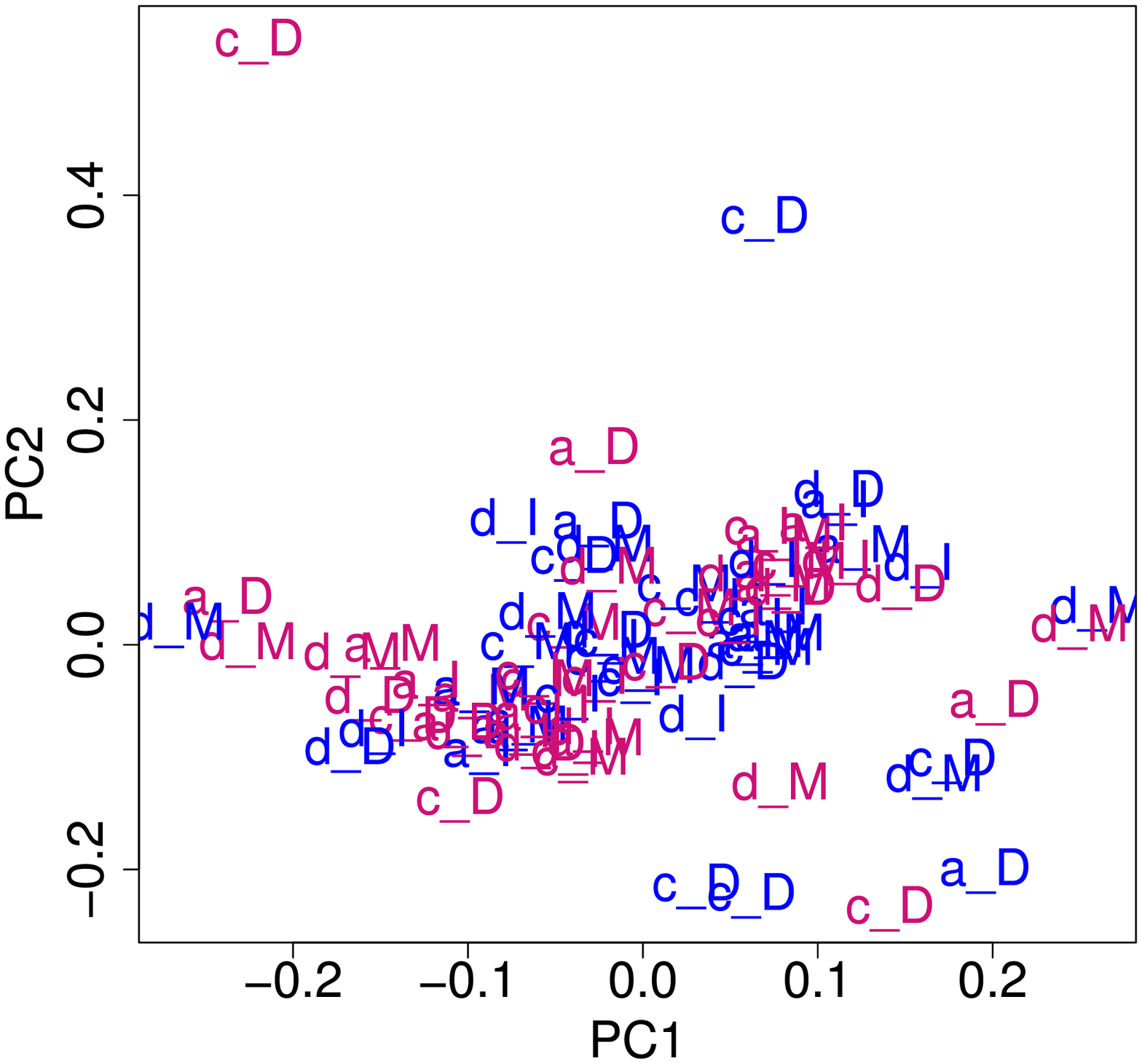}
  \caption{Samples of the gender study represented in the space of
    their first two principal components before correction (left
    panel) and after centering by lab plus brain region (right
    panel). Blue samples are males, pink samples are females. The
    labels indicate the laboratory and brain region of each
    sample. The capital letter is the laboratory, the lowercase one is
    the brain region.}
  \label{fig:gender_uncorrected}
\end{figure}

Since most genes are not affected by gender, and because $k$-means is
known to be sensitive to noise in high dimension, clustering by gender
gives better results in general when removing genes with low variance
before applying $k$-means. For each of the methods assessed, we
therefore try clustering after filtering different numbers of genes
based on their variance on the corrected. Figure~\ref{fig:gender_all}
shows the clustering error~\eqref{eq:clScore} for the methods against
the number of genes retained. The uncorrected and mean-centering cases
are not displayed to avoid cluttering the plot, but give values above
$0.95$ for all numbers of genes
retained. Figure~\ref{fig:gender_uncorrected} shows the samples in the
space of the first two principal components in these two cases,
keeping the $1260$ genes with highest variance. On the uncorrected
data (left panel), it is clear that the samples first cluster by lab
which is the main source of variance, then by brain region which is
the second main source of variance. This explains why the clustering
on uncorrected data is far away from a clustering by
gender. Mean-centering samples by region-lab (right panel) removes all
clustering per brain region or lab, but doesn't make the samples
cluster by gender. The gray line of Figure~\ref{fig:gender_all} shows
the performance of naive RUV-2 for $k=20$. Since naive RUV-2 is a
radical correction which removes all variance along some directions,
it is expected to be more sensitive to the choice of $k$. The
estimation is damaged by using $k=40$ (clustering error $0.99$). Using
$k=5$ also degrades the performances, except when very few genes are
kept.

The purple lines of Figure~\ref{fig:gender_all} represent the
replicate-based corrections. The solid line shows the performances of
the non-iterative method described in Section~\ref{sec:shot}. Its
performances are similar to the ones of naive RUV-2 in this case,
except when very few genes are selected and the replicate-based method
leads to a perfect clustering by gender. As for the naive RUV-2
correction, using a $k$ too large damages the performances. However,
the replicate-based correction is more robust to overestimation of
$k$~: using $k=40$ only leads to a $0.83$ error. This is also true for
the other benchmarks, and can be explained by the fact that the
correction is restricted to the variations observed among the
contrasts of replicate samples which are less likely to contain signal
of interest.

\begin{figure}
  \centering
  \includegraphics[width=.49\linewidth]{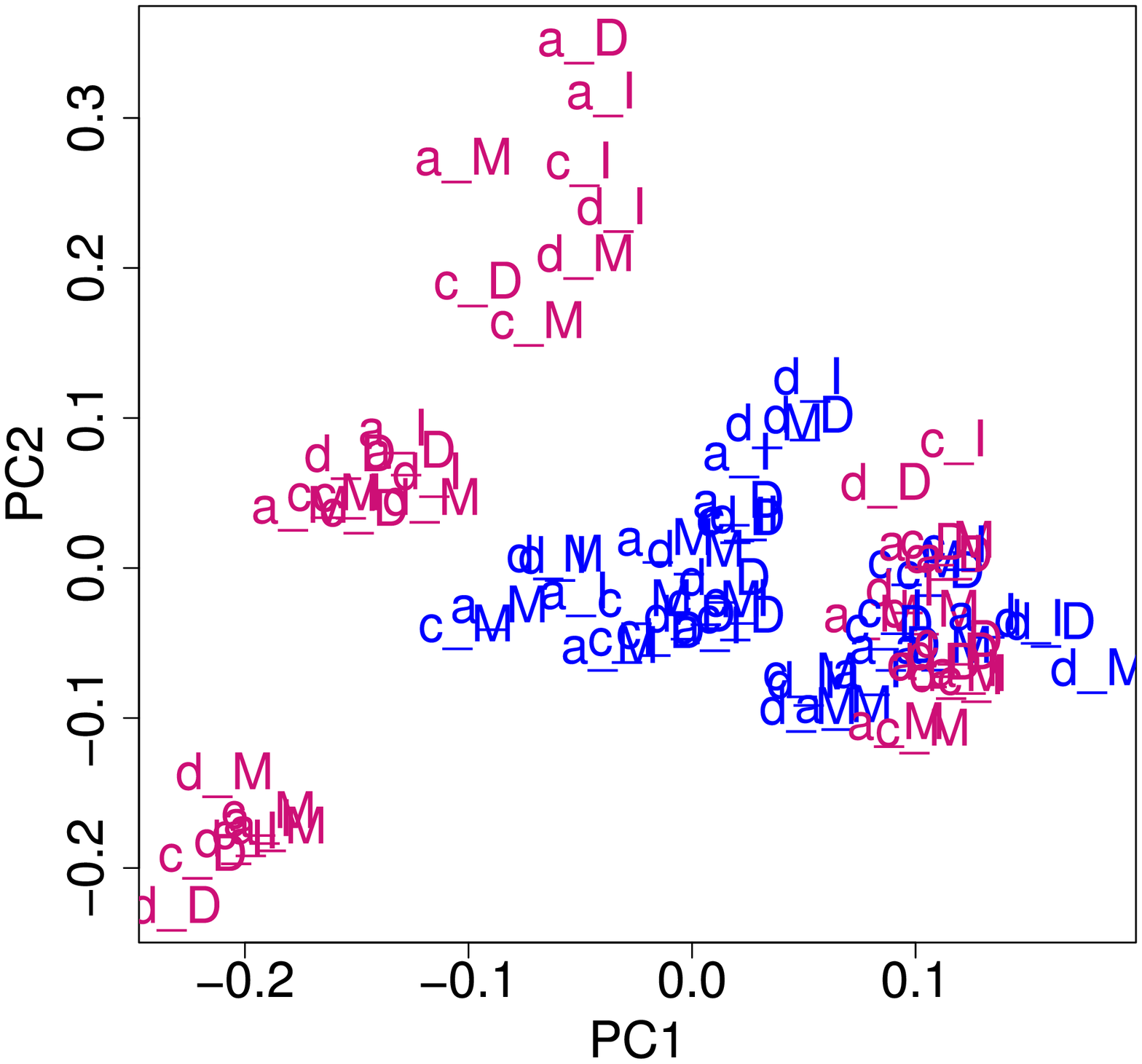}
  \includegraphics[width=.49\linewidth]{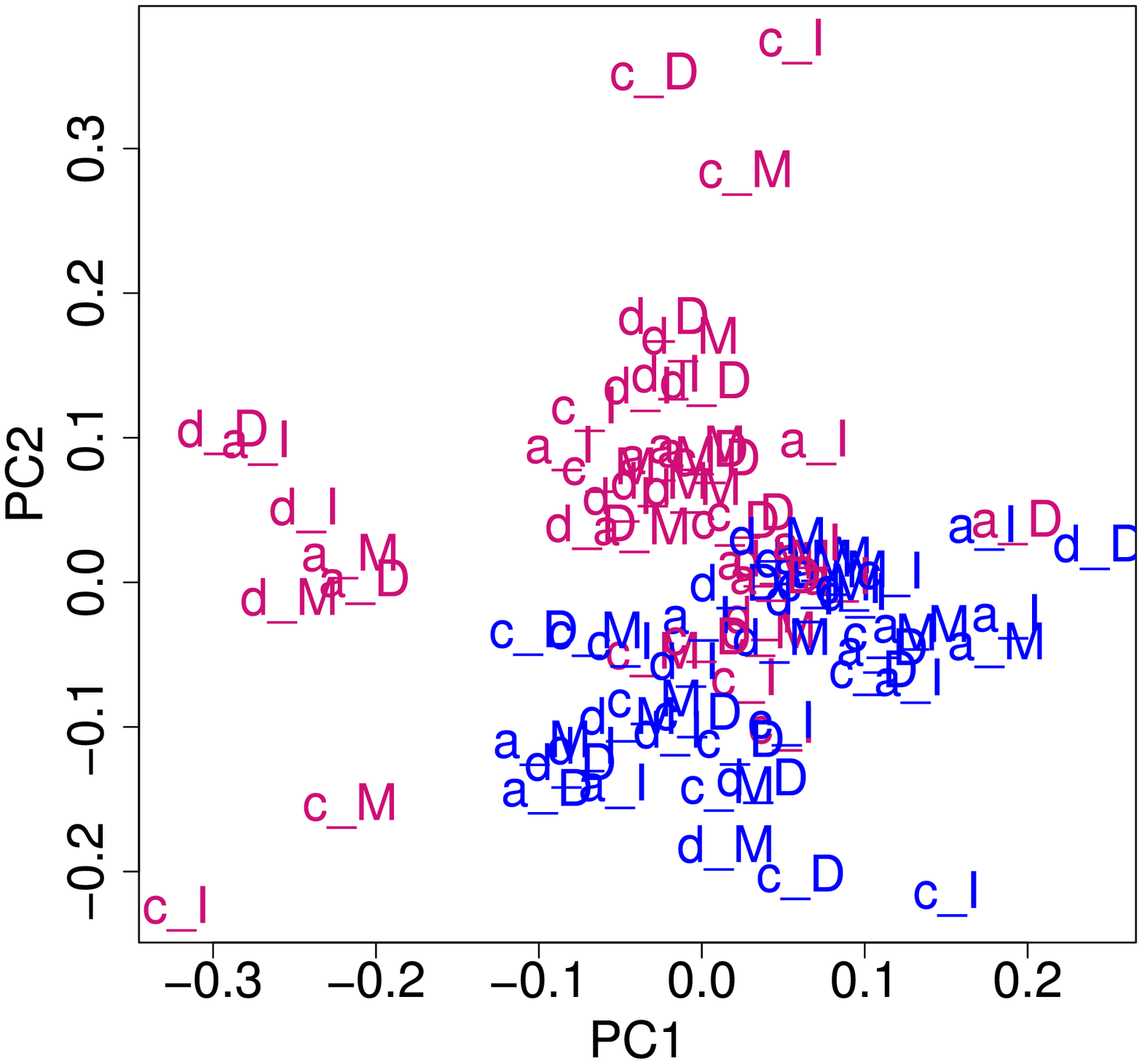}
  \caption{Using replicates. Left: no iteration, right: with iterations}
  \label{fig:gender_rep}
\end{figure}

The iterative version in dotted line leads to much better clustering
except again when very few genes are
selected. Figure~\ref{fig:gender_rep} shows the samples in the space
of the first two principal components after applying the non-iterative
(left panel) and iterative (right panel) replicate-based method. The
correction shrinks the replicates together, leading to a new variance
structure, more driven by gender although not separating perfectly
males and females. One may wonder whether this shrinking gives a
better partition structure than clustering each region-lab separately,
\emph{e.g.} clustering only the samples taken from the cerebellar
hemisphere and sent to UC Irvine. This actually leads to clustering
errors higher than $0.89$ for all region-lab pairs, even when keeping
only few genes based on their variance.

\begin{figure}
  \centering
  \includegraphics[width=.49\linewidth]{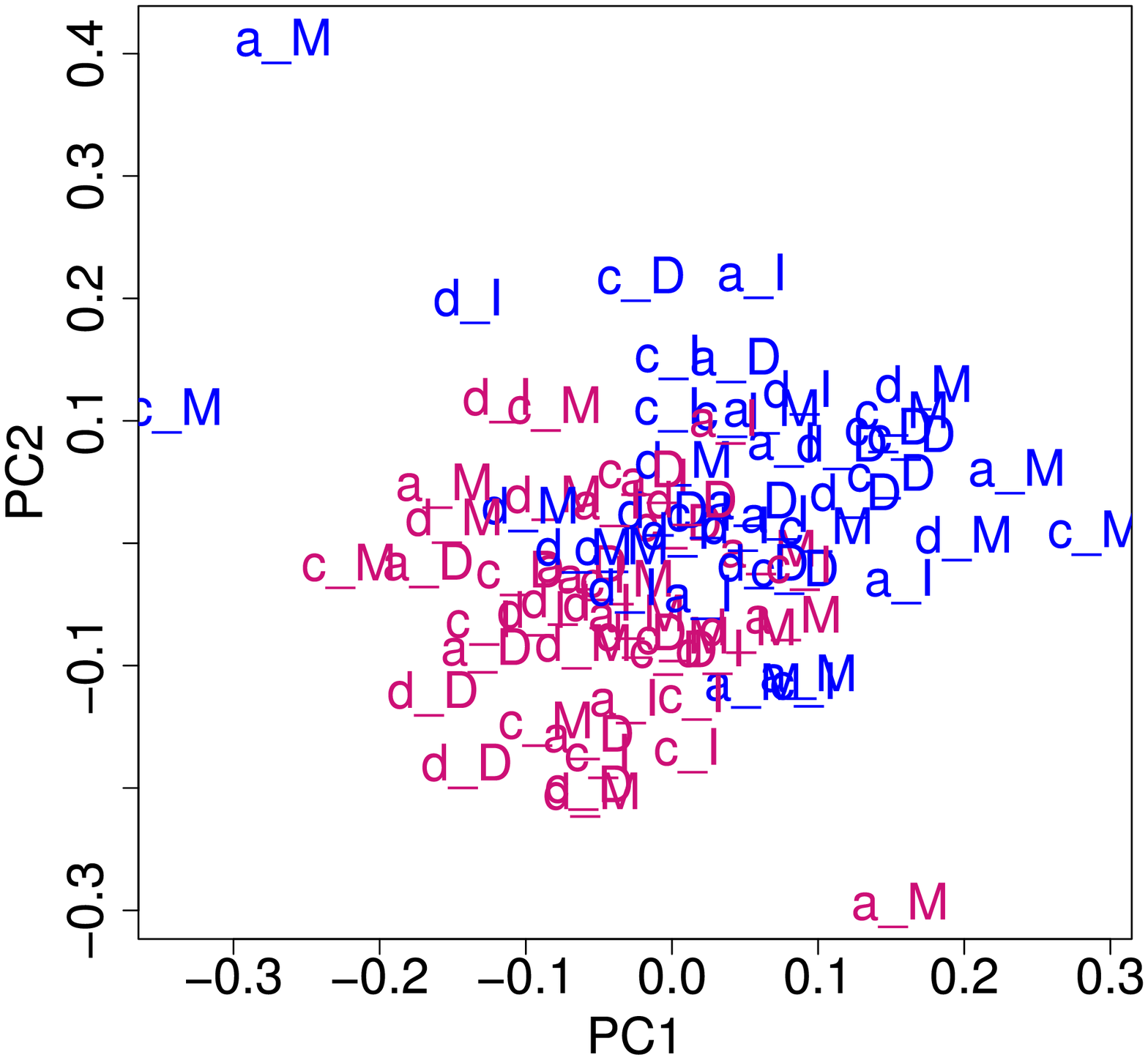}
  \includegraphics[width=.49\linewidth]{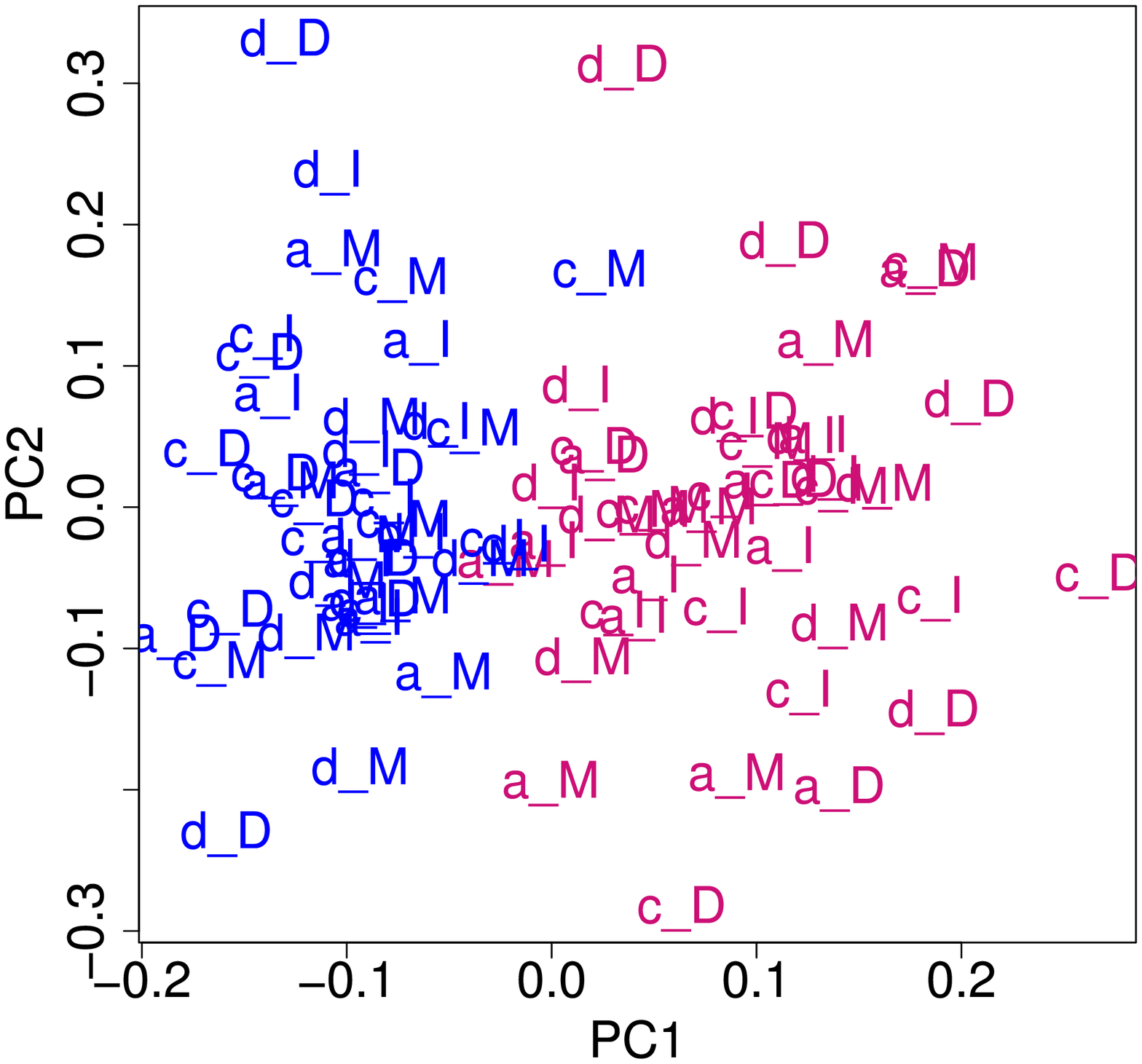}
  \caption{Random alpha with control genes only. Left: no iteration,
    right: with iterations}
  \label{fig:gender_rand}
\end{figure}

The green lines of Figure~\ref{fig:gender_all} correspond to the
random $\alpha$ based corrections using the estimate of $W$ computed
on control genes only. The solid line shows the results for the
non-iterative method. Note that in this case, the $\nu=\sigma_1(W^\top
W)\times 10^{-3}$ is not optimal, as larger $\nu$ lead to much lower
clustering errors. This correction yields good results on this
dataset, as illustrated by the reasonably good separation obtained in
the space of the first two principal components after correction on
the left panel of Figure~\ref{fig:gender_rand}. Note that the variance
$\|W\alpha\|_F^2$ removed by the random $\alpha$ correction is larger
that the one removed by naive RUV-2 and the replicate-based
procedure. However the advantage of ridge RUV-2 does not come from its
doing more deflation, it is rather a handicap here. As we discussed,
random $\alpha$ with a larger penalty $\nu$, and therefore a smaller
$\|W\alpha\|_F$, leads to better results. If we increase $k$ for the
naive and replicate based procedures to get similar $\|W\alpha\|$ as
the random $\alpha$ correction, we obtain clustering scores of $0.99$
and $0.83$ respectively.

In order for the random $\alpha$ correction to work, it is crucial to
have a good estimate of $\Sigma$ or equivalently a good estimate $W$
with the information of how much variance is caused by unwanted
variation along each direction. As a consequence, it is crucial in the
unsupervised case, even more than in the supervised case, to have good
control genes. This is probably why the random $\alpha$ based
correction works well on gender data~: few genes are actually
influenced by gender so in particular the housekeeping genes contain
little signal related to the factor of interest. The dotted green line
of Figure~\ref{fig:gender_all} corresponds to the random $\alpha$
based corrections with iterations plus sparsity, which in this case
leads to lower clustering errors. Here again, sparsity works well
because lots of genes are not affected by gender. As a sanity check,
we tried adding random genes to the control genes used in
non-iterative methods but this did not lead to a clear improvement.

Finally, the red lines of Figure~\ref{fig:gender_all} correspond to
the random $\alpha$ estimators using a combination $\Wtwo$ and
$\Wr$. The full line corresponds to the non-iterative method, and
shows that combining the two estimators of $W$ leads to a better
correction than the two other methods which each rely on only one of
these estimators. Recall however that the purple line uses $\Wr$ in a
fixed $\alpha$ model, and therefore illustrates the improvement
brought by using replicates to estimate $W$, compared to the gray line
which only uses negative control genes. Using $\Wr$ with a random
$\alpha$ model (not shown) leads to better performances than
$\Wtwo$. Combining the estimators further improves the performances,
although not by much. As for the previous methods, adding iterations
to better estimate $\alpha$ improves the quality of the correction.

\begin{figure}
  \centering
  \includegraphics[width=.49\linewidth]{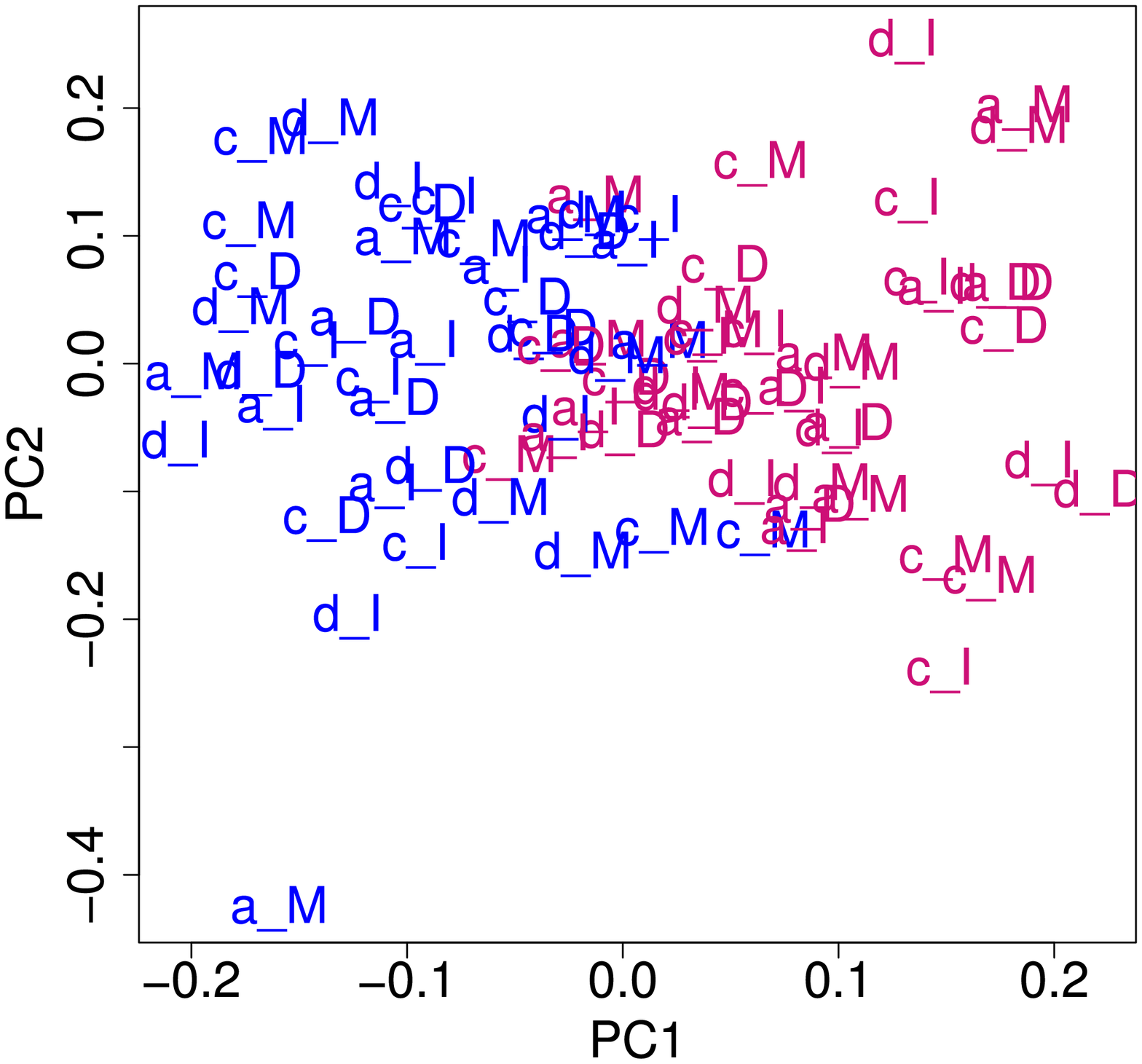}
  \includegraphics[width=.49\linewidth]{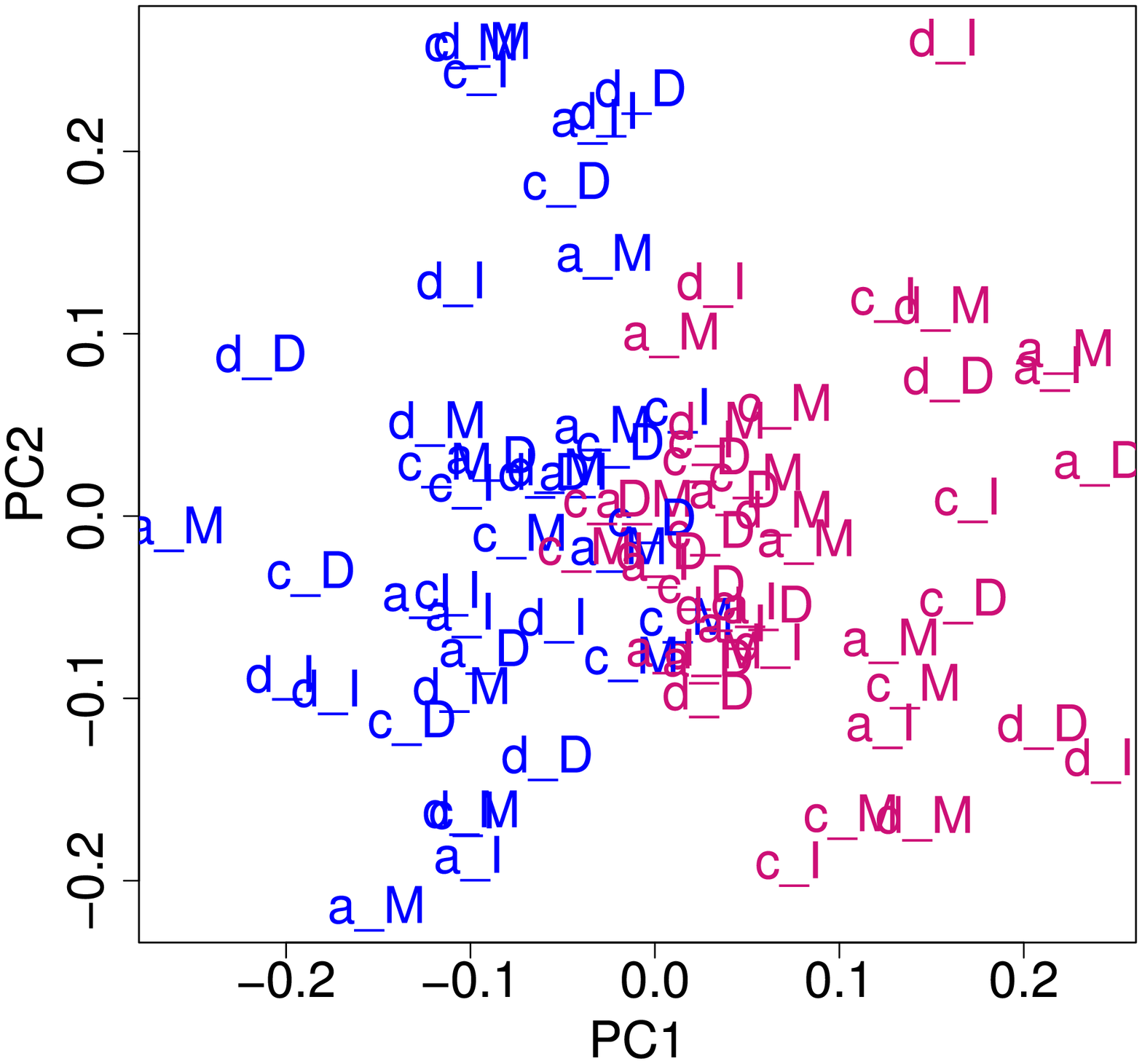}
  \caption{Random alpha with control genes only. Left: no iteration,
    right: with iterations}
  \label{fig:gender_comb}
\end{figure}

We also repeated the same experiment without centering the samples by
array type. The performances were essentially the same as in the
non-centered case for all methods, except that replicate based
corrections do not work anymore and give clustering error $1$ for all
numbers of filtered genes. This is expected since we don't have
replicates for array types and the replicate-based method has no way
to correct for something which does not vary across the
replicates. Using control genes however allows us to identify and
remove the array type effect.

\subsection{TCGA glioblastoma data}

\subsubsection{Data}

We now illustrate the performances of our method on the gene
expression array data generated in the TCGA project for glioblastoma
(GBM) tumors~\citep{Network2008Comprehensive}. These tumors were
studied in detail in \cite{Verhaak2010Integrated}. For each of the
$460$ samples, gene expression was measured on three different
platforms~: Affymetrix HT-HG-U133A Genechips at the Broad Institute,
Affymetrix Human Exon 1.0 ST Genechips at Lawrence Berkeley Laboratory
and Agilent 244K arrays at University of North
Carolina. \cite{Verhaak2010Integrated} selected $200$ tumors and $2$
normal samples from the dataset based on sample quality criterions and
filtered $1740$ genes based on their coherence among the three
platforms and their variability within each platform. The expression
values from the three platforms were then merged using factor
analysis. They identified four GBM subtypes by clustering analysis on
this restricted dataset~: Classical, Mesenchymal, Proneural and
Neural. We study these $202$ samples across the three platforms,
keeping all the $11861$ genes in common across the three
platforms. Among these $202$ samples, $173$ were identified by
\cite{Verhaak2010Integrated} as ``core'' samples~: they were good
representers of each subtypes. $38$ of them are Classical, $56$
Mesenchymal, $53$ Proneural and $26$ Neural. 

\subsubsection{Design}

For the purpose of the experiment, we study how well a particular
correction allows us to recover the correct label of the $147$
Classical, Mesenchymal and Proneural tumors, leaving the other ones
aside. Our objective is to recover the correct subtypes using a
$k$-means with $3$ clusters. We consider two settings. In the first
one, we use a full design with all $147$ samples from $3$
platforms. In the second one we build a confounding setting in which
we only keep the Classical samples on Affymetrix HT-HG-U133A arrays,
the Mesenchymal samples on Affymetrix Human Exon arrays and the
Proneural samples on Agilent 244K arrays. In each case, we use $5$
randomly selected samples that we keep for all $3$ platforms and use
as replicates. We do not use other samples as replicates even in the
full design when all samples could potentially be used as
replicates. Among the $5$ selected samples one was Neural, two
Proneural, and two were not assigned a subtype. The results presented
are qualitatively robust to the choice of these replicates.

In the confounded design, a correction which simply removes the
platform effect is likely to also lose all the subtype signal because
it is completely confounded with the platform, up to the replicate
samples. The reason why we only keep $3$ subtypes is to allow such a
total confounding of the subtypes with the $3$ platforms. However in
this design, applying no correction at all is likely to yield a good
clustering by subtype because we expect the platform signal to be very
strong. A good correction method should therefore perform well in both
the confounded and the full design. In the full design, the platform
effect is orthogonal to the subtype effect so we expect the correction
to be easier. Of course in this case, the uncorrected data is expected
to cluster by platform which this time is very different from the
clustering by subtype since each sample is present on each platform.

\subsubsection{Result}

\begin{table}[h]
  \centering
  \begin{tabular}{l|| l | l}
    Method & Full & Confounding\\\hline
    No correction & $2$ & $0$\\
    Mean-centering & $0.3$ & $1.93$\\
    Ratio method & $0.31$ & $0.79$\\
    Naive RUV-2 & $2$ & $0$\\\hline
    Random $\alpha$ & $0.21$ & $1.5$\\
    \hfil + iterations & $0.15$ & $1$\\\hline
    Replicate based & $0.2$ & $0.61$\\
    \hfil + iterations & $0.17$ & $0.16$\\\hline                                     
    Combined & $1.3$ & $1.2$\\
    \hfil + iterations & $1.3$ & $1$\\
  \end{tabular}
  \caption{Clustering error of TCGA GBM data with full and confounded designs for various
    correction methods. Since there are $3$ clusters, errors range
    between $0$ and $2$. }
  \label{tab:gbm}
\end{table}

\begin{figure}
  \centering
  \includegraphics[width=.49\linewidth]{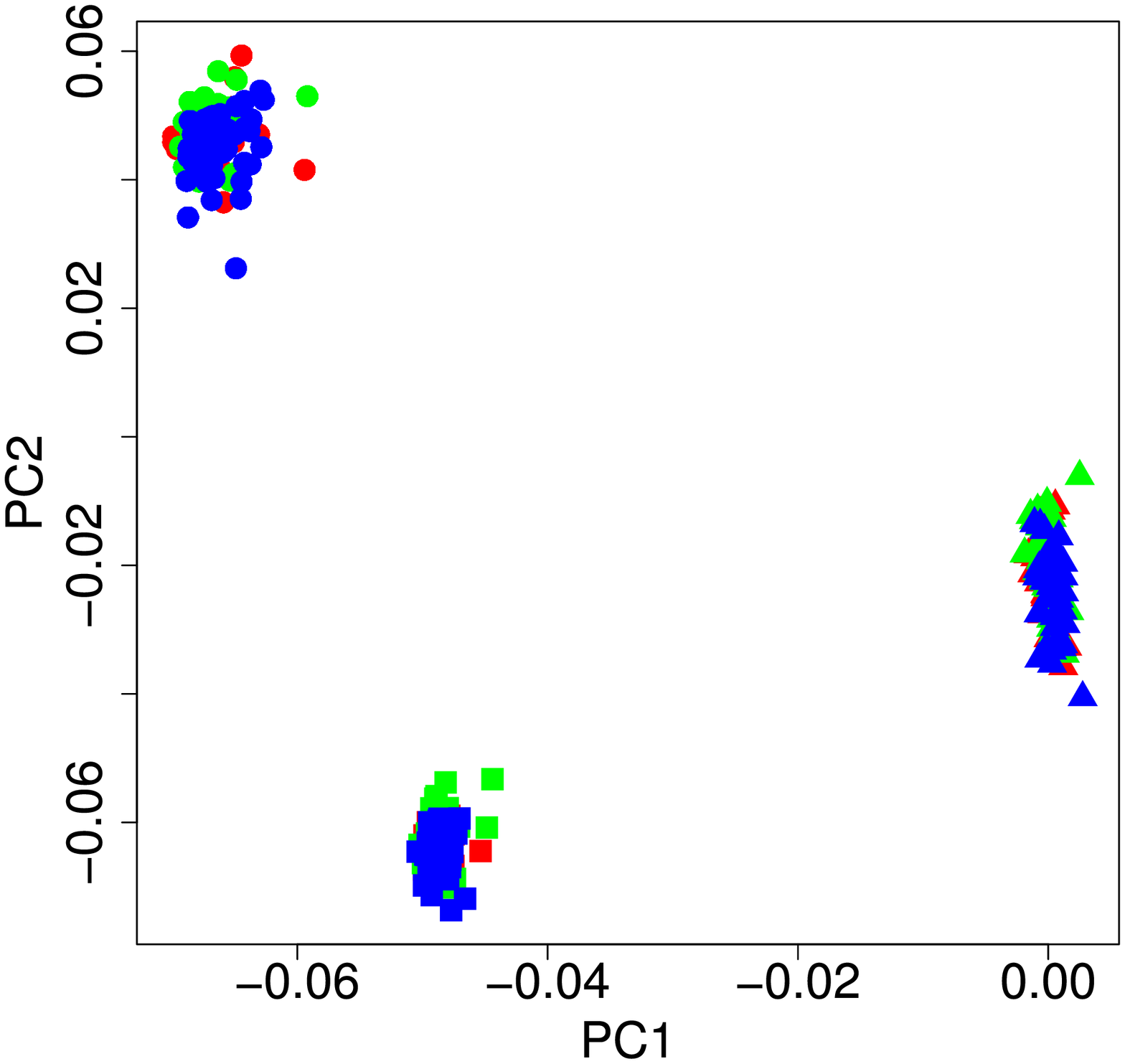}
  \includegraphics[width=.49\linewidth]{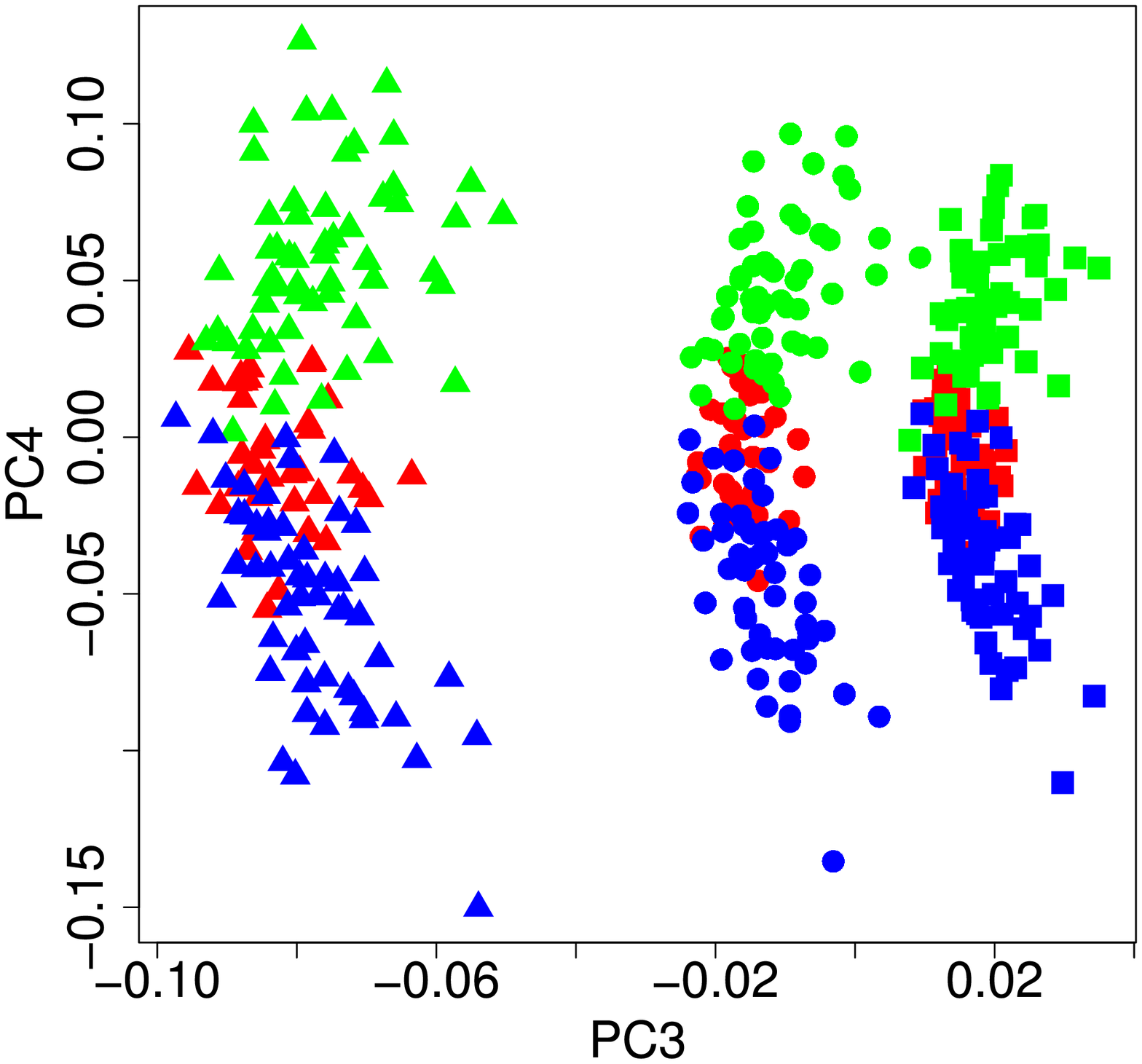}
  \caption{Uncorrected full design GBM data in the space of their
    first four principal components. Left panel~: PC 1 and 2, right
    panel~: PC 3 and 4. Colors denote subtypes, shapes denote platforms.}
  \label{fig:gbm_uncorr}
\end{figure}

Table~\ref{tab:gbm} shows the clustering error obtained for each
correction method on the two designs. Recall that since there are $3$
clusters, clustering errors range between $0$ and $2$. Most of the
plots of the corrected data in the space spanned by the first two
principal components are deferred to appendix~\ref{app:gbmPlots}. As
expected, the uncorrected data give a maximal error on the full design
and $0$ in the presence of confounding. This is because, as seen on
Figure~\ref{fig:gbm_uncorr}, the uncorrected data cluster by platform
which in the full design are orthogonal to the subtypes and in the
second design are confounded with subtypes. For similar reasons,
centering the data by platform works well in the full design but fails
when there is confounding because removing the platform effect removes
most of the subtype effect. When replicates are available, a variant
of mean centering is to remove the average of replicate samples from
each platform. This is known as the ratio
method~\citep{Luo2010comparison} and does improve on regular
mean-centering in the presence of confounding. A disadvantage of this
method is that it amounts to considering that $W$ is a partition of
the data by batch (in this case by platform) whereas as discussed in
Section~\ref{sec:knownW} the actual unwanted variation may be a non
linear function of the batch, possibly involving other factors. Note
that we do not explicitly assess the ratio method for the other
benchmarks because all samples are used as replicates so the ratio
methods becomes equivalent to mean centering.

Naive RUV-2 gives a maximal error in the full design, $0$
otherwise. This result is actually caused by the fact that naive RUV-2
is extremely sensitive to the choice of $k$. Since the total number of
differences formed on the $15$ replicates is $10$, we use $k=2$ as a
default for naive RUV-2 and the replicate based procedure. In the full
design, the platform effect is contained in the first three principal
components of the control gene matrix $Y_c$ and the fourth principal
component contains the subtype effect. This can clearly be seen on
Figure~\ref{fig:gbm_uncorr}. Removing only one or two directions of
variance leaves too much platform effect. Removing a third one ($k=3$)
gives a small error of $0.15$ and removing a fourth one gives an error
of $1.83$. When the platform is confounded with the subtype, removing
one or two components leads to a perfect clustering by subtype because
the third principal component still contains platform/subtype
signal. Removing more does not allow us to recover a clustering by
subtype. So if we used $k=3$ instead of $k=2$, the result would be
inverted~: good for the full design, bad in the presence of
confounding. The random $\alpha$ model works well in the full design,
less so in the presence of confounding, as illustrated on
Figure~\ref{fig:gbmRidge} and~\ref{fig:gbmConfRidge}
respectively. While it was reasonably robust to the choice of the
ridge parameter $\nu$ on the gender data, it is more sensitive on this
one. Using $\nu=\sigma_1(W^\top W)\times 5.10^{-2}$ instead of
$10^{-3}$ does not remove enough platform effect and leads to an error
of $2$ on the full design, $0$ in the presence of confounding. Using a
smaller factor of $5.10^{-4}$ leads to an error of $0.27$ ($1.65$ with
confounding), and $10^{-4}$ to an error of $1.94$ ($1.98$ with
confounding). Because the correction made by the random $\alpha$ model
is softer than the one of the fixed $\alpha$ naive RUV-2, using
$\nu=\sigma_1(W^\top W)\times 10^{-3}$ allows us to recover subtype
signal in both designs. The sensitivity to $\nu$ is likely to be
caused by the large difference of magnitude between $\sigma_1(W^\top
W)$ and the next eigen values~: the first one represents $98\%$ of the
total variance. This is to be expected in most cases in presence of a
strong technical batch effect. In both designs, using iterating
between estimation of $X\beta$ using sparse dictionary learning and
estimation of $\alpha$ using ridge regression further improves the
performances. The replicate-based correction gives good results for
both designs, as illustrated on Figure~\ref{fig:gbmRep}
and~\ref{fig:gbmConfRep}. Like for the gender data, it seems to be
robust to the choice of $k$. For $k=10$ it gives errors $0.2$ and
$0.53$ in the first and second design respectively. Here again, adding
iterations on $(X\beta, \alpha)$ improves the quality of the
correction in each case. Finally the combined method gives a
clustering error of $1.3$ and $1.2$ in the first and second design
respectively. This is not a very good performance but the method still
manages to retain some of the factor of interest and does so
consistently in the presence and in the absence of confounding. The
corresponding corrected data are shown in Figure~\ref{fig:gbmComb} and
\ref{fig:gbmConfComb} for the full and confounded design respectively.
Iterations improve the performance in the confounding design and do
not change the result in the full design. Figure~\ref{fig:gbmComb}
suggests that for the full design, the problem comes from the fact
that the correction does not remove all of the platform effect.

Note that, as with the gender data, the difference observed between
the correction methods cannot be only explained by the fact that some
of them remove more variance than others. For example in the full
design, naive RUV-2, the replicate based procedure and the random
$\alpha$ correction lead to similar $\|W\alpha\|_F$ but to very
different performances~: naive RUV-2 fails to remove enough platform
signal whereas the other corrections remove enough of it for the
arrays to cluster by subtype. In the confounding design, both naive
RUV-2 and the replicate based procedure lead to similar
$\|W\alpha\|_F$ and similar performances. The random $\alpha$
correction leads to a larger $\|W\alpha\|_F$ which explains its poor
behavior. Estimating what amount of variance should be removed is part
of the problem, so it would not be correct to conclude that the random
$\alpha$ correction works as well as the others in this case. It is
however interesting to check whether the problem of a particular
method is its removing too much or not enough variance or whether it
is a qualitative problem.

\subsection{MAQC-II data}

We finally assess our correction methods on a gene expression dataset
which was generated in the context of the MAQC-II
project~\citep{Shi2010MicroArray}. The study was done on rats and the
objective was to assess hepatotoxicity of $8$ drugs. For each drug,
three time points were done for three different doses. For each of
these $8 \times 3 \times 3$ combinations, $4$ animals were tested for
a total of $288$ animals. For each animal, one blood and one liver
sample were taken. Gene expression in blood and in the liver were
measured using Agilent arrays and gene expression in the liver was
also measured using Affymetrix arrays. The Agilent arrays were loaded
using the \texttt{marray} R package. Each array was loess normalized,
dye swaps were averaged and each gene was then assigned the median log
ratio of all probesets corresponding to the gene. The Affymetrix
arrays were normalized using the \texttt{gcrma} R package. Each gene
was then assigned the median log ratio of all probesets corresponding
to the gene. For this experiment we retain samples from all platforms
and tissues for the highest dose of each drug and for the last two
time points $24$ hours and $48$ hours. Most of these drugs are not
supposed to be effective for the earlier time points. This leads to a
set of $186$ arrays that we restrict to the $9502$ genes which are
common to all platforms. Each sample has a replicate for each tissue
and platform, but there is no replicate against the time effect. For
control genes, we used the same list of housekeeping genes as for the
other datasets but converted to their rat orthologs, leading to $210$
control genes.

The interest of this complex design is obvious for the purpose of this
paper~: the resulting dataset contains a large number of arrays
measuring gene expression influenced by the administered drug which we
consider to be our factor of interest and by numerous unwanted
variation factors. Array type, tissue, time and dose are likely to
influence gene expression, preventing the arrays from clustering by
drug. This clustering problem is much harder than the gender and
glioblastoma ones. First of all, the drug signal may not be as strong
as the gender which at least for a few genes is expected to be very
clear or as the glioblastoma subtypes which were defined on the same
dataset. Second and maybe more important, it is an $8$-class
clustering problem, which is intrinsically harder than $2$- or
$3$-class clusterings. Finally as we discuss in
Section~\ref{sec:ctlBenefit}, the control genes for this dataset do
not behave as expected.

\begin{table}[h]
  \centering
  \begin{tabular}{l|| l}
    Method & Error\\\hline
    No correction & $5.9$ \\
    Mean-centering & $5.1$\\
    Naive RUV-2 & $6.6$\\\hline
    Random $\alpha$ & $4.7$\\
    \hfil + iteration & $5.4$\\\hline
    Replicate based & $2.8$ -- $3.8$ \\ 
    \hfil + iterations & $2.8$ -- $3.8$\\\hline
    Combined & $5.7$\\
    \hfil + iterations & $5.7$\\
  \end{tabular}
  \caption{Clustering error of MAQC-II data for various correction methods.}
  \label{tab:maqc}
\end{table}

\begin{figure}
  \centering
  \includegraphics[width=.49\linewidth]{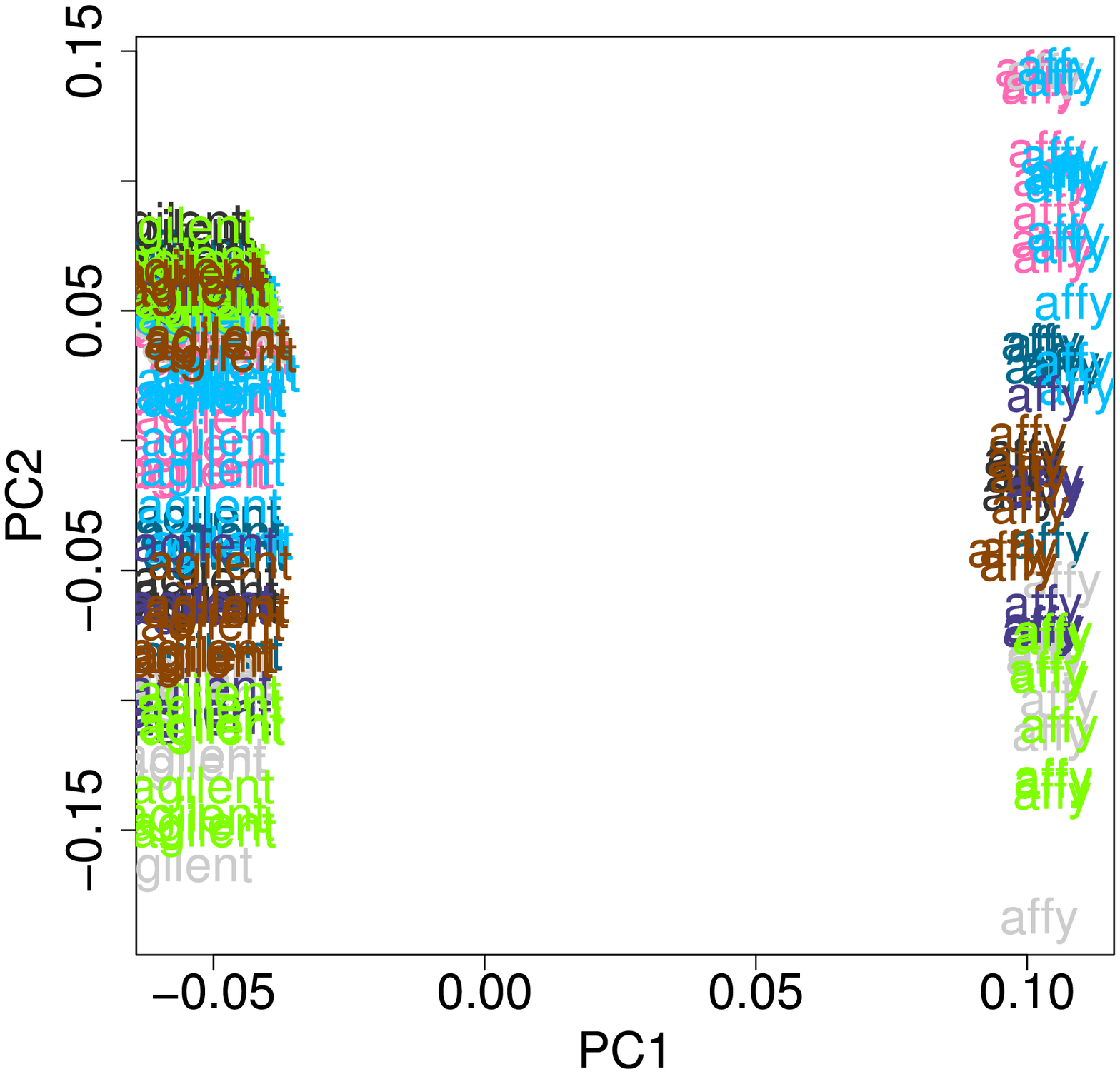}
  \includegraphics[width=.49\linewidth]{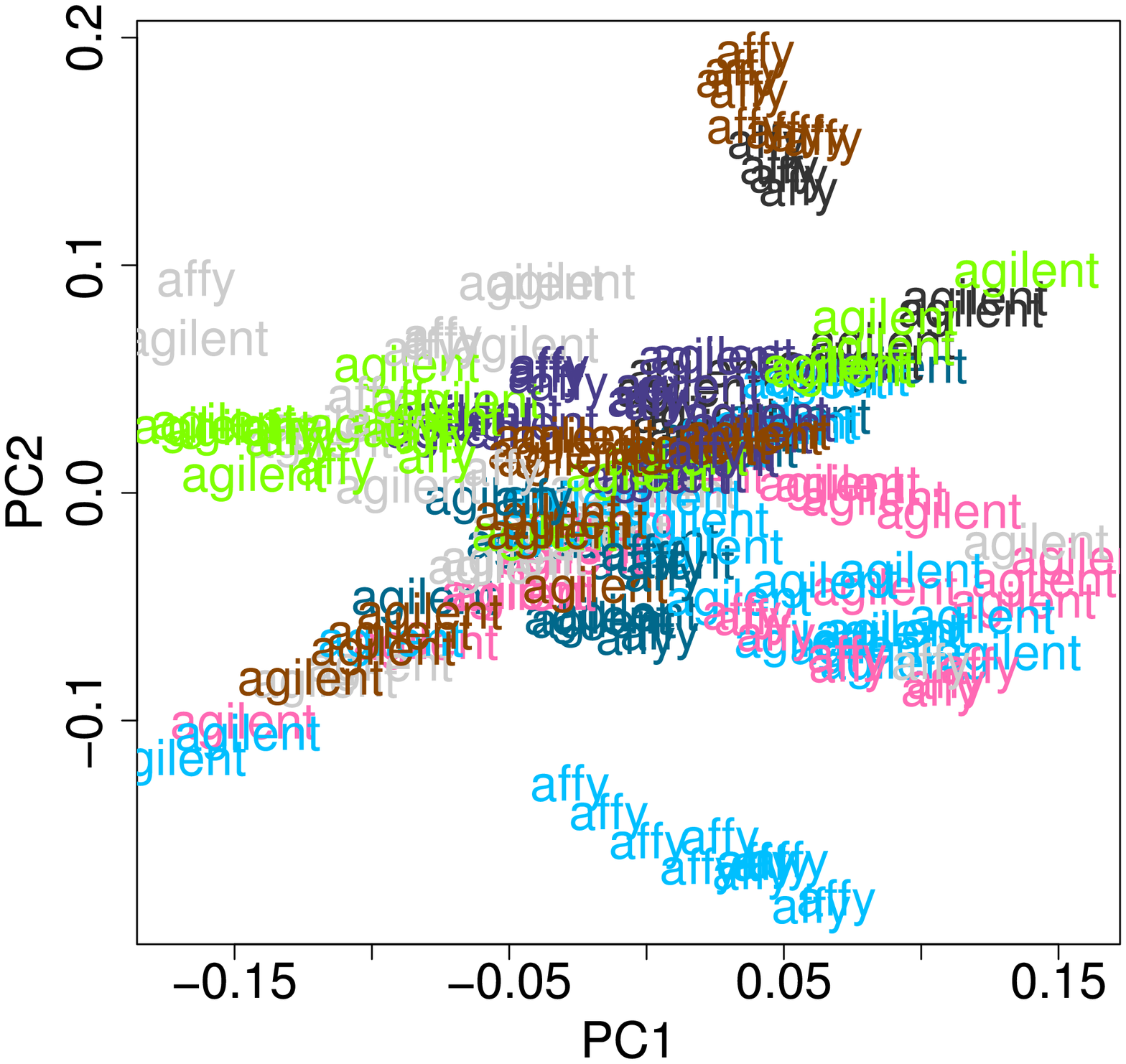}
  \caption{Samples of the MAQC-II study represented in the space of
    their first two principal components before correction (left
    panel) and after centering by tissue and platform region (right
    panel). Each color represents a different drug. The labels
    indicate the platform of each sample.}
  \label{fig:niehs_uncorr}
\end{figure}

\begin{figure}
  \centering
  \includegraphics[width=.49\linewidth]{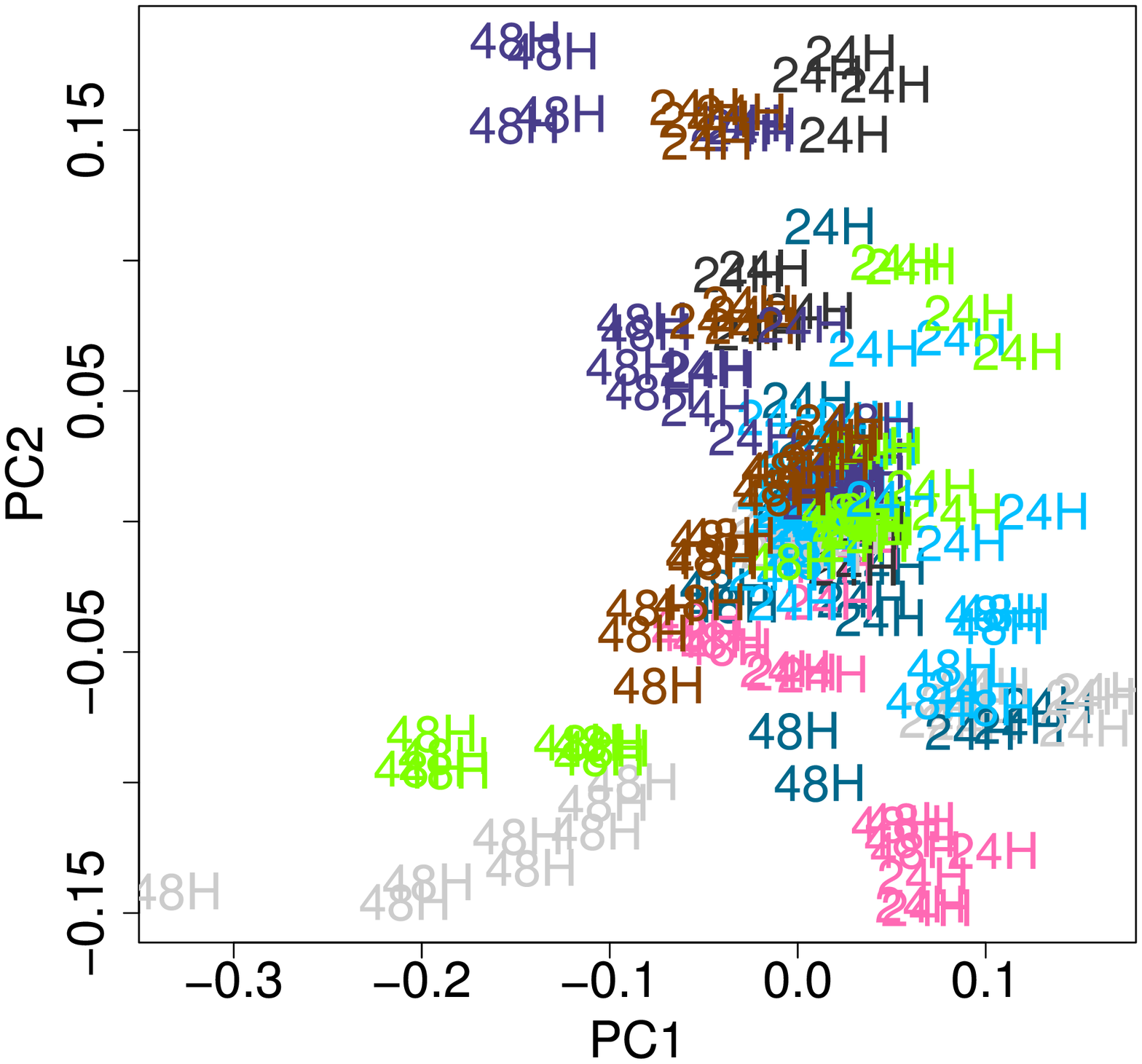}
  \includegraphics[width=.49\linewidth]{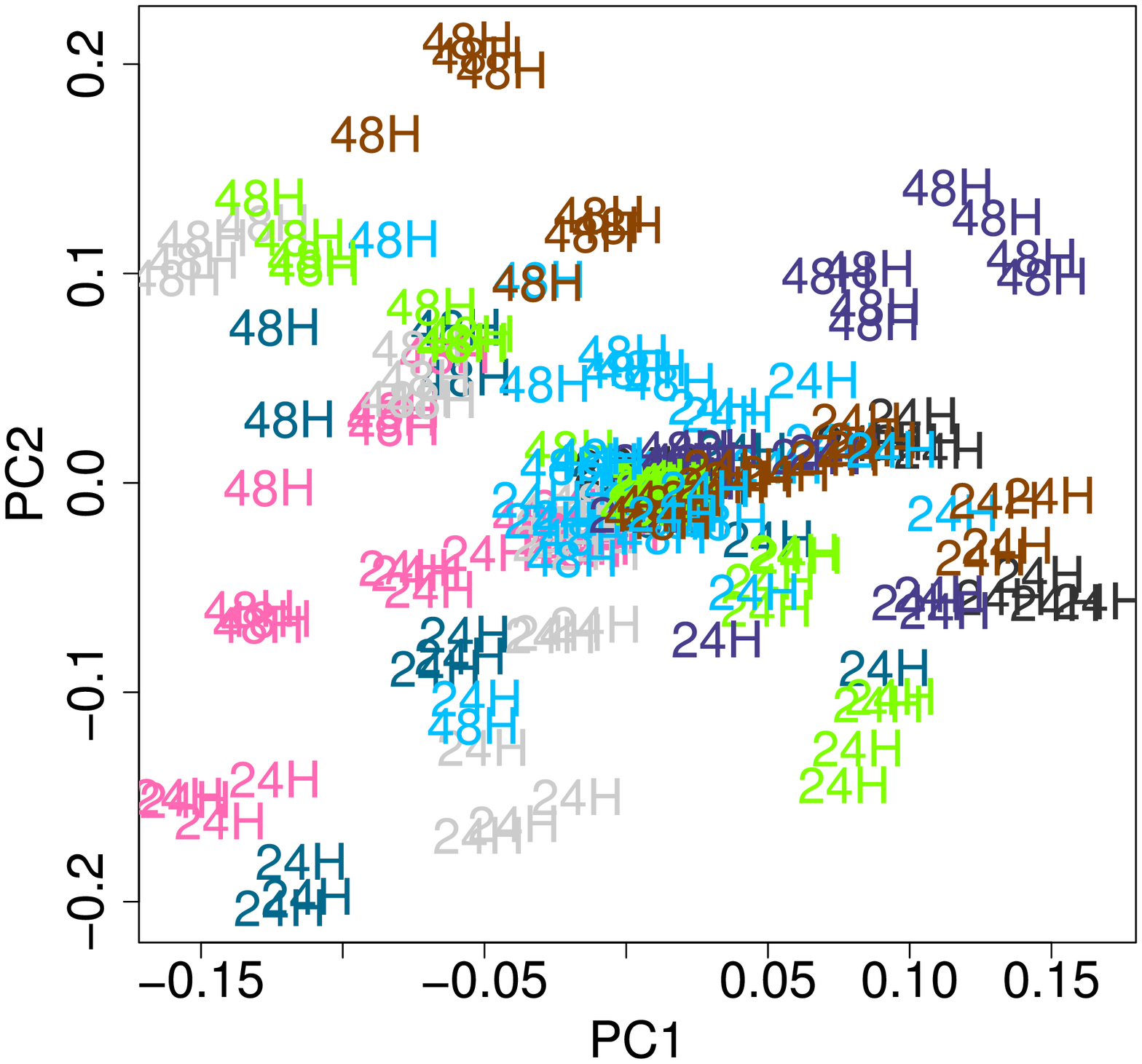}
  \caption{Samples of the MAQC-II study represented in the space of
    their first two principal components after applying the random
    $\alpha$ correction (left panel) and its iterative variant (right
    panel). Each color represents a different drug. The labels
    indicate the time of each sample.}
  \label{fig:niehs_ridge}
\end{figure}

\begin{figure}
  \centering
  \includegraphics[width=.49\linewidth]{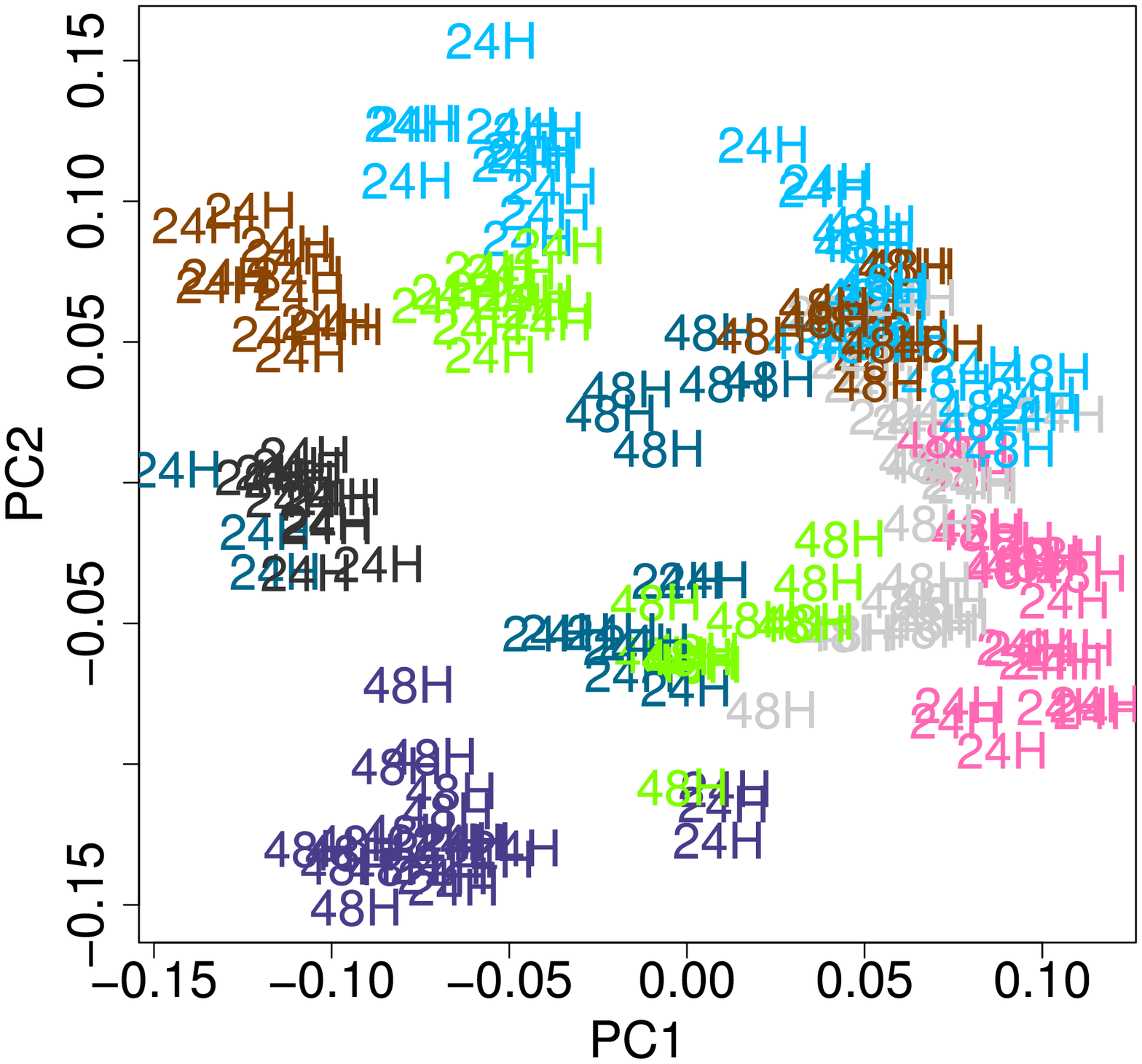}
  \includegraphics[width=.49\linewidth]{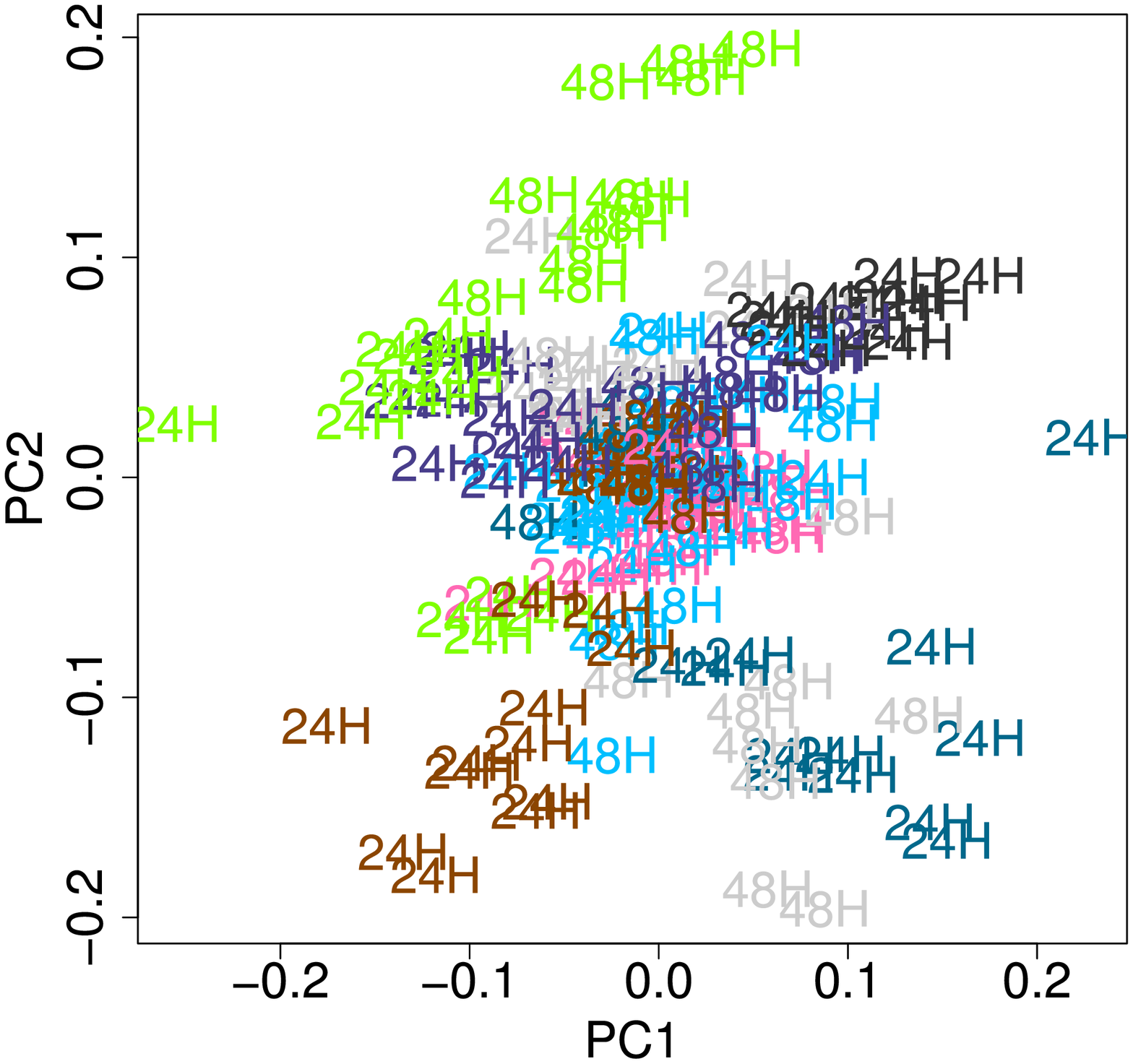}
  \caption{Samples of the MAQC-II study represented in the space of
    their first two principal components after applying the
    replicate-based correction (left panel) and its iterative variant
    (right panel). Each color represents a different drug. The labels
    indicate the time of each sample.}
  \label{fig:niehs_rep}
\end{figure}

\begin{figure}
  \centering
  \includegraphics[width=.49\linewidth]{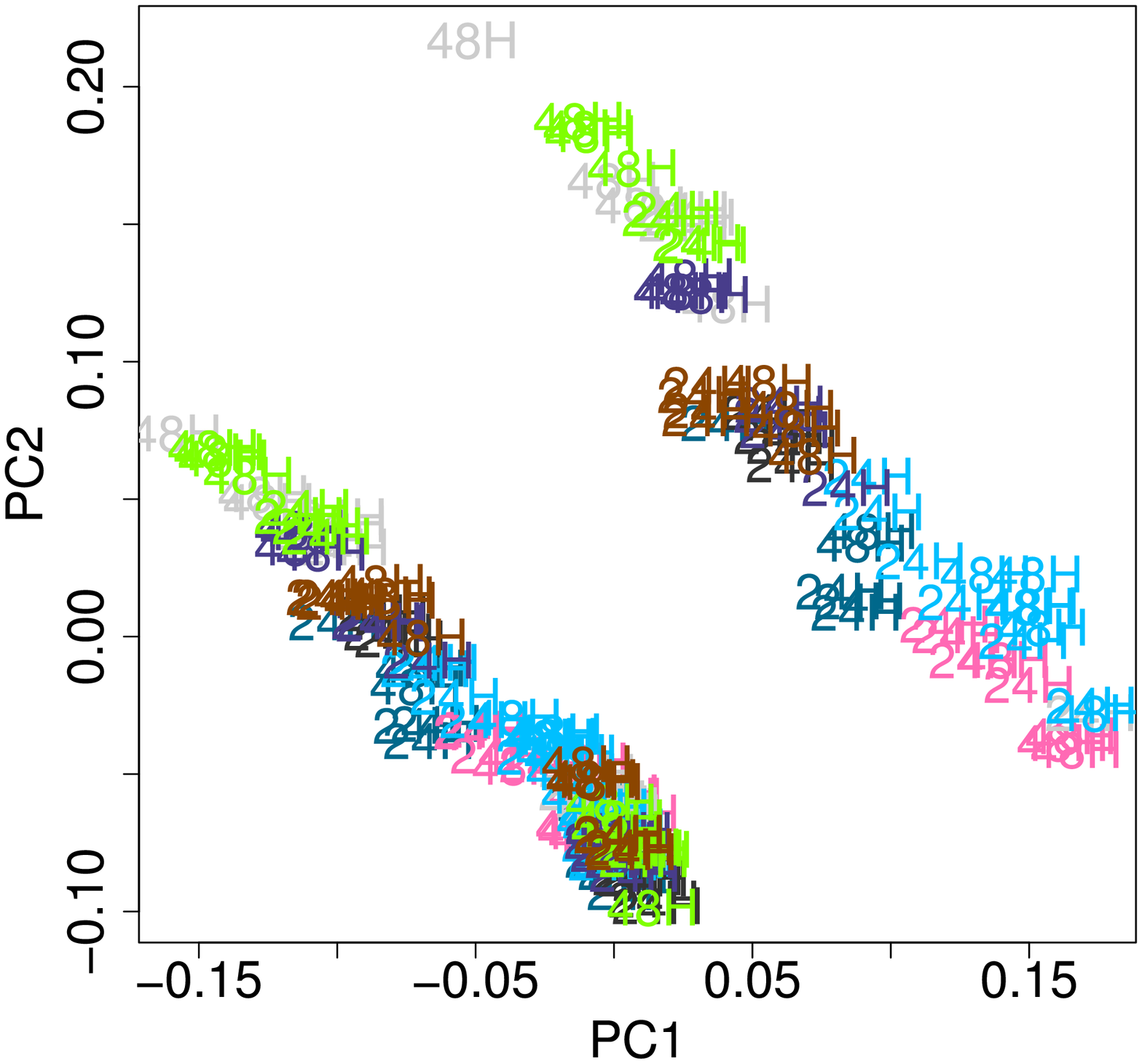}
  \includegraphics[width=.49\linewidth]{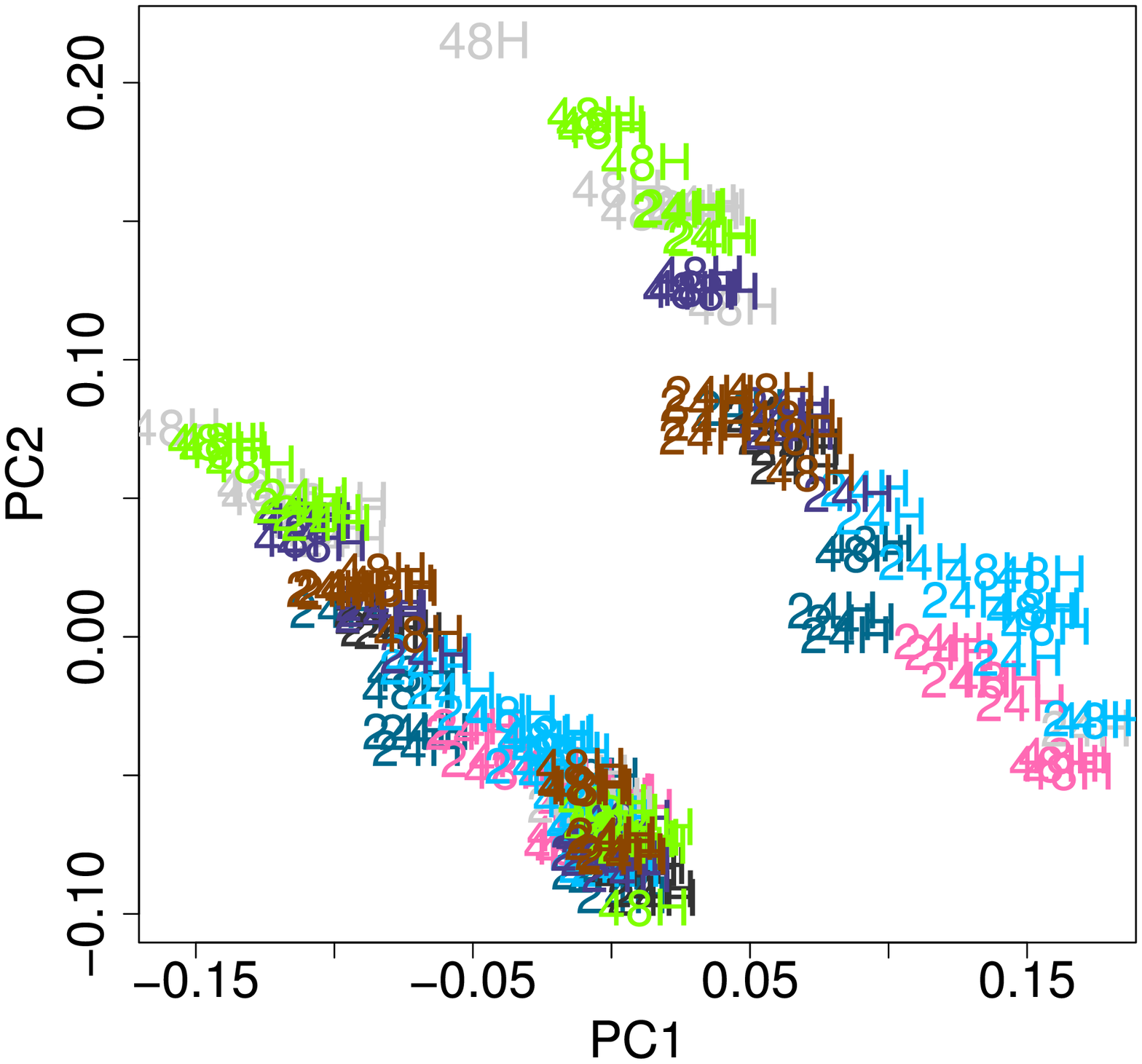}
  \caption{Samples of the MAQC-II study represented in the space of
    their first two principal components after applying the combined
    correction (left panel) and its iterative variant (right
    panel). Each color represents a different drug. The labels
    indicate the time of each sample.}
  \label{fig:niehs_comb}
\end{figure}

The errors obtained after applying each correction are displayed in
Table~\ref{tab:maqc}. Recall that for this dataset, we are trying to
recover a partition in $8$ classes corresponding to the $8$ drugs of
the study so the maximum clustering error is $7$. The left panel of
Figure~\ref{fig:niehs_uncorr} represents the uncorrected samples in
the space of the first two principal components. The first principal
components is clearly driven by the presence of two different types of
arrays. The clustering error in this case is $5.9$. Centering by
platform-tissue, \emph{i.e.} centering separately the Affymetrix
arrays, the Agilent liver and the Agilent blood, the data points do
not cluster by platform anymore but just like for the gender data this
does not lead to a clear clustering by drug. This can be seen on the
right panel of Figure~\ref{fig:niehs_uncorr}. The resulting clustering
error is $5.1$. The naive RUV-2 correction doesn't lead to any
improvement compared to the uncorrected data, leading to an error of
$6.6$. The random $\alpha$ estimator hardly improves the performances,
and its iterative variant even increases a little the
error. Figure~\ref{fig:niehs_ridge} shows that these methods lead to a
better organization of the samples by drug, but still far from a clean
clustering. The replicate-based method leads to better
performances. Even though we do $200$ runs of $k$-means to minimize
the within sum of square objective, different occurrences of the $200$
runs lead to different clusterings with close objectives. We choose to
indicate the range of clustering errors given by these different
clusterings ($2.8$--$3.8$). The iterative version of the estimator
gives the same range of errors. Figure~\ref{fig:niehs_rep} shows that
these corrections indeed lead to a better organization of the samples
by drugs in the space spanned by the first two principal components,
but fails to correct the time effect against which no replicate is
available. Finally the combined method gives poor
performances. Figure~\ref{fig:niehs_comb} shows that it does not
correct enough and that the data still cluster by platform.
Increasing the magnitude of the correction by using a smaller ridge
$\nu$ removes the platform effect but still doesn't lead to a good
clustering by drug. As it can be seen on Figure~\ref{fig:niehs_comb},
the iterative version of the estimator leads to a very similar and
equally poor estimate. The deflation $\|W\alpha\|_F$ obtained by the
naive RUV-2 and replicate based procedures are larger than the one
obtained by the random $\alpha$ correction, but this is not the reason
for the replicate based procedure to work better than the random
$\alpha$ correction~: the former is quite robust to changes in $k$ and
the latter does not improve when changing $\nu$.

\subsection{Benefit of control genes}
\label{sec:ctlBenefit}

We have assumed so far that control genes had little association with
$X$ and allowed proper estimation for the methods we introduced. In
this section, we assess this hypothesis on our three
benchmarks. Table~\ref{tab:ctlVsTotal} reproduces the results on the
first two benchmarks of the non-iterative methods that we considered
and which make use of control genes. In addition for each method and
each of our first two benchmarks, we show the performance of the same
method using all genes as control genes. For the gender data, we give
the clustering error when filtering in $1260$ genes, which correspond
to the last point of Figure~\ref{fig:gender_all}.

\begin{table}[h]
  \centering
  \begin{tabular}{l|| p{1.5cm} | p{1.5cm} || p{1.5cm} | p{1.5cm} || p{1.5cm} | p{1.5cm}}
    Method & Gender control & Gender all genes & GBM 1 control & GBM 1
    all genes & GBM 2 control & GBM 2 all genes\\\hline
    Naive RUV-2 & $0.75$ & $0.92$ & $2$ & $1.52$ & $0$ & $0.93$\\
    Replicate-based & $0.77$ & $0.77$ & $0.2$ & $0.25$ & $0.61$ & $0.37$\\ 
    Random $\alpha$ & $0.67$ & $0.99$ & $0.21$ & $0.24$ & $1.5$ &
    $1.8$\\
    Combined & $0.45$ & $0.54$ & $1.3$ & $1.3$ & $1.2$ & $1.8$\\
  \end{tabular}
  \caption{Clustering error of gender and glioblastoma data with full
    (1) or confounded (2) designs for various correction methods
    relying on control genes using either all genes or control genes.}
  \label{tab:ctlVsTotal}
\end{table}

The results of MAQC-II data are not presented in
Table~\ref{tab:ctlVsTotal} but the result of each method is the same
whether we use our control genes or all the genes for this
dataset. Overall, we can see that some methods are affected by the use
of control genes on the gender data, but using all the genes only
mildly affects the performances of most methods on the GBM dataset,
and as we said do not affect the performances on the MAQC-II dataset
at all. This suggests that the genes that we used as control genes
were indeed less affected by the factor of interest for the gender
data but were not for the glioblastoma and MAQC-II data. This is
consistent with the fact that methods which rely heavily on the
control genes like naive RUV-2 and random $\alpha$ are very sensitive
to the amplitude of the correction for the glioblastoma dataset and do
not work for the MAQC-II dataset. As one may expect from the
discussion of Section~\ref{sec:shotW}, methods using replicates,
\emph{i.e.}, the replicate-based one introduced in
Section~\ref{sec:shot} and the combined one seem less affected than
methods that solely rely on control genes, even on the gender
dataset. Remember that our replicate-based procedure estimates $W$ by
regressing the control genes $Y_c$ against the variations observed
among constrasts of replicates which can make it robust to the fact
that control genes are affected by the factor of interest.

\begin{figure}
  \centering
  \includegraphics[width=.32\linewidth]{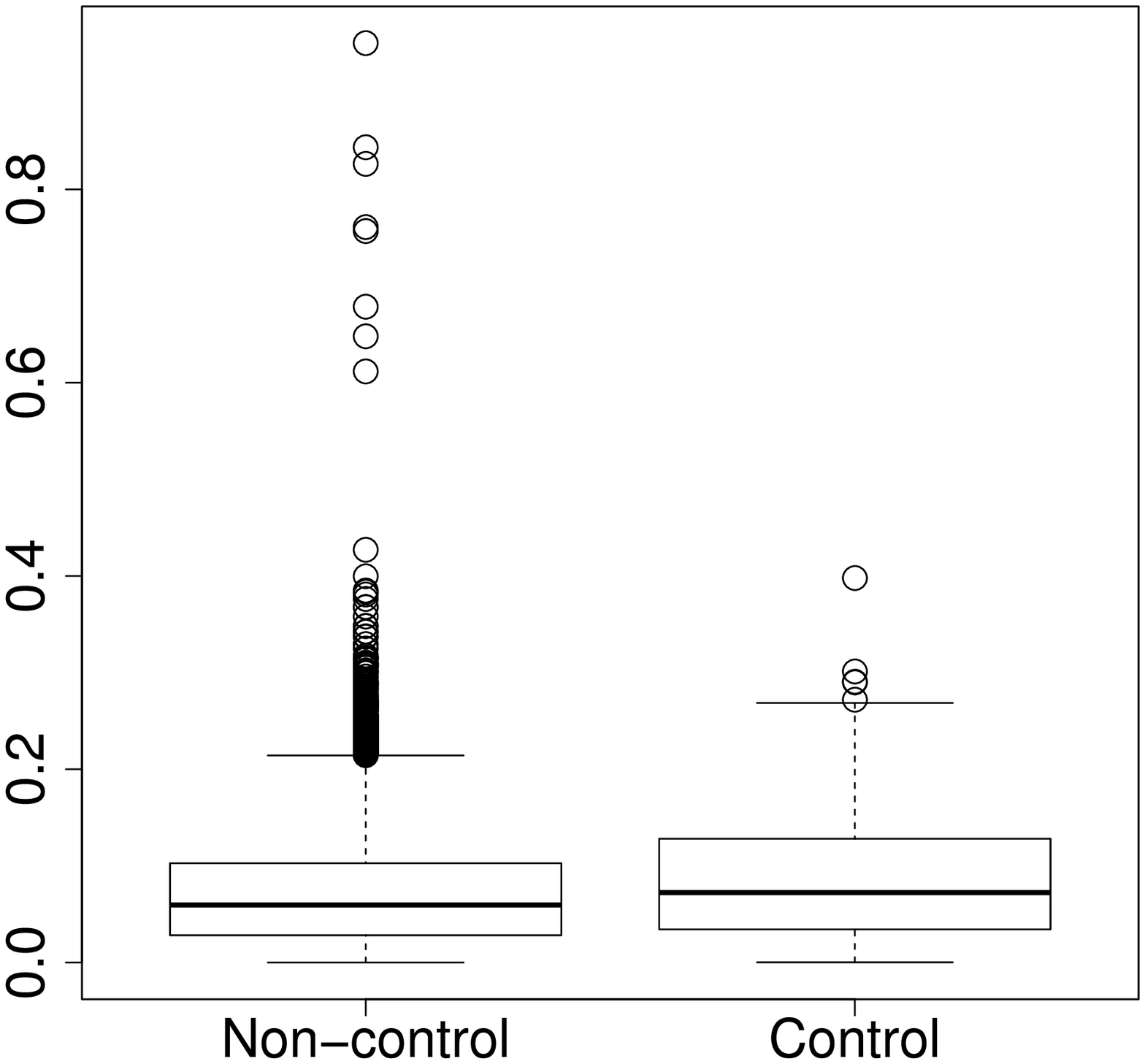}
  \includegraphics[width=.32\linewidth]{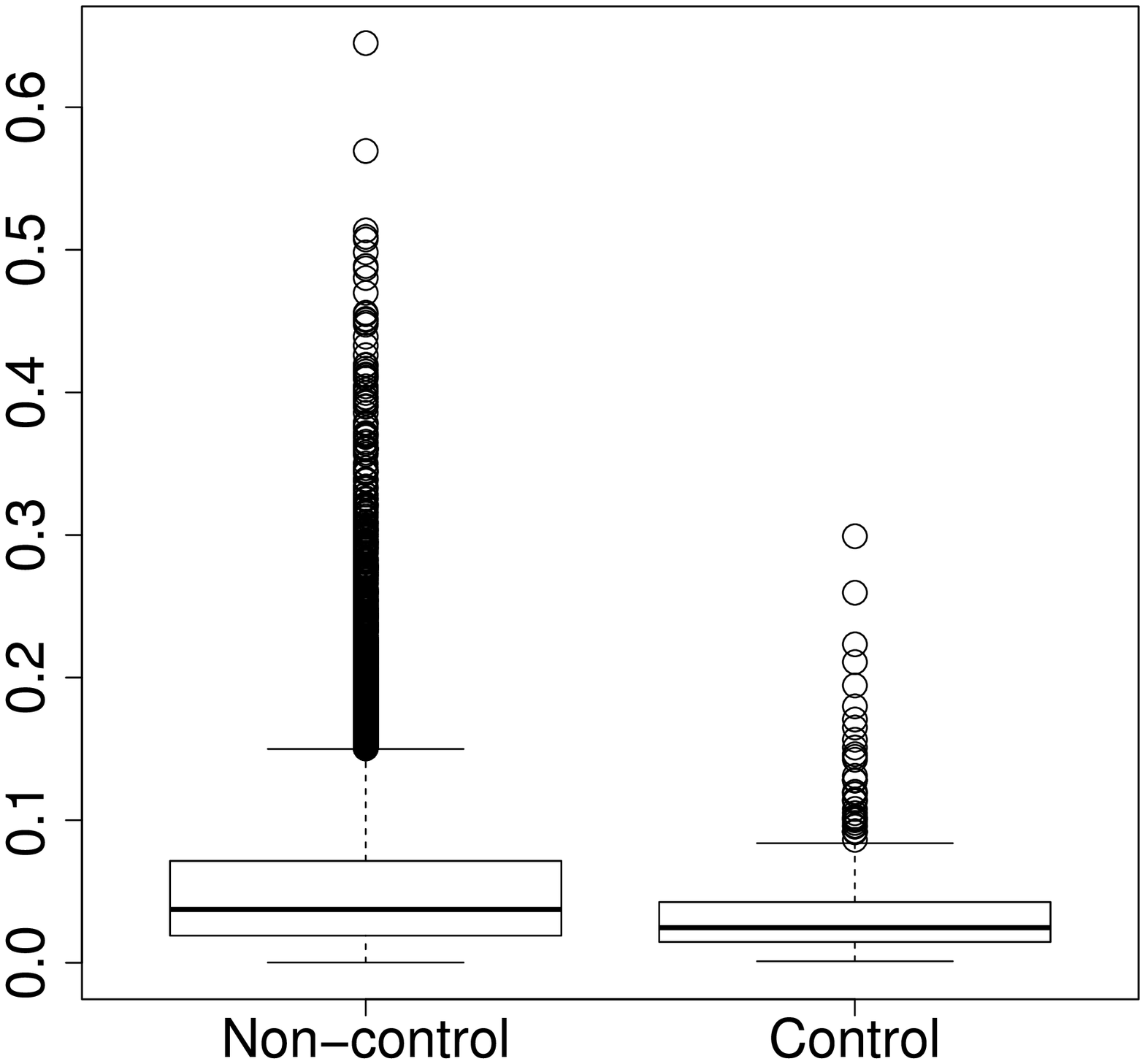}
  \includegraphics[width=.32\linewidth]{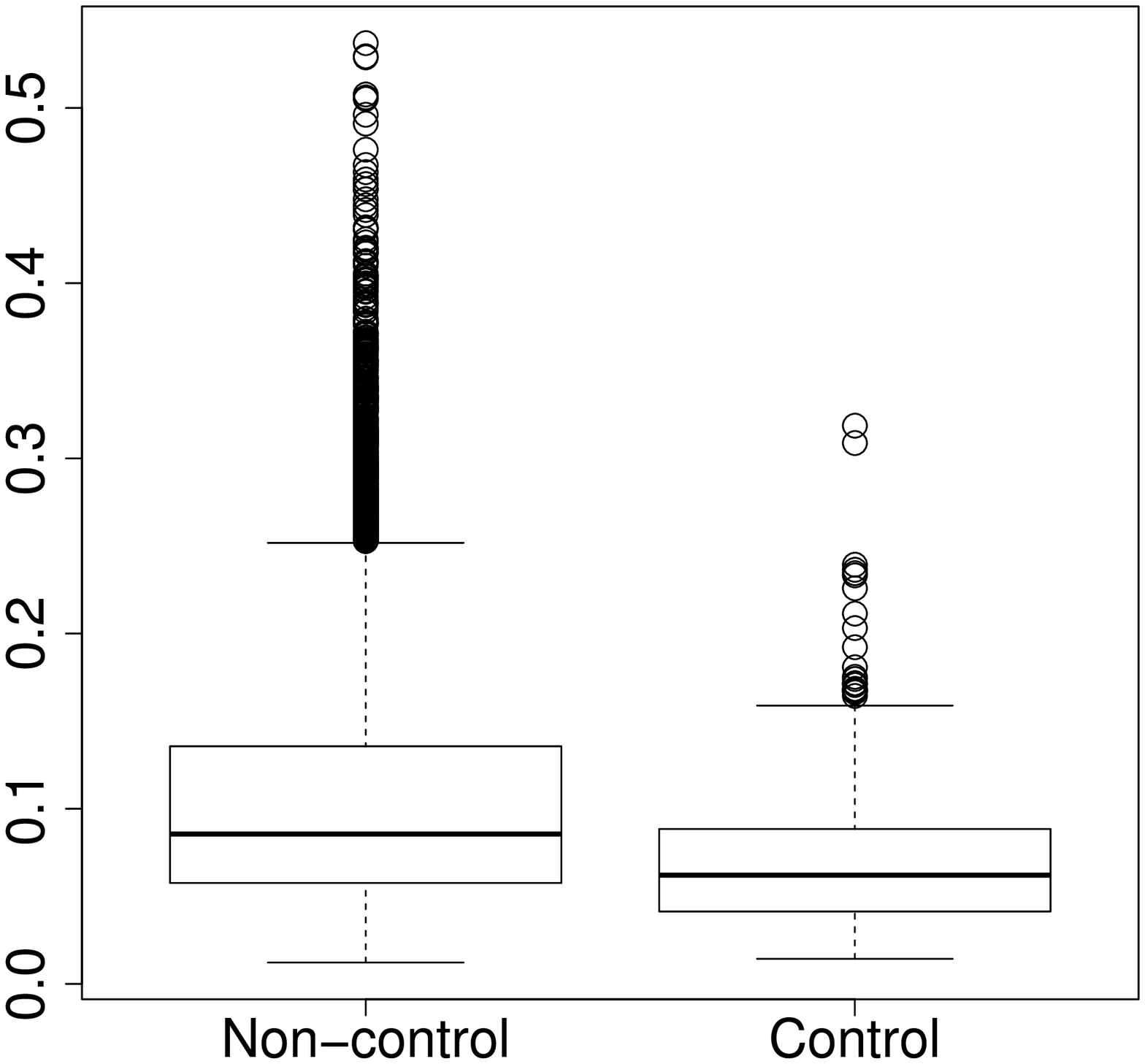}
  \caption{boxplot of the CCA of control and non-control genes with the factor of
    interest $X$ for the gender (left panel), glioblastoma (center panel)
    and MAQC-II (right panel) datasets.}
  \label{fig:cgQual}
\end{figure}

\begin{figure}
  \centering
  \includegraphics[width=.32\linewidth]{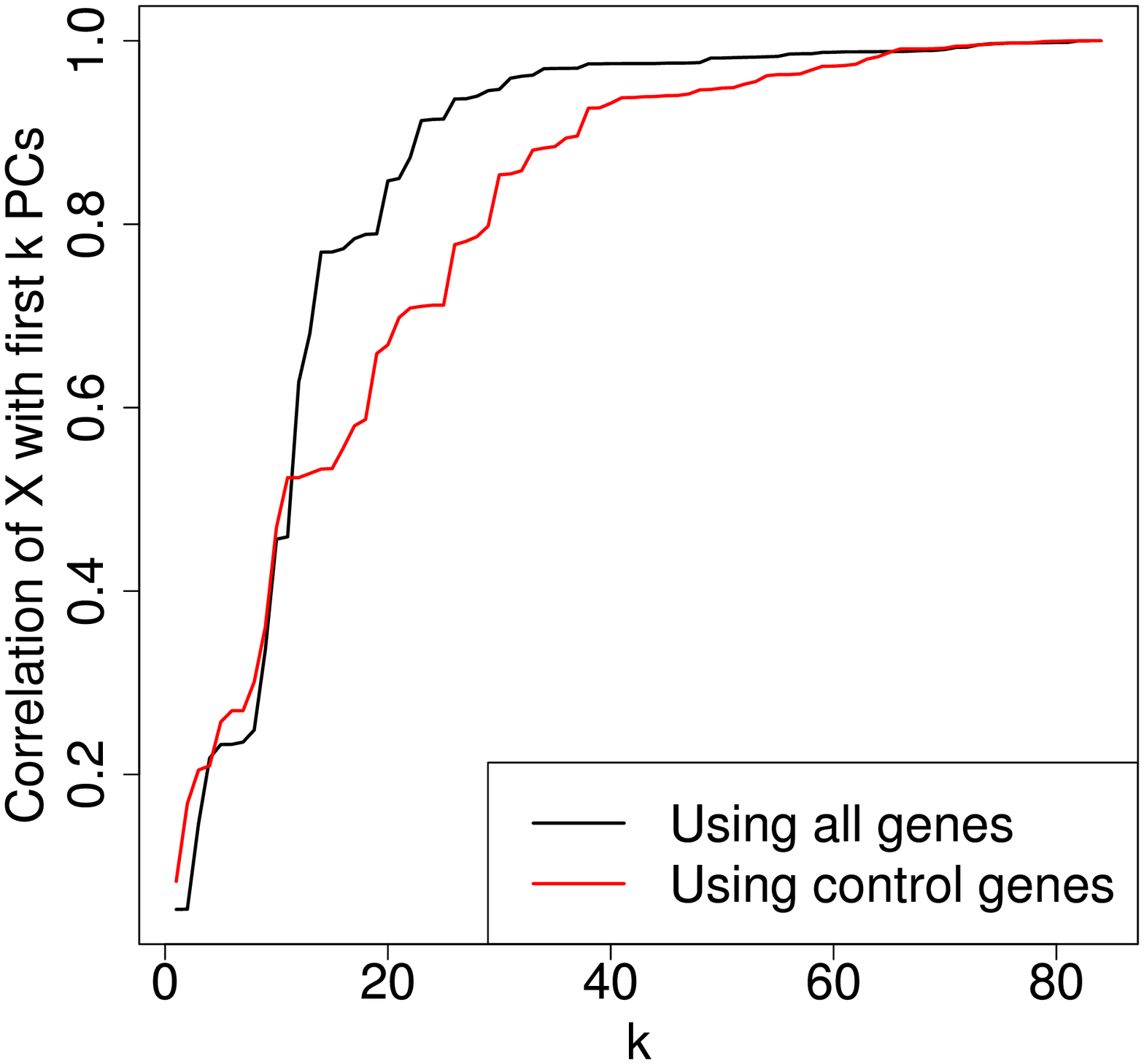}
  \includegraphics[width=.32\linewidth]{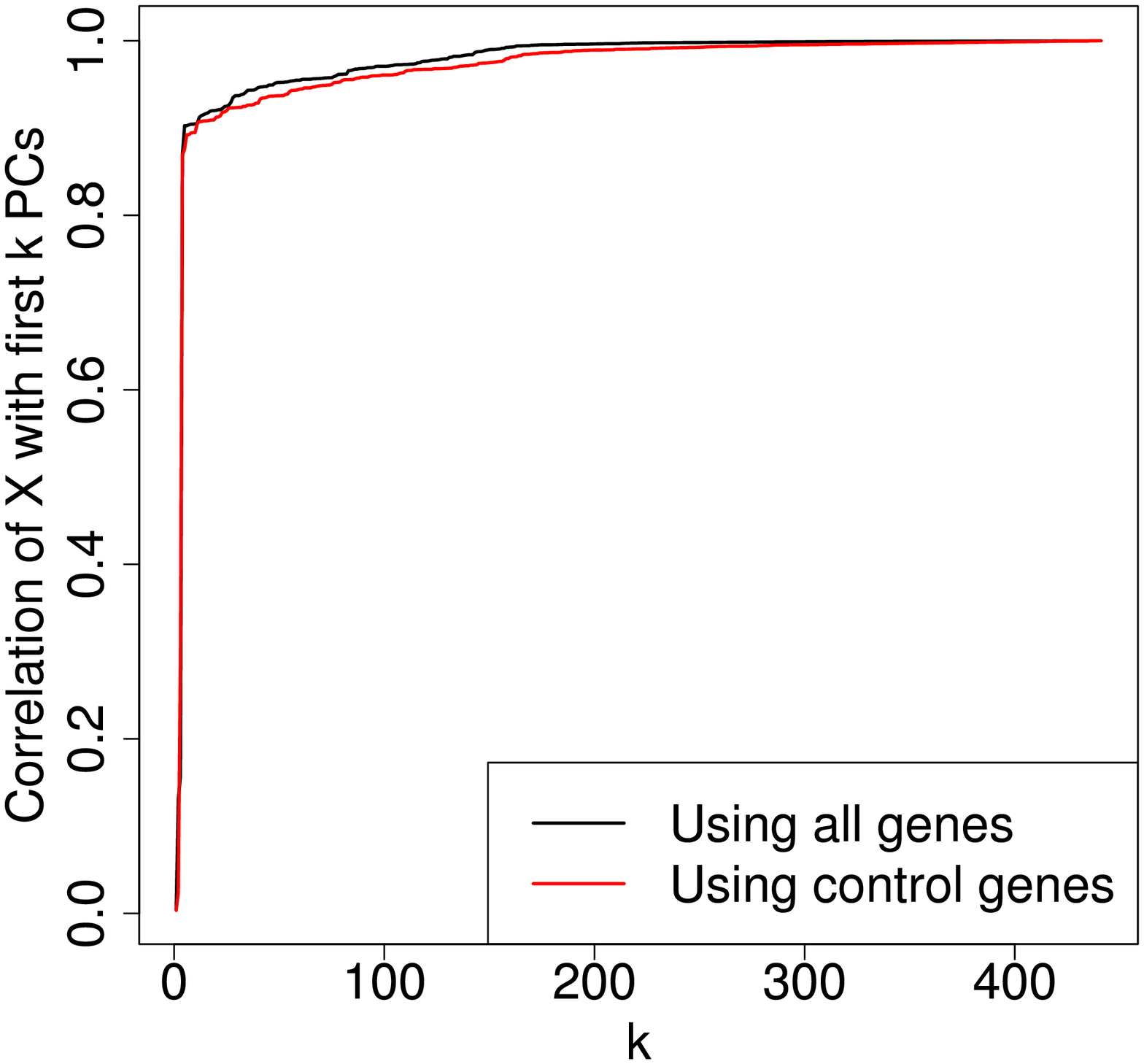}
  \includegraphics[width=.32\linewidth]{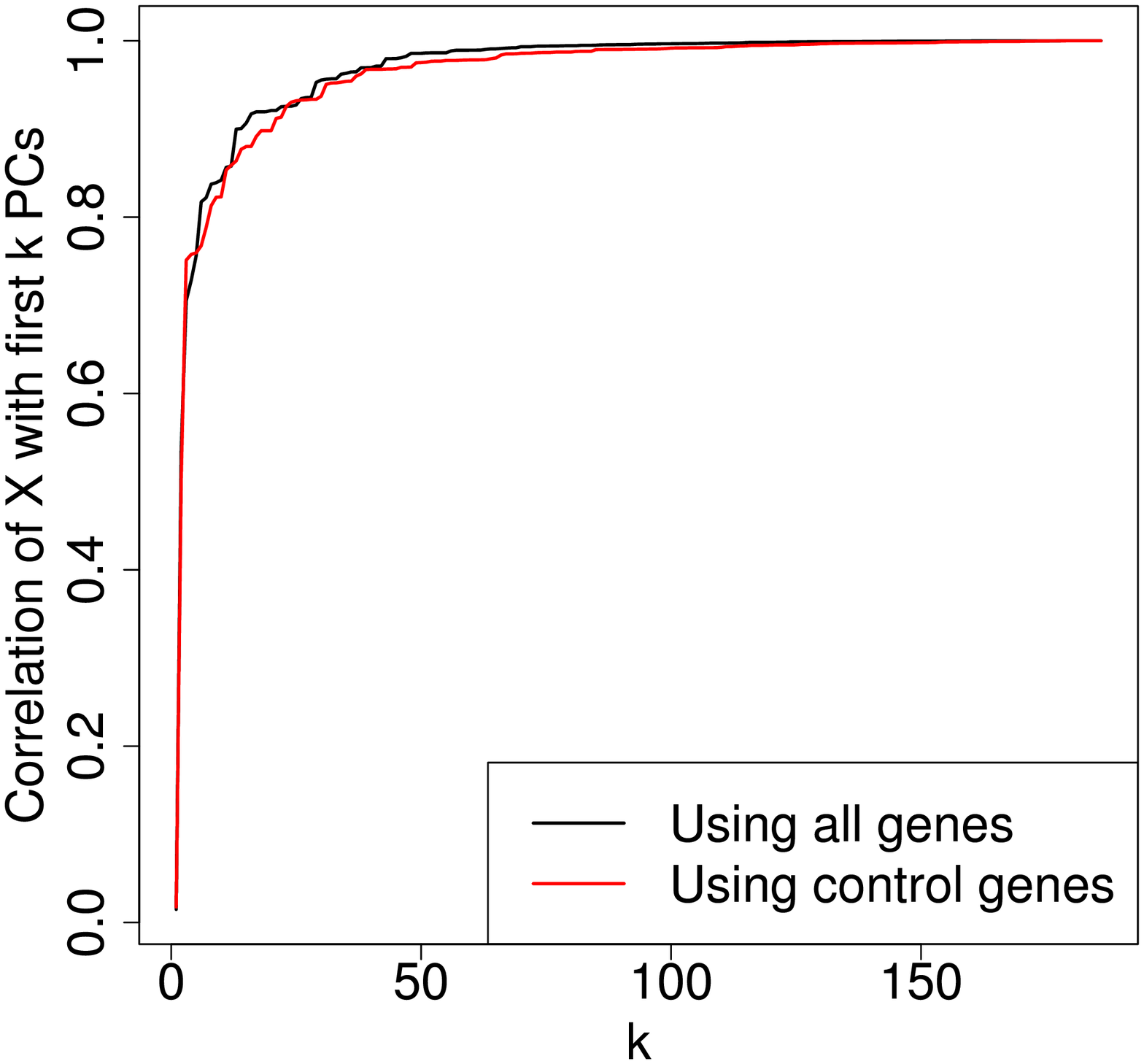}
  \caption{First canonical correlation of the factor of interest $X$
    with the space spanned by the $k$ first eigenvectors of the
    empirical covariance computed on control genes (in red) and
    non-control genes (in black) against $k$ for the gender (left
    panel), glioblastoma (center panel) and MAQC-II (right panel)
    datasets.}
  \label{fig:cgCorVsK}
\end{figure}

In order to verify the fact that the control genes used for the gender
data are good control genes whereas the ones used for the other
datasets are not good control genes, we show the CCA of all control
genes and all non-control genes with the factor of interest $X$ as a
boxplot for each dataset on Figure~\ref{fig:cgQual}. Interestingly,
the control genes used in the gender data are typically more
associated with $X$ than the non-control genes whereas the opposite is
observed for the glioblastoma and MAQC-II datasets. This seems to
contradict the fact that control genes help identifying $W$ in the
gender data and does not in the two others. Since $W$ is essentially
estimated using PCA on $Y_c$ which is a multivariate procedure, we
represent the first canonical correlation of $X$ with the eigen space
corresponding to the $k$ first eigenvectors as a function of $k$ on
Figure~\ref{fig:cgCorVsK}. It is clear from the figure that for the
gender dataset the eigen space built using control genes has a smaller
association with $X$ than the one built using non-control genes
whereas this is much less clear for the two other datasets.

To conclude, the case of gender data suggests that when good control
genes are available they do help estimate and remove unwanted
variation, especially for estimators which do not use replicate
samples. The notion of good control samples seems to have more to do
with the fact that the directions of maximum variance among these
genes are not associated with $X$ than with individual univariate
association of the genes with $X$. When control genes are as
associated as the other genes with $X$, methods using replicate
samples still give reasonable estimates and other methods become
either ineffective or very sensitive to the amplitude of the
correction.

\section{Discussion}
\label{sec:discussion}

We proposed methods to estimate and remove unobserved unwanted
variation from gene expression data when the factor of interest is
also unobserved. In particular, this allows us to recover biological
signal by clustering on datasets which are affected by various types
of batch effects. The methods we introduced generalized the ideas
of~\cite{Gagnon-Bartsch2012Using} and used two types of controls~:
negative control genes, which are assumed not to be affected by the
factor of interest and can therefore be used to estimate the structure
of unwanted variation, and replicate samples, whose differences are by
construction not affected by the factor of interest. Differences of
replicate samples can therefore also be used to estimate unwanted
variation.

On three gene expression benchmarks, we verified that when good
negative control genes were available, the correction techniques that
used these negative controls were able to successfully estimate and
remove unwanted variation, leading to the expected clustering of the
samples according to the factor of interest. When the negative control
genes are affected by the factor of interest, these techniques become
less efficient, but replicate-based methods are successful at removing
unwanted variation with respect to which replicates are
available. Correcting for unwanted variation with respect to which
replicates are given may sound less useful, because it implies that
the unwanted variation is an observed factor such as a batch or a
platform. However, even in this case, our techniques which estimate
the unwanted variation factor outperform methods which use the known
factor. This can be explained by the fact that the actual factor
affecting the samples may be a non-linear function of the observed
factor. In addition, replicates with respect to an observed unwanted
variation factor may embed unknown unwanted variation. This suggests
that when both types of controls are available, replicate based
correction is a safer option, as it removes some unwanted variation and
in the worst case scenario leaves some other unwanted variation in the
data. Negative control gene based methods can remove more unwanted
variation but run the risk to remove some signal of interest if the
control genes were actually affected by the factor of interest. While
the quality of negative control genes did affect the relative ranking
of the two families of techniques, both gave reasonable performances
on the first two benchmarks, indicating that they could be helpful on
real data for this difficult estimation problem.

For both families of methods, we also proposed iterative versions of
the techniques, which alternate between solving an unsupervised
estimation problem for a fixed estimate of the unwanted variation, and
estimating unwanted variation using the current estimate of the signal
of interest. This approach often improved the latter estimate and
never damaged the performances.

When the objective is to do clustering, the correction may benefit
from more appropriate surrogates than the sparse dictionary learning
problem. Better techniques to choose the hyperparameter $\nu$ for the
random $\alpha$ based methods could also improve the performances.

\section*{Acknowledgements}

This work was funded by the Stand Up to Cancer grant
SU2C-AACR-DT0409. The authors thank Julien Mairal, Anne Biton, Leming
Shi, Jennifer Fostel, Minjun Chen and Moshe Olshansky for helpful
discussions.

\bibliographystyle{plainnat-lj}

\clearpage
\appendix

\section{Proof of Proposition~$1$}
\label{app:prop1}

The left hand side of~\eqref{eq:prop1} is a standard ridge
regression and has a closed form solution~:
\[
\alpha^* \stackrel{\Delta}{=} \argmin_{\alpha\in\RR^{k\times n}} \left\{\|R - W\alpha\|^2_F + \nu \|\alpha\|^2_F\right\}
= \left( W^\top W + \nu I_k \right)^{-1} W^\top R,
\]

so for any $R\in\RR^{k\times n}$,

\begin{align*}
&\min_{\alpha\in\RR^{k\times n}} \left\{\|R - W\alpha\|^2_F + \nu \|\alpha\|^2_F \right\} =
\left\|\left(I_m - W\left(W^\top W + \nu
      I_k\right)^{-1}W^\top\right)R\right\|^2_F \\
&+ \nu\left\|\left( W^\top
  W + \nu I_k \right)^{-1} W^\top R\right\|^2_F\\
=& \tr R^\top \biggl(I_m - 2W\left(W^\top W + \nu I_k\right)^{-1}W^\top
+ W\left(W^\top W + \nu I_k\right)^{-1}W^\top W\left(W^\top W + \nu
  I_k\right)^{-1}W^\top \\
&+ \nu W\left(W^\top W + \nu
  I_k\right)^{-2}W^\top \biggr) R,
\end{align*}
where we used the fact that $\|A\|_F^2 = \tr A^\top A$. This is $\tr
R^\top \left(I_m + WBW^\top\right) R$ with~:
\begin{align*}
  B &\stackrel{\Delta}{=} \left(W^\top W + \nu I_k\right)^{-1}W^\top
  W\left(W^\top W + \nu I_k\right)^{-1} + \nu \left(W^\top W + \nu
    I_k\right)^{-2} -2\left(W^\top W + \nu I_k\right)^{-1}\\
  &= \left(W^\top W + \nu I_k\right)^{-1} \left(W^\top W\left(W^\top W
      + \nu I_k\right)^{-1}+\nu \left(W^\top W + \nu I_k\right)^{-1} -
    2I_k
  \right)\\
  &= \left(W^\top W + \nu I_k\right)^{-1} \left( \left(W^\top W + \nu
      I_k\right)\left(W^\top W + \nu I_k\right)^{-1} - 2I_k \right)\\
  &= -\left(W^\top W + \nu I_k\right)^{-1},
\end{align*}
so
\begin{align*}
\min_{\alpha\in\RR^{k\times n}} \left\{\|R - W\alpha\|^2_F + \nu \|\alpha\|^2_F\right\}  &=
\tr R^\top \left(I_m - W\left(W^\top W+\nu
    I_k\right)^{-1}W^\top\right) R\\
&= \|R\|^2_{S(W, \nu)}.
\end{align*}

\section{Estimator of $\beta$ for a particular $\Sigma'$ defined on
  the columns of the residuals}
\label{app:mtl}

We consider the following model~:
\begin{equation}
  \label{eq:storey}
  Y = X\beta + \tilde{\varepsilon},
\end{equation}
where $Y, \tilde{\varepsilon} \in \RR^{m\times n}$, $X\in \RR^{m\times
  p}$,
$\beta\in \RR^{p\times n}$. We further assume that each row of
$\tilde{\varepsilon}$ is distributed as $\mathcal{N}(0,\Sigma')$ where
$\Sigma'\in \RR^{n\times n}$ is a covariance matrix. Note that this is
different from~\eqref{eq:ruvRand} where the $m\times m$ covariance was
defined on the rows of $\tilde{\varepsilon}$. 

\eqref{eq:storey} is equivalent to
\begin{equation}
  \label{eq:storey2}
  Y = X\beta + W\alpha + \varepsilon,
\end{equation}
where each column $W_j$ of $W$ is such that
$W_j\stackrel{iid}{\sim}\mathcal{N}(0,\sigma_W^2I_m)$,
$\varepsilon_{ij}\stackrel{iid}{\sim}\mathcal{N}(0,\sigma^2)$ and
$\alpha\in\RR^{k\times n}$ for some $k\leq m$ is such that $\alpha^\top
\alpha + \sigma^2I_m = \Sigma'$.

Assume $k=1$ and $\alpha=\mathbf{1}^\top$, where $\mathbf{1}$ is the
all-one vector in $\RR^n$. Then $\Sigma' = \sigma^2I_n +
\mathbf{1}\mathbf{1}^\top$, \emph{i.e.} a constant matrix plus some
additional constant on the diagonal. In this special case, $W$ is a
single standard normal column vector and \eqref{eq:storey2} can be
written~:
\begin{align*}
  Y &= X\beta + W\alpha + \varepsilon\\
  &= X\beta + W\mathbf{1}^\top + \varepsilon\\
  &= X\left(\beta + (X^\top X)^{-1}X^\top W\mathbf{1}^\top\right) + \varepsilon + R,
\end{align*}
where $R$ is the projection of $W\mathbf{1}^\top$ to the orthogonal
space of $X$. We can disregard it because a noise orthogonal to $X$
has no effect on a regression against $X$. Denoting
$V\stackrel{\Delta}{=}(X^\top X)^{-1}X^\top W \mathbf{1}^\top$, we see
that \eqref{eq:storey} with this particular covariance is equivalent
to assuming $Y = Xb + \varepsilon$, where $b=\beta+V$. Each column of
$V$ is equal to $v = (X^\top X)^{-1}X^\top W$, \emph{i.e.}, to the
projection of $W$ on $X$. If $\beta$ is assumed to be non-random and
we estimate it by maximum likelihood we recover the regular OLS~:
$\beta+V$ is not identifiable. If we add a normal prior on $\beta$,
the MAP equation is~:
\begin{align}
  \label{eq:MAP}
\max_{\beta,v} L(\beta,v|X,Y) &\propto \max_{\beta,v}
L(X,Y|\beta,v)p(\beta)p(v)\\
&= \max_{\beta,v}\left\{ \log
  L(X,Y|\beta,v)+ \log p(\beta)+ \log p(v)\right\}\\
&=  \min_{\beta,v} \|Y-X(\beta+V)\|_F^2  + \lambda\|\beta\|^2_F + \nu \|v\|^2_F,
\end{align}
where $\lambda,\nu$ depend on the prior variances of $\beta$ and
$\alpha$. Then plugging $b = \beta+V$ and denoting its columns by
$b_i$,
\begin{align*}
&\min_{\beta,v}\left\{ \|Y-X(\beta+V)\|_F^2  + \lambda\|\beta\|^2_F + \nu
\|v\|^2_F\right\}\\
 &= \min_{b,v} \left\{\|Y-Xb\|_F^2  + \lambda\sum_{i=1}^n\|b_i - v \|^2_F + \nu
\|v\|^2_F\right\}\\
&= \min_{b}\left\{ \|Y-Xb\|_F^2  + \lambda\min_v\left(\sum_{i=1}^n\|b_i - v \|^2_F + \frac{\nu}{\lambda}
\|v\|^2_F\right)\right\}\\
&= \min_{b}\left\{ \|Y-Xb\|_F^2  + \lambda \sum_{i=1}^n\|b_i - \bar{b} \|^2_F + \nu
\|\bar{b}\|^2_F\right\},
\end{align*}
where $\bar{b}\stackrel{\Delta}{=}\argmin_v\left(\sum_{i=1}^n\|b_i - v \|^2_F + \frac{\nu}{\lambda}
\|v\|^2_F\right)$ is a shrinked average of the $b_i$. The first equality is
replacing $\beta+V$ by $b$ and $\beta$ by $b-V$. The second equality
is moving the $\min_v$ to the part which depends on $v$. The last
equality carries out the minimization over $v$.

Adding a block structure to $\Sigma'$, or equivalently rows to
$\alpha$ which are $1$ for some genes and $0$ for others, leads to an
additional regularizer which penalizes the sum of squares among the
$\beta_i$ within each block.

\section{PCA plots for the GBM data}
\label{app:gbmPlots}

\begin{figure}
  \centering
  \includegraphics[width=.49\linewidth]{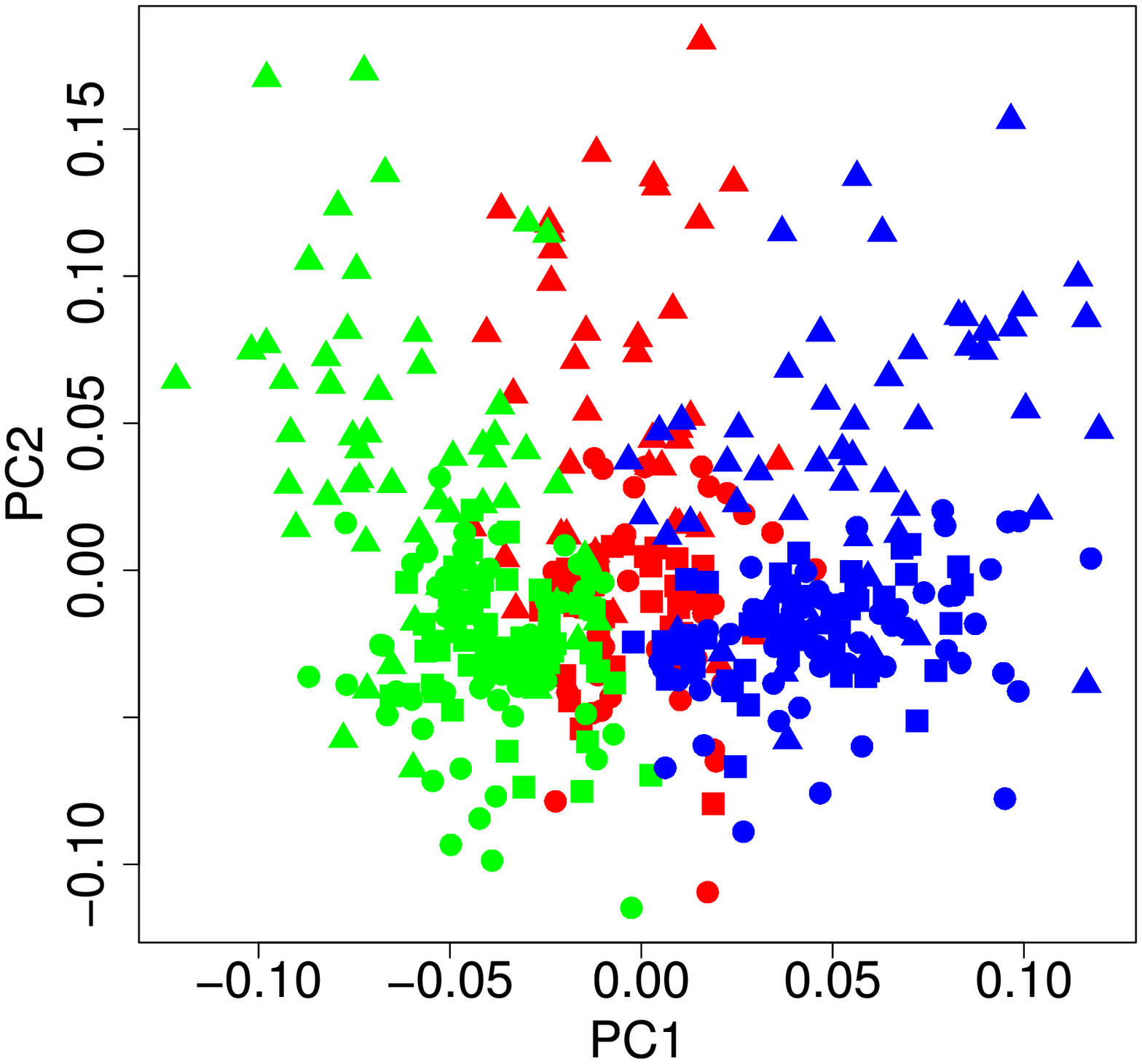}
  \includegraphics[width=.49\linewidth]{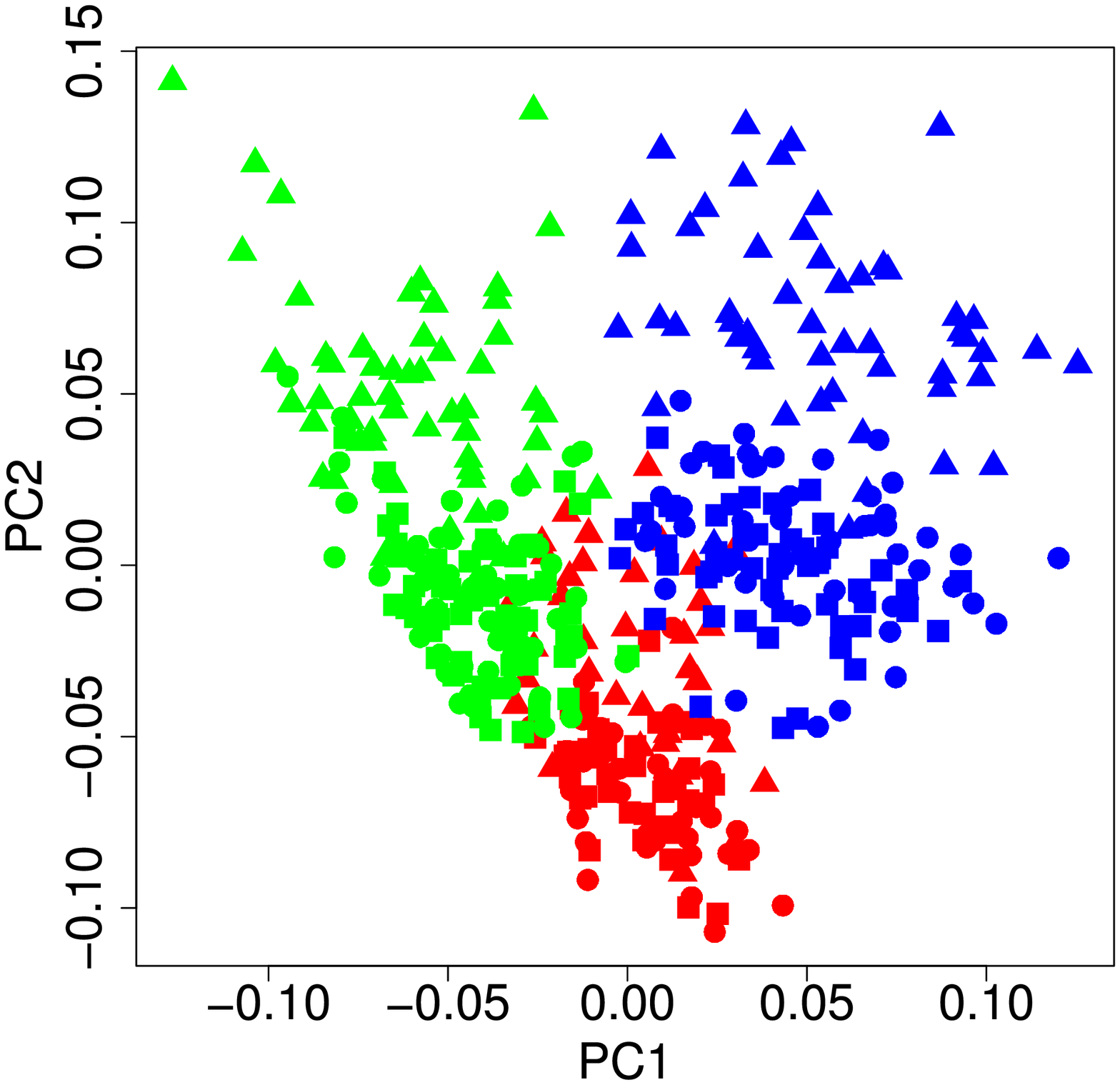}
  \caption{GBM full design random $\alpha$ with and without
    iterations. Colors represent subtypes, shapes represent platforms.}
  \label{fig:gbmRidge}
\end{figure}

\begin{figure}
  \centering
  \includegraphics[width=.49\linewidth]{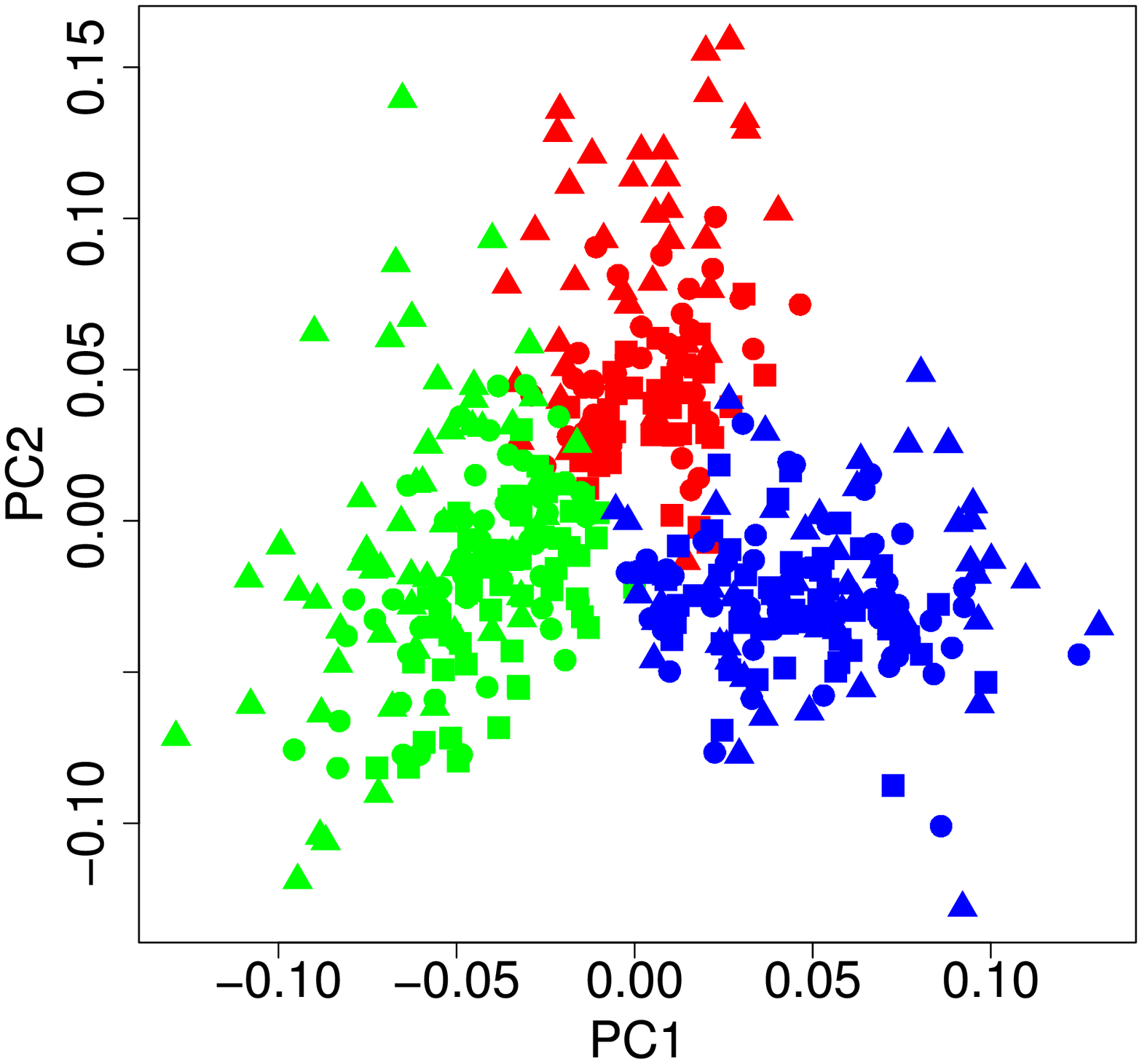}
  \includegraphics[width=.49\linewidth]{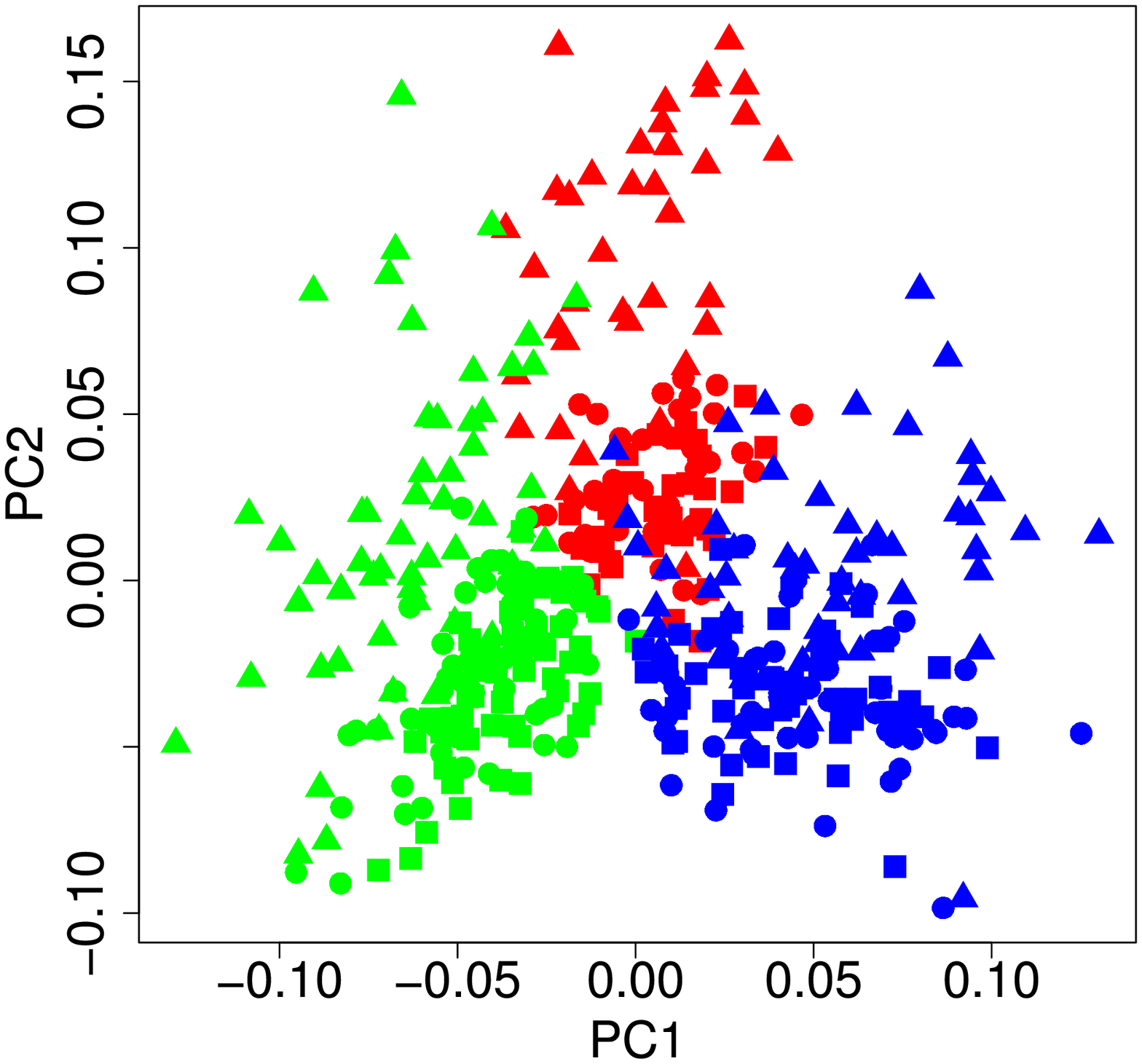}
  \caption{GBM full design replicate-based with and without
    iterations. Colors represent subtypes, shapes represent
    platforms.}
  \label{fig:gbmRep}
\end{figure}

\begin{figure}
  \centering
  \includegraphics[width=.49\linewidth]{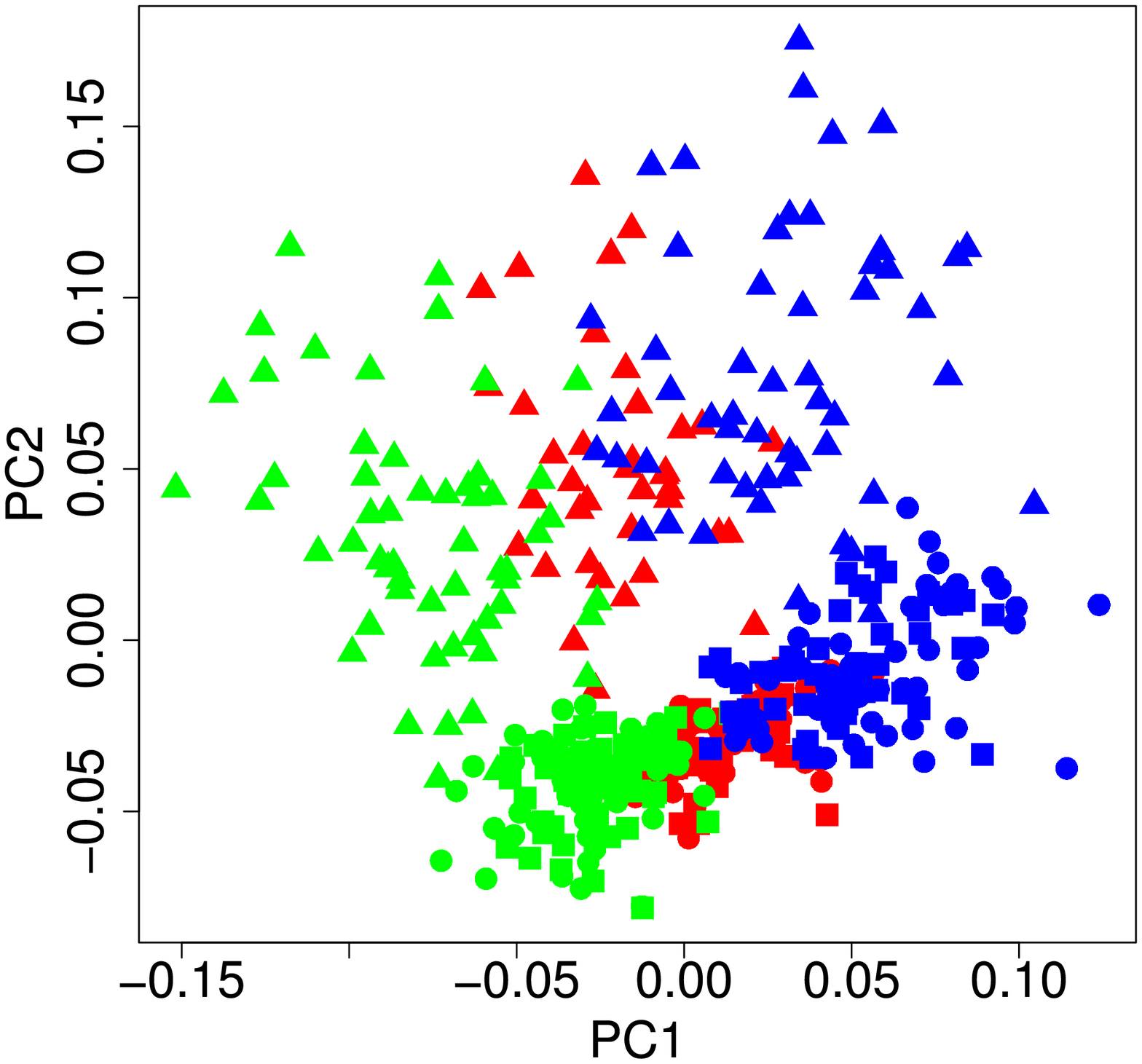}
  \includegraphics[width=.49\linewidth]{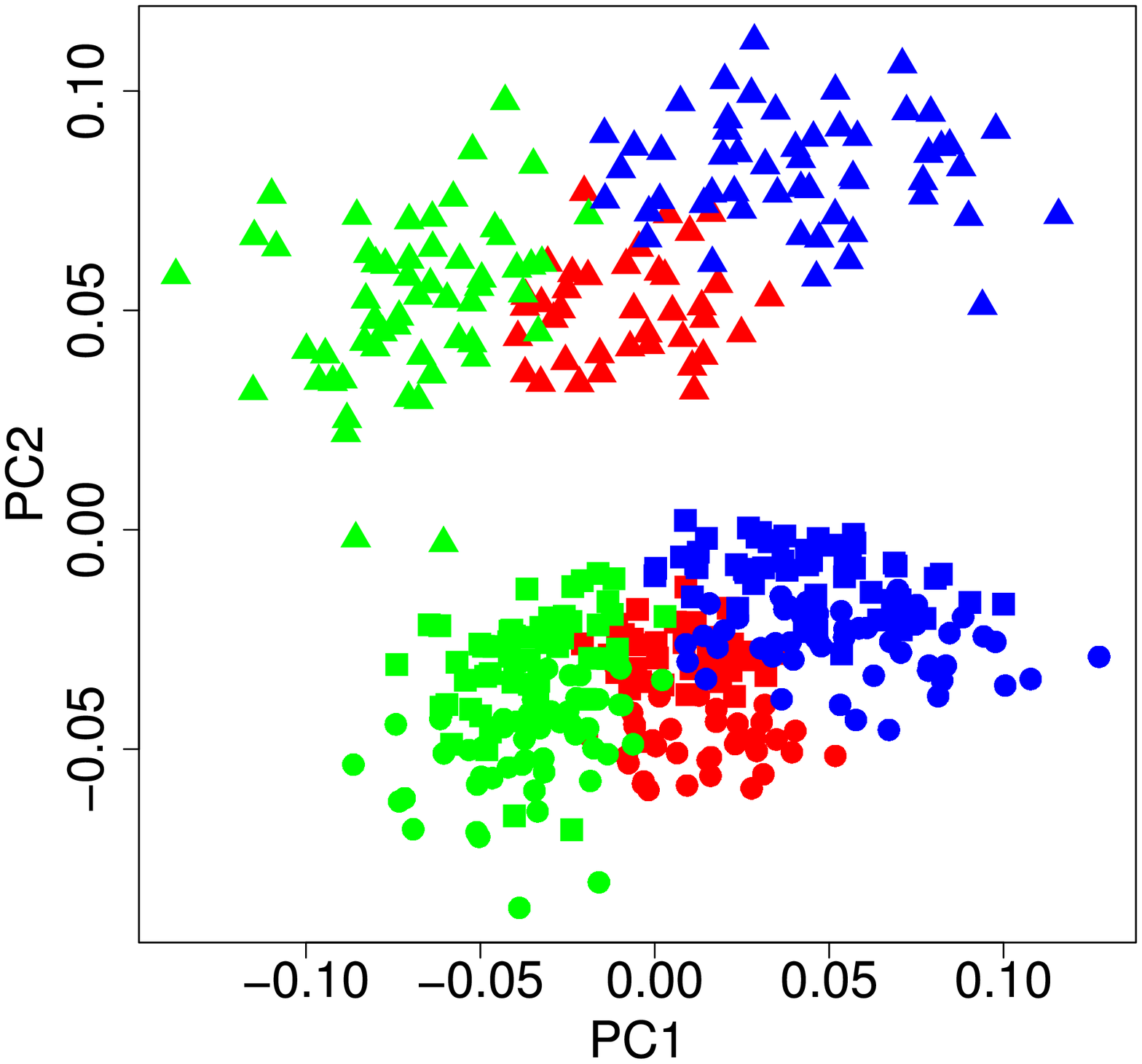}
  \caption{GBM full design combined with and without
    iterations. Colors represent subtypes, shapes represent
    platforms.}
  \label{fig:gbmComb}
\end{figure}

\begin{figure}
  \centering
  \includegraphics[width=.49\linewidth]{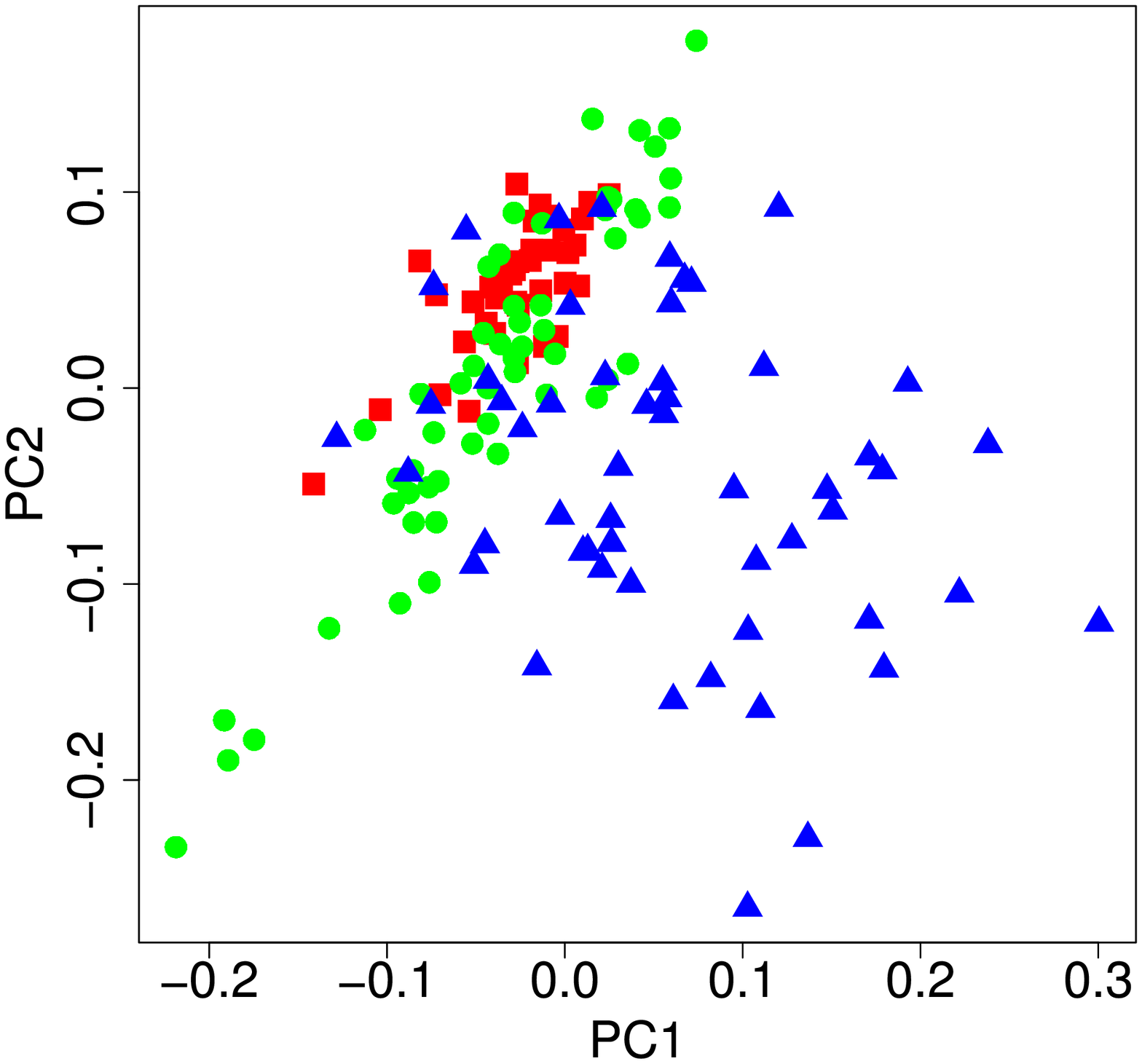}
  \includegraphics[width=.49\linewidth]{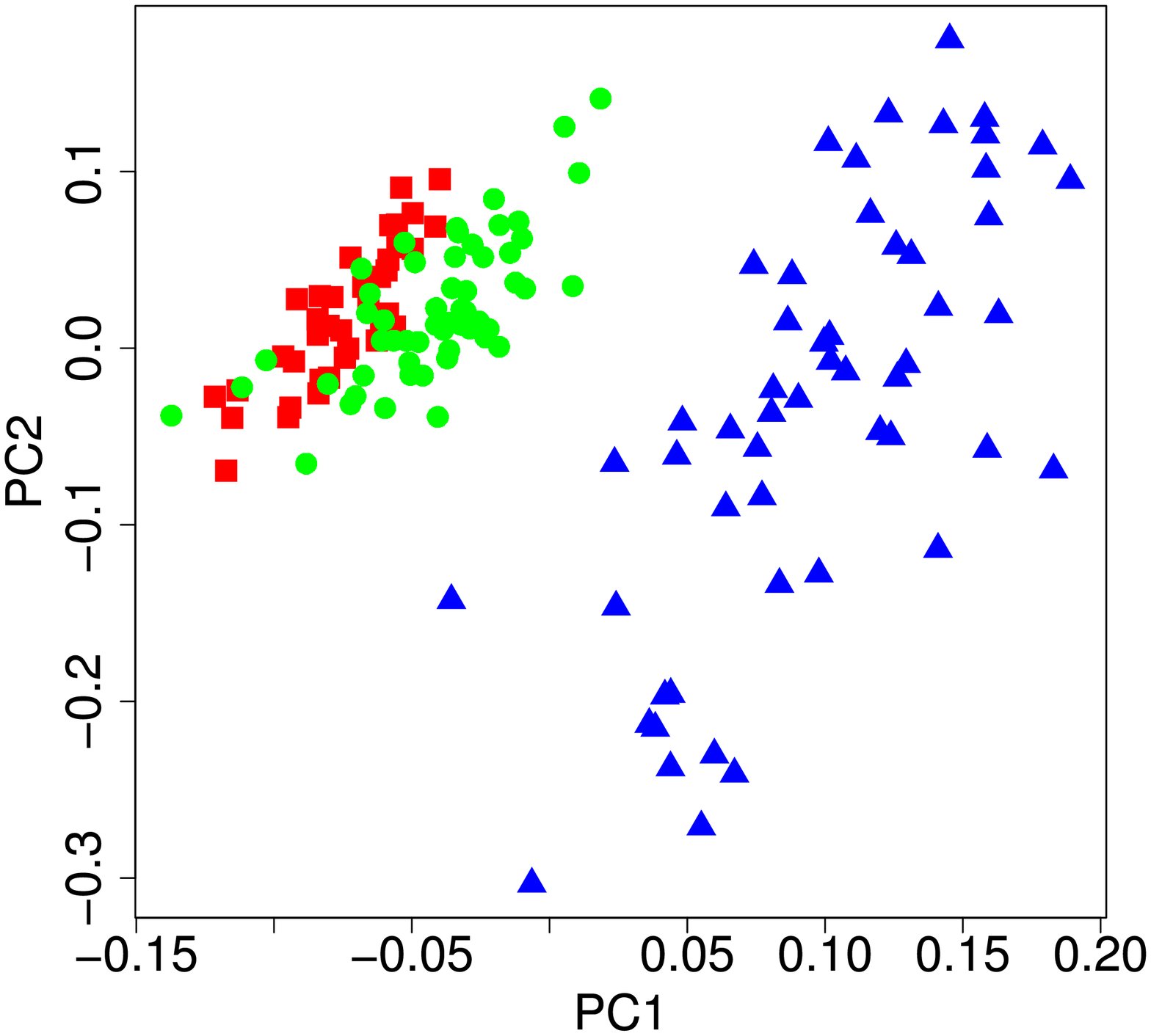}
  \caption{GBM confounded design random $\alpha$ with and without
    iterations. Colors represent subtypes, shapes represent platforms.}
  \label{fig:gbmConfRidge}
\end{figure}

\begin{figure}
  \centering
  \includegraphics[width=.49\linewidth]{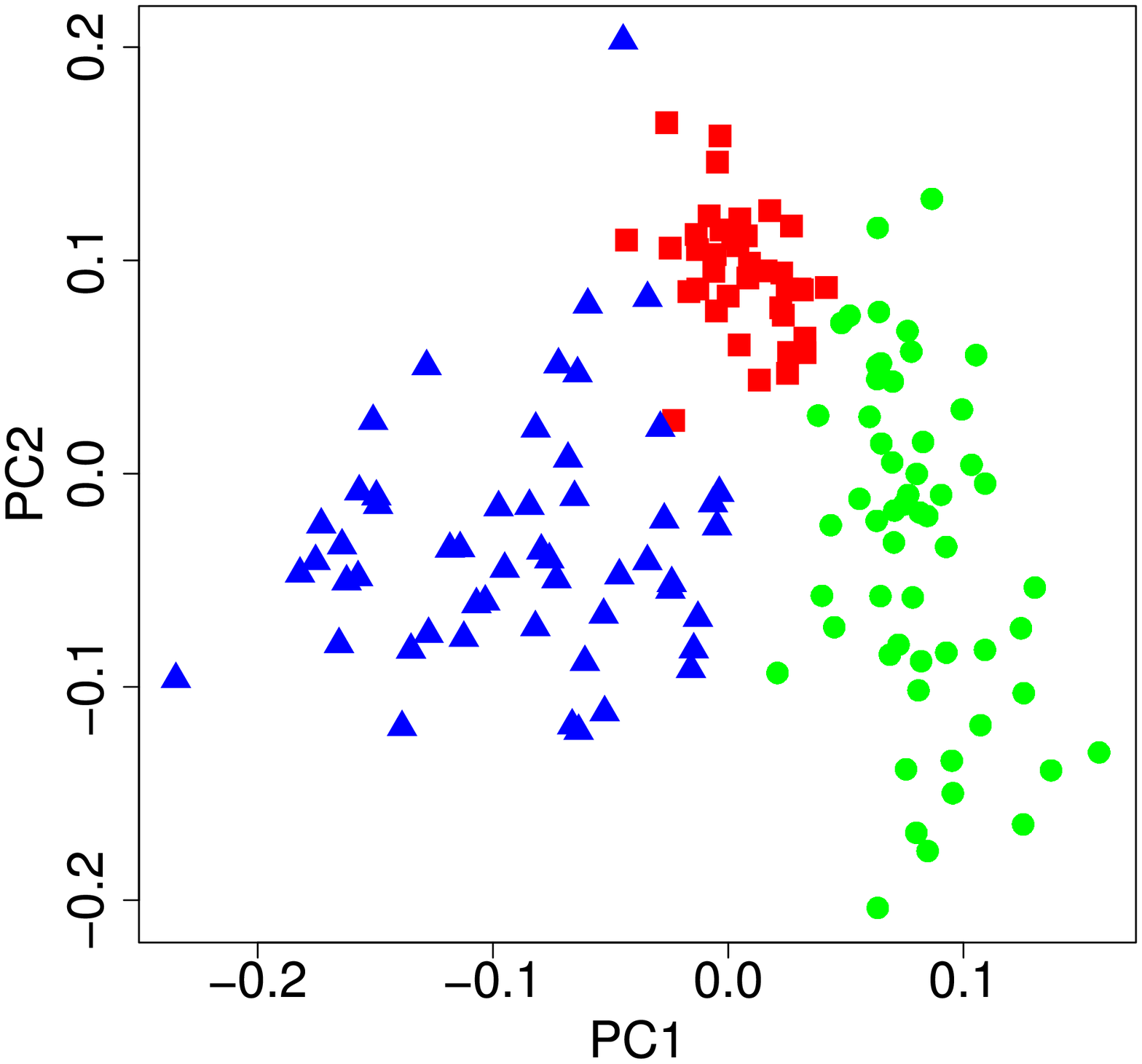}
  \includegraphics[width=.49\linewidth]{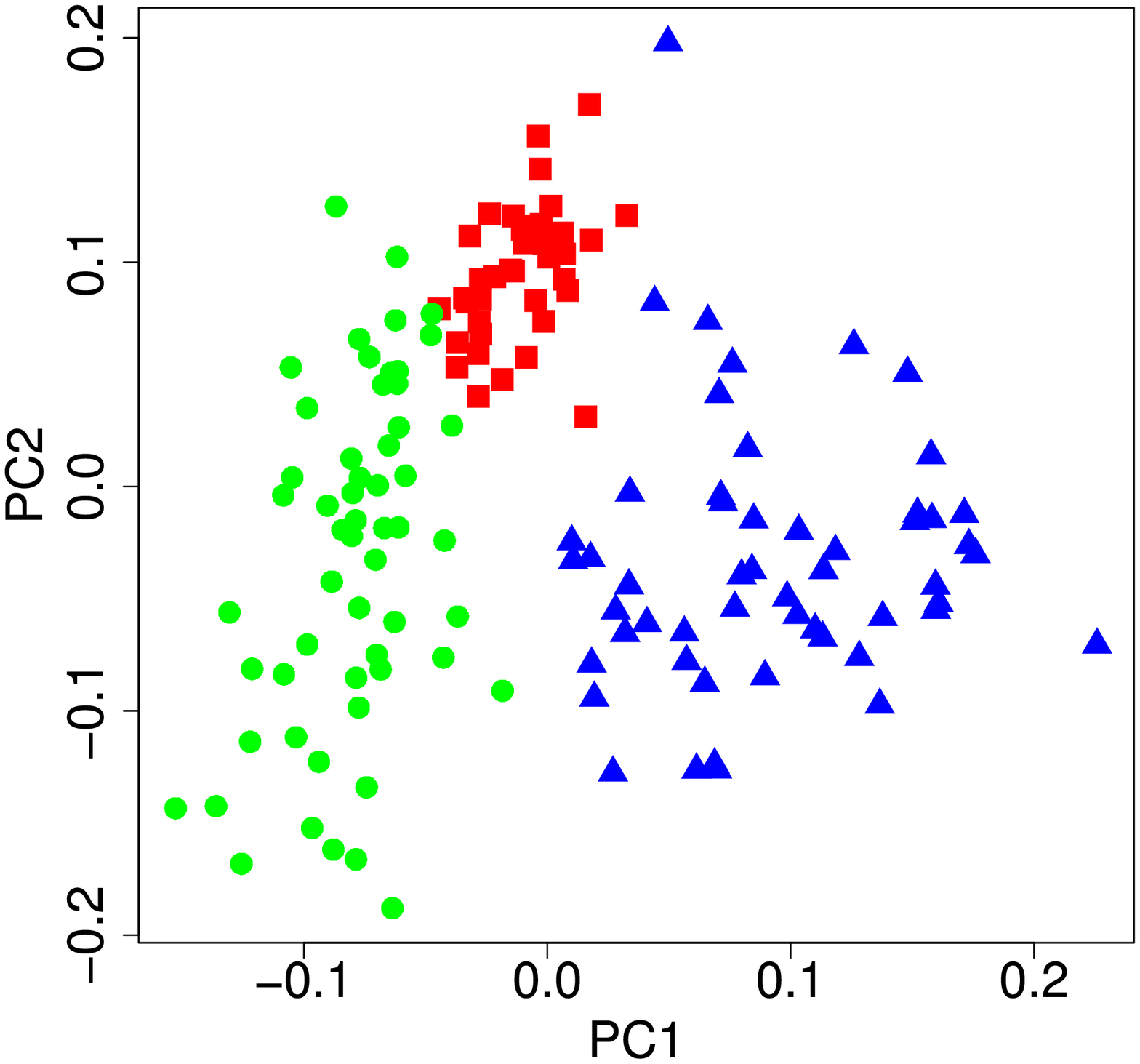}
  \caption{GBM confounded design replicate-based with and without
    iterations. Colors represent subtypes, shapes represent
    platforms.}
  \label{fig:gbmConfRep}
\end{figure}

\begin{figure}
  \centering
  \includegraphics[width=.49\linewidth]{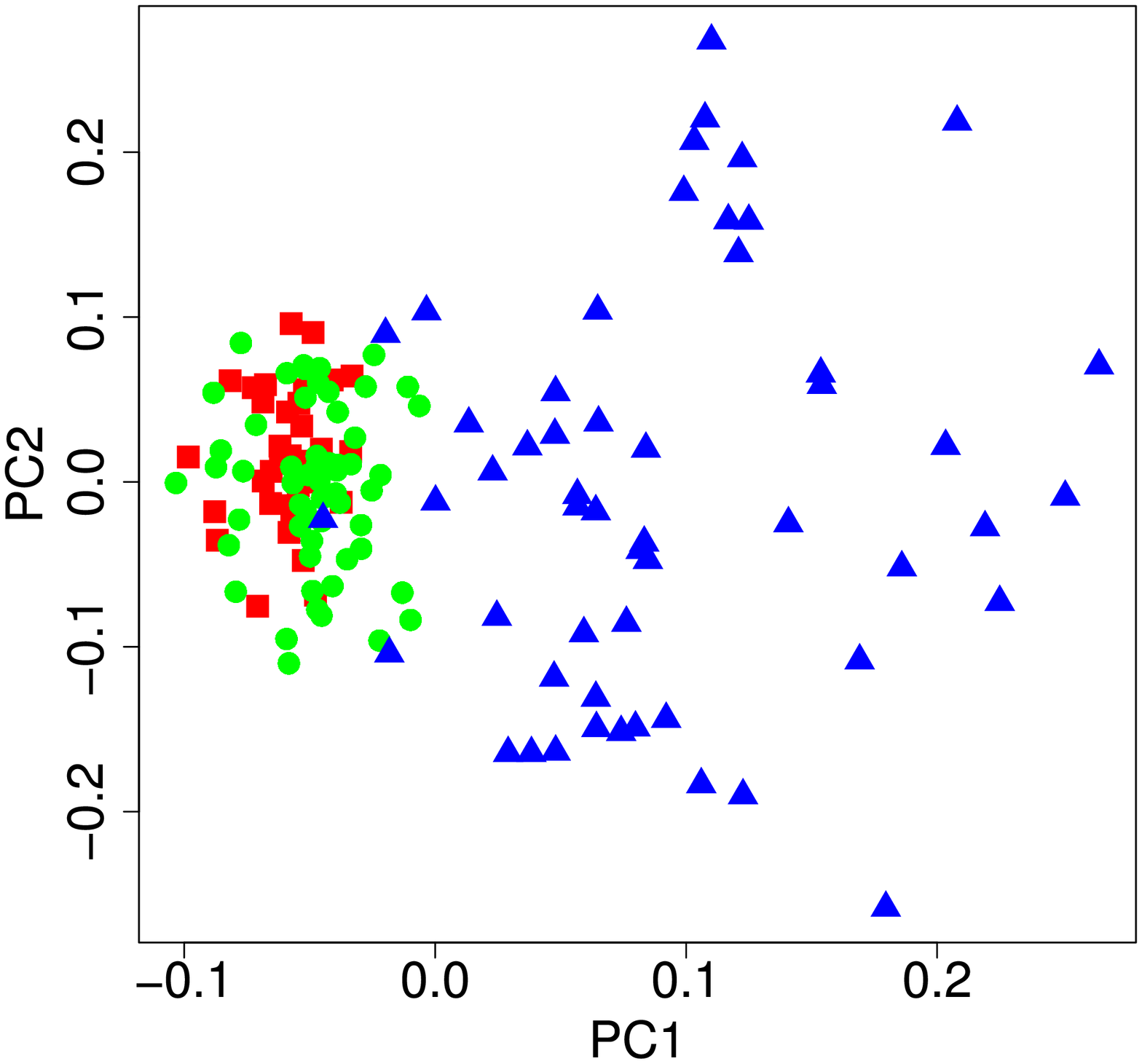}
  \includegraphics[width=.49\linewidth]{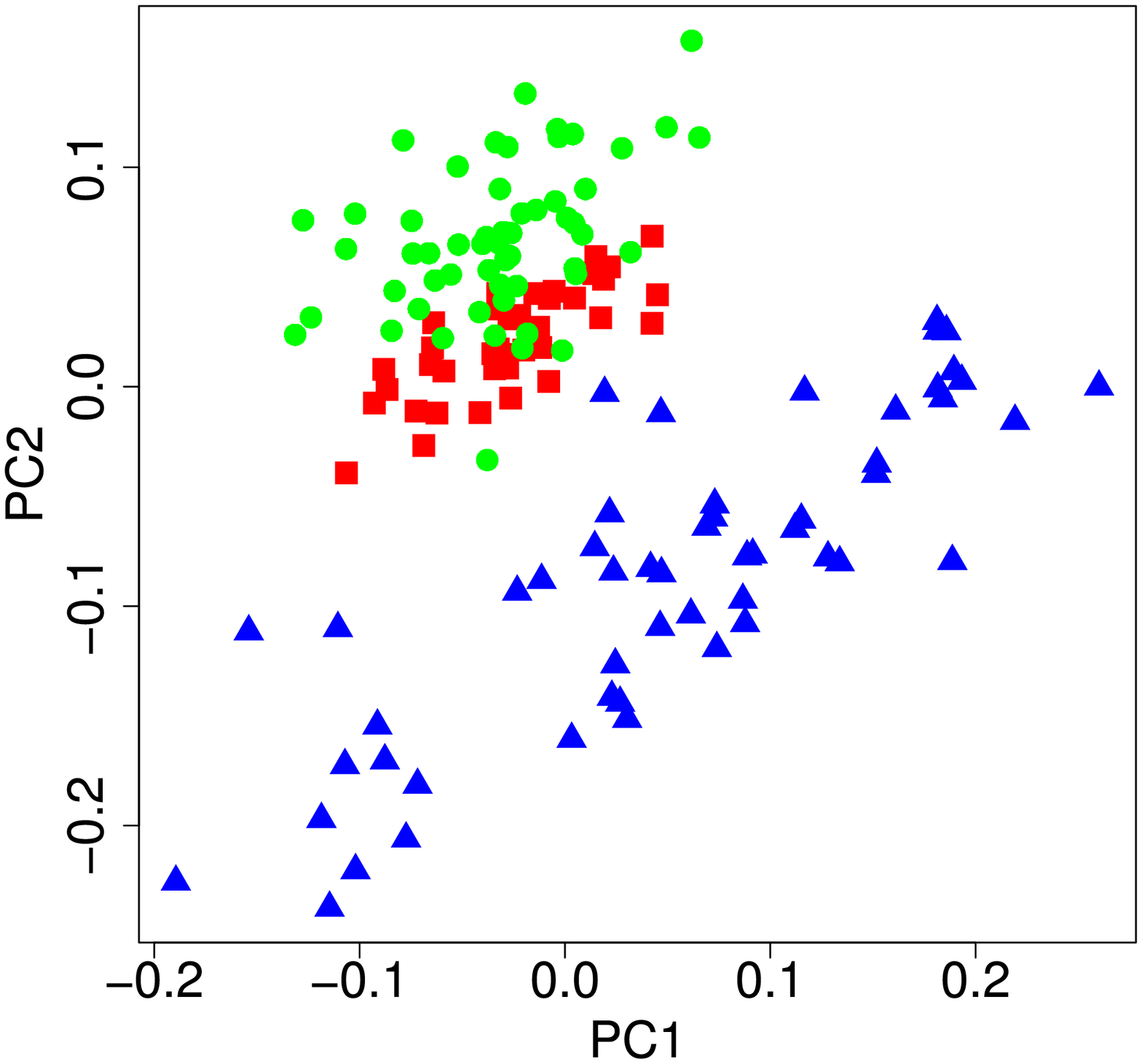}
  \caption{GBM confounded design combined with and without
    iterations. Colors represent subtypes, shapes represent
    platforms.}
  \label{fig:gbmConfComb}
\end{figure}

\end{document}